\documentclass[iop]{emulateapj}
\usepackage{epsfig, natbib}
\usepackage{mathrsfs,amssymb}
\usepackage{graphicx,latexsym}
\usepackage[]{subfigure}
\newcommand{\ang}{{\rm \AA}}

\newcommand{\cmcub}{\mbox{$\rm cm^{-3}$}}
\newcommand{\etal}{{\rm et~al.\/}}

\newcommand{\Hline}[1]{\mbox{H{\footnotesize {#1}}}}
\newcommand{\Halpha}{\Hline{\mbox{$\alpha$}}}
\newcommand{\Hbeta}{\Hline{\mbox{$\beta$}}}
\newcommand{\heii}{He{\sc ii}}
\newcommand{\hei}{He{\sc i}}
\newcommand{\HI}{{\sc \rm HI}}
\newcommand{\hi}{H{\sc i}}
\newcommand{\hii}{H{\sc ii}}
\newcommand{\iraf}{{\sl \rm IRAF\/}}

\newcommand{\Lsun}{\mbox{$\rm L_{\odot}$}}
\newcommand{\mum}{\mbox{$\mu \rm m$}}

\newcommand{\neiii}{[Ne{\sc iii}]}

\newcommand{\nii}{[N{\sc ii}]}
\newcommand{\oiii}{[O{\sc iii}]}
\newcommand{\oii}{[O{\sc ii}]}
\newcommand{\oi}{[O{\sc i}]}
\newcommand{\Qha}{\mbox{$Q_{0,\Halpha}$}}
\newcommand{\Qsed}{\mbox{$Q_{0,\rm SED}$}}

\newcommand{\Rin}{\mbox{$R_{\rm inner}$}}
\newcommand{\Rout}{\mbox{$R_{\rm nebula}$}}
\newcommand{\sii}{[S{\sc ii}]}
\newcommand{\siii}{[S{\sc iii}]}
\newcommand{\Te}{\mbox{$T_e$}}
\newcommand{\Teff}{\mbox{$T_{\rm eff}$}}

\slugcomment{*** Submitted 4 Oct 2012, resubmitted \today ***}

\shorttitle{Single-Star \hii\ Regions}
\shortauthors{Zastrow \etal}

\begin{document}

\title{Single-Star \hii\ Regions as a Probe of Massive Star Spectral Energy Distributions}

\author{Jordan \ Zastrow\altaffilmark{1},
        M.S.\ Oey\altaffilmark{1},
        E.W.\ Pellegrini\altaffilmark{2}}

\altaffiltext{1}{Astronomy Department, University of Michigan, 830 Dennison Building, 500 Church Street, Ann Arbor, MI, 48109-1042}
\altaffiltext{2}{Department of Physics and Astronomy, University of Toledo, 2801 West Bancroft Street, Toledo, OH 43606}

\begin{abstract}

The shape of the OB-star spectral energy distribution is a critical component in many diagnostics of the ISM and galaxy properties.  We use single-star HII regions from the LMC to quantitatively examine the ionizing SEDs from widely available CoStar, TLUSTY, and WM-basic atmosphere grids.  We evaluate the stellar atmosphere models by matching the emission-line spectra that they predict from CLOUDY photoionization simulations with those observed from the nebulae.  The atmosphere models are able to reproduce the observed optical nebular line ratios, except at the highest energy transitions $\geq 40$\ eV, assuming that the gas distribution is non-uniform.  Overall we find that simulations using WM-basic produce the best agreement with the observed line ratios. The rate of ionizing photons produced by the model SEDs is consistent with the rate derived from the \Halpha\ luminosity for standard, log($g$) = 4.0 models adopted from the atmosphere grids.  However, there is a systematic offset between $Q_0$ from different atmosphere models that is correlated with the relative hardness of the SEDs.  In general WM-basic and TLUSTY atmosphere models predict similar \Teff, while CoStar predicts \Teff\ that are cooler by a few thousand degrees. We compare our effective temperatures, which depend on the nebular ionization balance, to conventional photospheric-based calibrations from the literature.  We suggest that in the future, SpT-\Teff\ calibrations can be constructed from nebular data.

\end{abstract}

\section{Introduction} \label{s:intro}

Through the effects of radiative, mechanical and chemical feedback, massive stars play a critical role in shaping the galaxies in which they reside.  These stars ionize their local interstellar medium (ISM) and form bright \hii\ regions that can be observed in distant galaxies, even where the individual stars are not resolved.  Since the \hii\ region emission-line spectrum is sensitive to both the spectral energy distribution (SED) of the ionizing source and the properties of the nearby gas, they are used as diagnostics for galaxy properties, such as star formation rates and histories \citep[e.g.,][]{b:Leitherer_apjs99,b:Hunter_aj90,b:Kennicutt_apj00}, properties of the ionizing stellar population and the slope of the IMF \citep[e.g.,][]{b:Baldwin_pasp81,b:Kaler_apj78,b:Rigby_apj04,b:Stasinska_apjs96,b:Copetti_aap86,b:Dufour_apj75}, and the chemical abundances and chemical evolution of galaxies \citep[e.g.,][]{b:Kewley_ApJS02,b:Edmunds_mnras84,b:Bresolin_apj99}. 

The shape of the stellar SED is particularly important for these diagnostics. For example, it determines the rate of ionizing photons ($Q_0$) emitted by massive stars, upon which the commonly used \Halpha\ star formation rate indicator depends \citep{b:Kennicutt_apj83}.  Other popular diagnostics use flux ratios of lines with different ionization potential as diagnostics for the effective temperature (\Teff) of the ionizing stars \citep[e.g.,][]{b:Vilchez_mnras88,b:Stoy_mnras33,b:Zanstra_apj27}.  Thus, the shapes of massive star SEDs play key roles in deriving the physical properties of nebulae and stars from observed \hii\ region spectra.

Since massive stars radiate most of their flux at FUV wavelengths that are relatively inaccessible to observations, we are dependent on the predictions generated by stellar atmosphere models to describe the properties and SEDs of massive stars.  To reproduce the stellar SED, the models need to take into account the effects of non-LTE conditions, stellar winds, and line-blanketing \citep[e.g.,][]{b:Kudritzki_araa90,b:Schaerer_aap94,b:Stasinska_aap97,b:Lanz_apjs03,b:Pauldrach_aap01}.  Calculating  these in detail is both challenging and computationally time-consuming.  Therefore, the model atmospheres currently available incorporate these processes with different approximations, by balancing an exact treatment of the physics against faster computation.  The different treatments affect the shape of the SED and the properties derived from them \citep[e.g.,][]{b:Simon-Diaz_mnras08,b:Voges_aj08,b:Giveon_apj02}.  It is therefore crucial to know how well these different atmosphere models represent the true SED of these stars.

\hii\ regions can be used as a test of these atmosphere models.  As mentioned earlier, \hii\ region emission-line spectra strongly depend on the shape of the ionizing SED.  Thus, a direct comparison between the emission lines from observed \hii\ regions and those predicted by photoionization simulations will reveal how well the different atmosphere models represent the SEDs of massive stars \citep[e.g.,][]{b:Oey_apjs00,b:Morisset_aap04,b:Esteban_aap93,b:Crowther_aap99,b:Stasinska_aap97}.  For example, previous studies of both Galactic and Magellanic Clouds HII regions have shown that plane-parallel atmospheres are too soft and that line blanketing is a crucial process to include in the models \citep[e.g.,][]{b:Martin-Hernandez_aap02, b:Morisset_aap02, b:Giveon_apj02, b:Bresolin_apj99}. This approach, however, is not always straightforward.  To make this comparison, the metallicity and morphology of the gas in the nebula must be constrained.  Furthermore, most studies use nebulae that contain many stars that are distributed throughout the ionized region.   Properly accounting for these multiple ionizing sources is a major challenge to this approach and vastly limits the constraints that one can put on the SEDs \citep[e.g.,][]{b:Ercolano_mnras07}. Our study circumvents this challenge by using single-star \hii\ regions in the nearby Large Magellanic Cloud (LMC).  The \hii\ regions are spatially resolved and spherical, which makes the modeling more straightforward.  We use photoionization simulations to evaluate how well atmosphere models represent the shape of the massive star SED.   

\subsection{Description of the model atmosphere codes}\label{s:moddesc}

Our goal in this work is to understand how well the OB atmosphere-model grids that are available to general users represent ionizing stars.  For this reason, we use publicly available grids, rather than fitting each stellar spectrum in detail to obtain the appropriate model.  We use the O-star grid presented in \citet{b:Smith_mnras02}, hereafter SNC02, in addition to the CoStar \citep{b:Schaerer_aap97}, TLUSTY \citep{b:Lanz_apjs03}, and WM-basic \citep{b:Pauldrach_aap01} grids that are already available in the stellar atmosphere library of CLOUDY \citep{b:Ferland_pasp98}. These massive-star atmosphere codes include important, but complex, physical processes such as non-LTE conditions and the effects of metal lines and winds on the transmitted spectrum.  Including these processes in detail is computationally expensive.  Therefore, research groups use various methods and algorithms to approximate some of these processes.  The differences in the algorithms used result in non-negligible differences in the SEDs produced by different atmosphere codes \citep[e.g.,][]{b:Simon-Diaz_mnras08}.  

All the atmosphere codes considered in this work solve for non-LTE radiation transfer. Typically this is accomplished by grouping lines of similar excitation energies together and applying the same non-LTE correction to the populations in that group \citep[e.g.,][]{b:Hubeny_apj95,b:Pauldrach_aap01}.  In some atmosphere codes, such as CoStar and WM-basic, the non-LTE solution includes the effects of spherically expanding winds, while others, such as TLUSTY, assume a plane-parallel geometry.  

Metal lines in the UV impact the emergent SED in two ways, line blocking and line blanketing.  Line blocking refers to the absorption and scattering of the emergent flux due to the higher opacity in the line.  Line blocking will increase the temperature in the deeper layers of the star because of the scattering of radiation back towards the star, an effect known as backwarming \citep{b:Pauldrach_aap01}.  Line blanketing refers to the redistribution of energy to regions where the metal lines are not so densely packed \citep{b:Pauldrach_aap01}.  The main method used to include line blocking and blanketing is the opacity sampling method. For this method, the opacities are evaluated for a grid of frequency points.  The approximation approaches the exact solution as the code increases the number of frequency points it samples.  CoStar incorporates the opacity sampling using $\sim50$\ \ang-wide bands in the Monte Carlo radiative transfer solution \citep{b:Schaerer_aap94}.  Of the three codes considered here, CoStar has the most approximate treatment. The TLUSTY atmosphere code includes 180,000-200,000 frequency points in the opacity sampling grid \citep{b:Lanz_apjs03}. The WM-basic atmosphere code is solved in two parts. First, the radiative transfer is solved with a fast approximate treatment that samples $\sim 1,000$ frequency points in the Lyman continuum.  This fast solution is repeated iteratively and generates starting values for the final solution.  The final solution consists of fewer iterations that solve the radiative transfer with an exact treatment of the line-blanketing \citep{b:Pauldrach_aap01}.

In addition to line-blanketing by metal lines, stellar winds change the shape of the ionizing SED at high energies \citep[e.g.,][]{b:Sellmaier_aap96}.  This is particularly important near the  54.4eV ionization potential of \heii\ \citep{b:Gabler_aap89}.  Both CoStar and WM-basic calculate the non-LTE solution for expanding atmospheres.  In the WM-basic atmosphere code, the standard opacity sampling is modified to take into account line shifts due to the expanding winds.  These line shifts effectively increase the frequency range that can be blocked by a given line \citep{b:Pauldrach_aap01}.

While SNC02 is generated with the WM-basic code, they use a different set of assumptions for the stellar parameters in calculating the SEDs. Specifically, they determine the appropriate mass-loss rate and terminal velocities a priori using the empirical relations from \citet{b:Prinja_apj90}, \citet{b:Lamers_apj95} and \citet{b:Kudritzki_araa00}.  To ensure the final atmospheres have these parameters, they manually adjust the radiative acceleration by changing the force multipliers used in the code \citep{b:Smith_mnras02}. In contrast, the WM-basic grid included in CLOUDY obtains the the mass-loss rate and wind terminal velocity from the solution of the atmosphere code itself \citep{b:Pauldrach_98}.  We note that the SNC02 grid was previously implemented in STARBURST99 \citep{b:Leitherer_apjs99}.


\section{Observations and Method} \label{s:method}

\subsection{Observations}\label{s:obs}

{
\begin{deluxetable*}{lllcccccccc}
\tablewidth{0pt}
\tabletypesize{\footnotesize}
\tablecaption{Observed Nebular and Stellar Properties \label{t:obsprops}}  
\tablehead{\colhead{MCELS} & 
           \colhead{Other ID$^a$} &
           \colhead{Stellar ID$^c$} &
  	       \colhead{R.A.$^e$} &
	       \colhead{Dec$^e$} &
	       \colhead{SpT} &
	       \colhead{$V^f$} &
	       \colhead{$M_{\rm V}$} &
	       \colhead{\Rout} &
	       \colhead{error} &
	       \colhead{log($L_{\Halpha}$)}\\
	       \colhead{} & 
	       \colhead{} & \colhead{} & \colhead{(J2000)} & \colhead{(J2000)} &
	       \colhead{} & \colhead{mag} & \colhead{mag} & \colhead{pc} & \colhead{pc} & \colhead{} }
\startdata
L 28&DEM L 08c&8225&04:52:11.4&-66:54:29&O5.5 V&14.75&-4.28&6.7&1.0&36.75\\ 
L 32&N4c$^b$ &8229&04:52:23.3&-66:55:16&B0 V&15.44&-3.54&3.9&1.0&35.70\\
L 35&\nodata&8203&04:52:35.2&-66:55:42&B0 Ib&14.16&-4.69&6.5&1.0&35.99\\ 
L 43&DEM L 20 &17251$^d$&04:53:30.7&-67:23:21&O8 V&14.85&-3.72&6.1&1.0&36.26\\ 
L 52&DEM L 26 &19696$^d$&04:54:12.0&-68:21:53&O6.5 V&14.07&-4.53&8.5&1.3&36.70\\ 
L 344&DEM L 276&50092&05:40:08.5&-71:11:02&O7--6.5 V((f))&13.98&-5.03&9.2&1.3& 36.82\\ 
L 345&DEM L 278&50093&05:40:09.5&-71:12:27&B0.5 Iab&13.15&-6.30&5.1&1.0&36.27 \\
L 346&DEM L 275&45830&05:40:11.8&-69:55:01&O9 V$^g$&14.69&-4.16&4.8&0.7&35.97\\
L 351&DEM L 281&43846&05:40:43.1&-70:02:30&O6.5 V&14.13&-5.51&11.3&1.6&37.17\\ 
L 390&DEM L 320&28307&05:48:02.0&-69:53:53&O9 V &14.46&-4.78&5.1&0.8&36.45\\ 
L 394&DEM L 324&28329&05:48:43.6&-69:50:39&O9 V &14.68&-4.17&6.5&1.0&35.79\\ 
\nodata&DEM L 283b&44979&05:40:48.1&-69:43:18&O6.5 V((f))&15.16$^h$&-5.19&5.5&1.0&36.30
\enddata
\tablenotetext{a}{Unless otherwise noted, ID from \citet{b:Davies_memras76}} 
\tablenotetext{b}{ID from \citet{b:Henize_apjs56}}
\tablenotetext{c}{Unless otherwise noted, ID from OGLE-III \citep{b:Udalski_actaa08}}
\tablenotetext{d}{ID from \citet{b:Massey_apjs02}}
\tablenotetext{e}{R.A. and Dec correspond to that of the ionizing star.}
\tablenotetext{f}{Measurement error on \emph{V} is $~0.05$ mag and the systematic error is discussed in the text.}
\tablenotetext{g}{SpT is inferred but not directly observed, see \S\ref{s:indob}.} 
\tablenotetext{h}{Magnitude from the OGLE-III \citep{b:Udalski_actaa08}}
\end{deluxetable*}}

To evaluate the atmosphere grids described above, we first construct a sample of single-star \hii\ regions.  We use the narrow-band emission-line images from the Magellanic Clouds Emission Line Survey \citep[MCELS;][]{b:Smith_05} to select small, spherical \hii\ regions that are likely ionized by a single star.  Table \ref{t:obsprops} lists the properties of these \hii\ regions.  The first five columns give the MCELS \citep{b:Pellegrini_apj12}, DEM \citep{b:Davies_memras76} and stellar designations, and the (J2000) positions. The next three columns list observed properties of the ionizing stars.  The ninth column has the observed radius of the \hii\ region in parsecs as measured from the \Halpha\ MCELS image.  Figure \ref{mcelsim} shows both \Halpha\ and three-color composites in \oiii, \sii, and \Halpha\ of the \hii\ regions in our sample.

{
\begin{figure*}[ht]
\centering
\subfigure[MCELS~L~28, MCELS~L~32, MCELS~L~35]{
\includegraphics[width=3.3cm,angle=270]{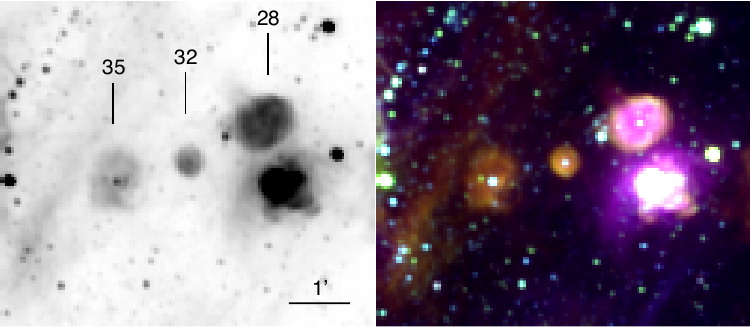}
\label{im456}}
\subfigure[MCELS~L~351]{
\includegraphics[width=3.3cm,angle=270]{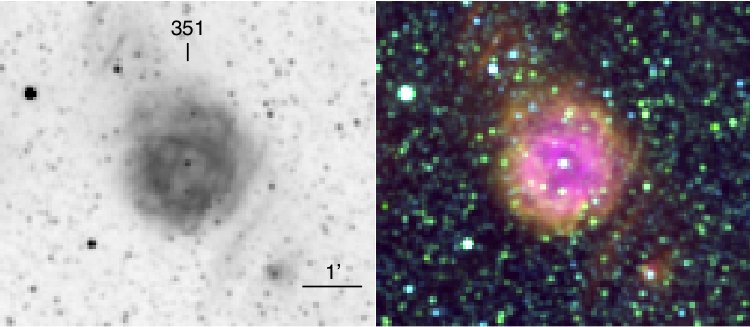}
\label{im25}} \\
\subfigure[DEM~L~283b]{
\includegraphics[width=3.3cm,angle=270]{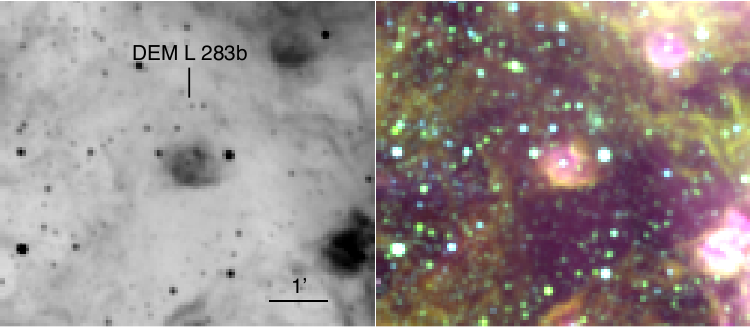}
\label{im34}}
\subfigure[MCELS~L~344, MCELS~L~345]{
\includegraphics[width=3.3cm,angle=270]{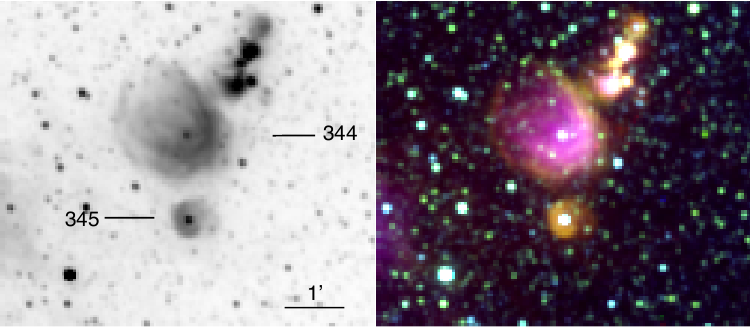}
\label{im3133}} \\
\subfigure[MCELS~L~43]{
\includegraphics[width=3.3cm,angle=270]{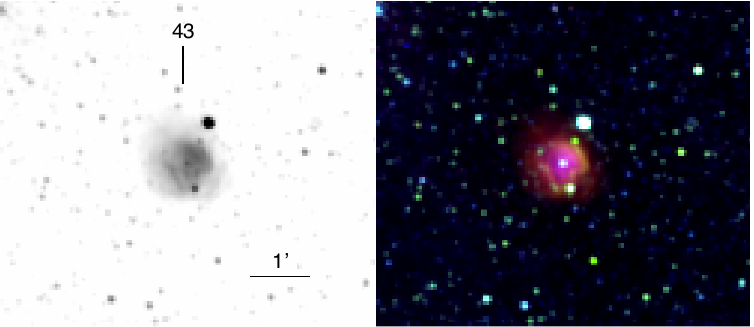} 
\label{im7}}
\subfigure[MCELS~L~52]{
\includegraphics[width=3.3cm,angle=270]{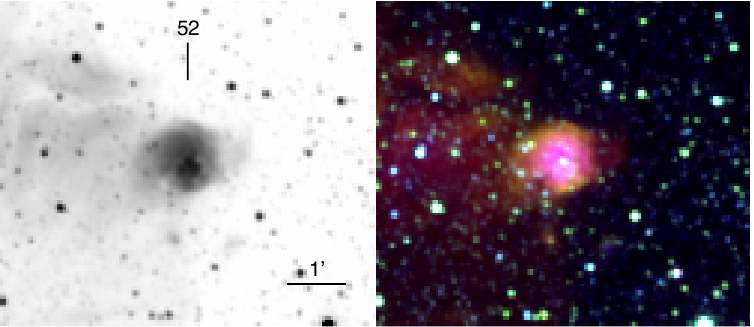}
\label{im9}} \\
\subfigure[MCELS~L~394, MCELS~L~390]{
\includegraphics[width=3.3cm,angle=270]{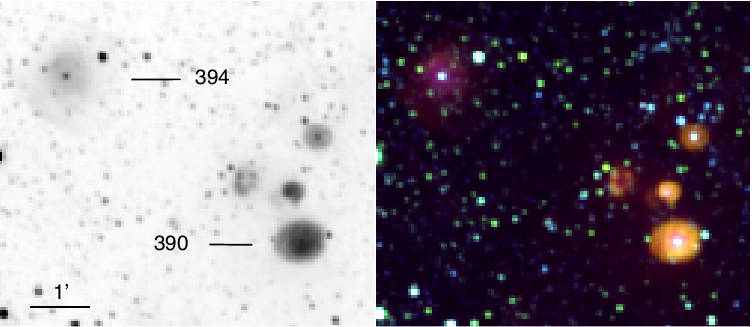}
\label{im2122}}
\subfigure[MCELS~L~346]{
\includegraphics[width=3.3cm,angle=270]{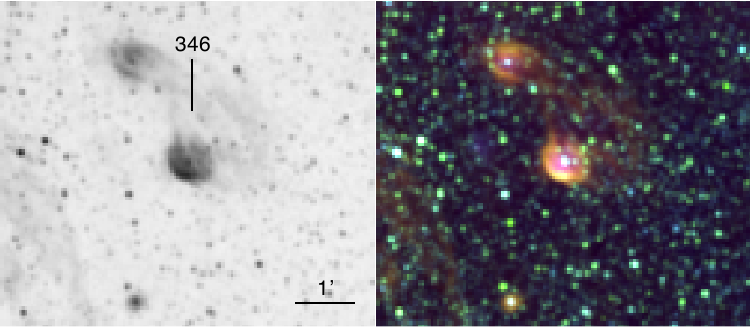}
\label{im24}}
\caption{\footnotesize MCELS images for the objects in our sample. In each image, the left side shows the \Halpha\ MCELS image, while the right side is a three-color composite of the MCELS bands; Red, blue, and green are \Halpha, \oiii\ $\lambda5007$, \sii\ $\lambda6720$, respectively. North is up and east is to the left.  The subfigures show the following objects: (a)MCELS~L~28, MCELS~L~32, and MCELS~L~35, (b) MCELS~L~351, (c) DEM~L~283b, (d) MCELS~L~344 and MCELS~L~345, (e) MCELS~L~43, (f) MCELS~L~52, (g) MCELS~L~390 and MCELS~L~394, (h) MCELS~L~346 \label{mcelsim}}
\end{figure*}
}

We used the Inamori-Magellan Areal Camera \& Spectrograph (IMACS) on the Magellan Baade Telescope at Las Campanas Observatory to obtain both long slit spectra and Bessell \emph{B} and \emph{V} images. Our data were collected on the nights of 2008 Jan 29--31 using the f/4 configuration.  In this setup, IMACS has an eight-chip mosaic CCD that has a total of 8,000$\times$8,000  15 \mum\ pixels, each of which corresponds to 0.11\arcsec.  The seeing was good over the observing run, resulting in a final spatial resolution of $\sim 1.6$\arcsec.  

The long slit observations consist of 3 $\times$ 1200s exposures using a 0.7\arcsec\ slit. We use the 600 l/mm grating, which has a spectral resolution $R \sim 2730$ at \Halpha. The data are binned by 2 and 4 in the dispersion and spatial directions, respectively, which results in $\Delta \sim 0.76$\ \AA\ per pixel. For the first night of the observing run the spectral coverage is 3700-5900~\ang, while for the nights of 2008~Jan~30-31 the wavelength coverage is 3700 -- 6740~\ang\ due to a different grating tilt.  

{
\begin{figure*}[ht]
\centering
\includegraphics[width=8.5in,angle=270]{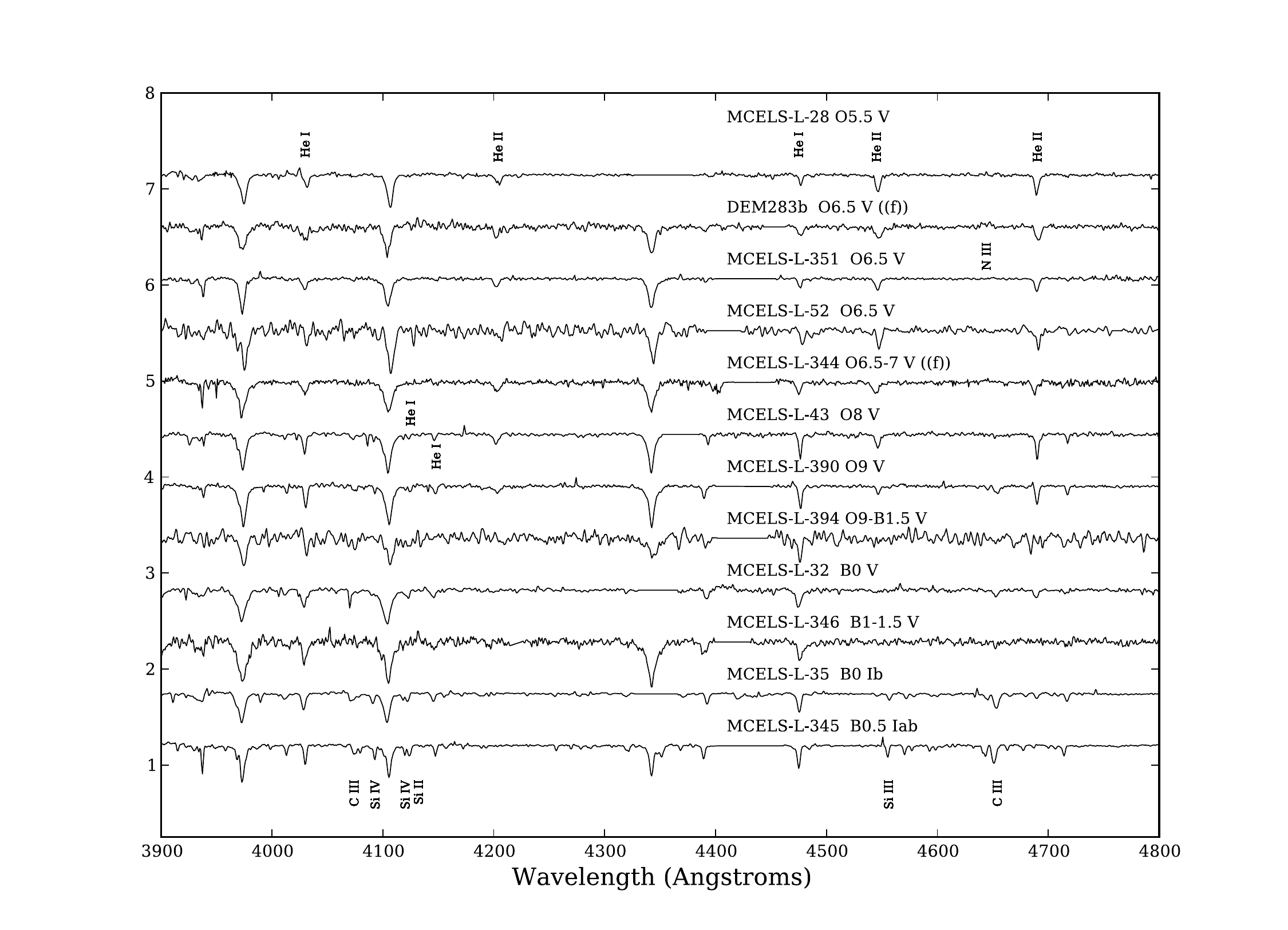}
\caption{\footnotesize Stellar spectra from the ionizing stars of the \hii\ regions.  Spectral type changes toward later type from the top to bottom. The stellar spectra for MCELS~L~52 and MCELS~L~394 are boxcar smoothed with a smoothing length of 3 pixels. The flat, noiseless regions are the chip gaps, for which we assigned a value of unity. MCELS~L~346 is an eclipsing binary, and the stellar spectrum shown here is not the ionizing star, see \S\ref{s:indob} \label{stellarspec}} 
\end{figure*}}

We use standard \iraf\footnote{\iraf\ is distributed by NOAO, which is operated by AURA, Inc., under cooperative agreement with the National Science Foundation.} procedures for the data reduction.  The spectra are extracted with the \iraf\ task {\tt apextract} using separate apertures for the nebula and star.  The extraction includes a local background subtraction.  We extract the nebular spectra in two or more apertures.  These apertures are selected to exclude any stars along the slit, and they otherwise span the entire nebula. We sum those spectra to obtain the total emission from the nebula along the slit.  Figures \ref{stellarspec} and \ref{HIIspec2} show the stellar and nebular spectra for the objects in our sample.  The intensities are scaled to an arbitrary value for presentation, and the gaps between the chips are assigned a value equal to the continuum level.  We flux calibrate the spectra using standard stars LTT~3218, LTT~1788, LTT~2415, LTT~2754 and EG~21 \citep{b:hamuy_pasp94}. 

We measure the emission line fluxes from the nebular spectra (Figure \ref{HIIspec2}) with the \iraf\ task {\tt splot} assuming gaussian line profiles.  We use the reddening equation:  
\begin{equation} \label{eq:redd}
\frac{I(\lambda)}{I(\Hbeta)} = \frac{I_0(\lambda)}{I_0(\Hbeta)}\rm 10^{-\emph{c}(\emph{f}(\lambda) - \emph{f}(\Hbeta))}%
\end{equation} to find the reddening coefficient, \emph{c}(\Hbeta).  Here, $I(\lambda)/I(\Hbeta)$ and $I_0(\lambda)/I_0(\Hbeta)$ are the observed and intrinsic Balmer ratios, respectively, and we use the reddening law of \citet{b:Cardelli_apj89} to determine $f_{\lambda}$\ and $f_{\Hbeta}$.  For the reddening solution we set the ratio of total to selective extinction, $R_V$ = 3.45, which is appropriate for the LMC \citep{b:Gordon_apj03}.  The de-reddened line strengths, relative to \Hbeta, and \emph{c}(\Hbeta) values for each nebula are listed Table \ref{t:flux}.  In Table \ref{t:flux}, we also include the \Hbeta\ flux that we measure in the slit.  Note that the slit width is 0.7\arcsec\ relative to the 0.5-2\arcmin\ sizes of the nebulae. Therefore, the \Hbeta\ flux in Table \ref{t:flux} is not representative of the \Hbeta\ flux in the nebula.  We calculate the emission-line measurement error as $\sqrt{\sigma^2_{\rm cont} N + \sigma^2_{\rm cont}\rm EW/\Delta}$ \citep{b:Gonzalez-Delgado_apj94}.  Here, $\sigma_{\rm cont}$ is the rms in the continuum near the line, \emph{N} is width of the measured line in pixels, EW is the equivalent width of the line and $\Delta$ is the dispersion of the spectra in \AA\ per pixel \citep{b:Gonzalez-Delgado_apj94}.  The reddening error and the flux calibration error together contribute $<10\%$ error for all our sample, and $\sim5\%$ for most of our sample.  These sources of error are combined in quadrature to obtain the flux error listed in Table \ref{t:flux}.

\begin{deluxetable*}{l|c|c|c|c|c|c|c|c|c|c|c|c|c|c|c}
  \tablewidth{0pt}
  \tabletypesize{\footnotesize}
  \tablecaption{De-reddened Emission Lines Fluxes$^{a}$ and Derived Properties \label{t:flux} }
  \tablehead{\colhead{     } & \colhead{L28 n1$^{b}$} &\colhead{error} & \colhead{L28 n2$^{b}$} &\colhead{error} & \colhead{L 32 n1} & \colhead{error} & \colhead{L 32 n2} & \colhead{error} & \colhead{L 35 n2$^c$} & \colhead{error} &\colhead{L 35 n2$^c$} & \colhead{error} }
\startdata     
\oii\ 3726  & 1.503 & 0.115 &  ...  &  ...  & 0.546 & 0.061 &  ...  & ...   &  ...  &  ...  &  ...  &  ...   \\
\oii\ 3726  & 3.690 & 0.390 & 3.446 & 0.081 & 1.464 & 0.211 & 2.283 & 0.168 & 3.705 & 0.416 & 4.407 &  ...   \\
\oii\ 3729  & 2.187 & 0.159 &  .... &  ...  & 0.918 & 0.074 &  ...  & ...   &  ...  &  ...  &  ...  &  ...   \\
\neiii\ 3869& 0.103 & 0.009 & 0.125 & 0.026 &  ...  &  ...  &  ...  &  ...  &  ...  &  ...  &  ...  &  ...  \\ 
\oiii\ 4363 & 0.009 & 0.002 &  ...  &  ...  &  ...  &  ...  &  ...  &  ...  &  ...  &  ...  &  ...  &  ...   \\
\hei\ 4471  & 0.043 & 0.003 & 0.043 & 0.004 &  ...  &  ...  & 0.042 & 0.018 & 0.048 & 0.005 &  ...  &  ...  \\
\oiii\ 5007 & 2.015 & 0.092 & 2.141 & 0.027 &$<0.10$& 0.009 & 0.049 & 0.009 & 0.068 & 0.005 & 0.069 & 0.008  \\
\hei\ 5876  &  ...  &  ...  & 0.118 & 0.002 &  ...  &  ...  & 0.026 & 0.003 & 0.045 & 0.003 & 0.037 & 0.006  \\
\oi\ 6300   &  ...  &  ...  & 0.018 & 0.002 &  ...  &  ...  &  ...  &  ...  & 0.021 & 0.003 & 0.015 & 0.004 \\
\siii\ 6312 &  ...  &  ...  & 0.014 & 0.001 &  ...  &  ...  &  ...  &  ...  &  ...  &  ...  &  ...  &  ...  \\
\nii 6584   &  ...  &  ...  & 0.265 & 0.005 &  ...  &  ...  & 0.360 & 0.014 & 0.495 & 0.017 & 0.482 & 0.044 \\
\sii 6716   &  ...  &  ...  & 0.221 & 0.004 &  ...  &  ...  & 0.479 & 0.020 & 0.487 & 0.017 & 0.470 & 0.046 \\
\sii 6731   &  ...  &  ...  &  ...  &  ...  &  ...  &  ...  & 0.333 & 0.014 &  ...  & 0.325 & 0.033 &  ...  \\
c(\Hbeta)   & 0.053 & 0.058 & 0.220 & 0.062 &  ...  &  ...  & 0.150 & 0.044 & 0.198 &  ...  & 0.182 &  ...  \\
log(O/H)    & -3.51 &  0.10 &  ...  &  ...  &  ...  &  ...  &  ...  &  ...  &  ...  &  ...  &  ...  &  ...  \\
log(N/O)    &  ...  &  ...  & -1.34 &  ...  &  ...  &  ...  & -1.20 &  ...  &  ...  &  ...  &  ...  &  ...  \\
log(He/H)   & -1.06 &  0.01 & -1.06 &  0.01 &  ...  &  ...  & -1.07 & 0.02  &  ...  &  ...  &  ...  &  ...  \\
\Hbeta\ flux$^d$& 15.22 &  ...  & 23.57 &  ...  & 2.156 &  ...  & 2.741 &  ...  & 3.196 &  ...  & 3.313 &  ...  \\
\hline \hline
            &L35 n1$^c$&error&L35 n1$^c$&error& L43 & error &  L52  & error &L344 n1& error &L344 n2& error \\
\hline
\oii\ 3726  & 1.835 & 0.173 & 1.306 & 0.136 & 1.344 & 0.101 & 1.493 & 0.061 & 1.358 & 0.121 & 1.839 & 0.135 \\
\oii\ 3726  & 3.190 &  ...  & 3.293 & 0.325 & 3.692 & 0.197 & 3.676 &  ...  & 3.299 &  ...  & 3.046 & 0.352 \\
\oii\ 3729  & 2.572 & 0.240 & 1.884 & 0.182 & 1.949 & 0.124 & 2.200 & 0.076 & 1.941 & 0.158 & 1.207 & 0.108 \\
\neiii\ 3869&  ...  &  ...  & 0.037 & 0.014 & 0.040 & 0.008 & 0.033 & 0.006 & 0.070 & 0.009 & 0.129 & 0.027 \\ 
\oiii\ 4363 &  ...  &  ...  &  ...  &  ...  &  ...  &  ...  & 0.016 & 0.050 &  ...  &  ...  &  ...  &  ...  \\
\hei\ 4471  & 0.008 & 0.003 & 0.011 & 0.004 & 0.040 & 0.005 & 0.028 & 0.003 & 0.042 & 0.004 & 0.044 & 0.008 \\
\oiii\ 5007 & 0.034 & 0.003 &$<0.03$&  ...  & 1.356 & 0.032 & 1.377 & 0.023 & 1.623 & 0.075 & 1.841 & 0.043 \\
\hei\ 5876  &  ...  &  ...  &  ...  &  ...  & 0.115 & 0.005 & 0.110 & 0.030 &  ...  &  ...  & 0.121 & 0.005 \\
\oi\ 6300   &  ...  &  ...  &  ...  &  ...  & 0.034 & 0.008 & 0.014 & 0.002 &  ...  &  ...  &  ...  &  ...  \\
\siii 6312  &  ...  &  ...  &  ...  &  ...  & 0.016 & 0.008 & 0.012 & 0.002 &  ...  &  ...  & 0.013 & 0.003  \\
\nii\ 6584  &  ...  &  ...  &  ...  &  ...  & 0.258 & 0.011 & 0.268 & 0.005 &  ...  &  ...  & 0.280 & 0.010 \\
\sii\ 6716  &  ...  &  ...  &  ...  &  ...  & 0.276 & 0.010 & 0.265 & 0.005 &  ...  &  ...  & 0.222 & 0.008 \\
\sii\ 6731  &  ...  &  ...  &  ...  &  ...  &  ...  &  ...  & 0.181 & 0.004 &  ...  &  ...  & 0.151 & 0.006  \\
c(\Hbeta)   & 0.128 &  ...  & 0.133 &  ...  & 0.014 & 0.036 & 0.032 & 0.083 & 0.484 & 0.056 & 0.210 & 0.014  \\     
log(O/H)    &  ...  &  ...  &  ...  &  ...  &  ...  &  ...  &$>-4.10$& ...  &  ...  &  ...  &  ...  &  ...   \\ 
log(N/O)    &  ...  &  ...  &  ...  &  ...  & -1.34 &  ...  & -1.40 &  ...  &  ...  &  ...  & -1.27 &  ...   \\ 
log(He/H)   &  ...  &  ...  &  ...  &  ...  & -1.09 & 0.01  & -1.10 & 0.01  &  ...  &  ...  & -1.05 &  0.02 \\
\Hbeta\ flux$^d$& 3.080 &  ...  & 2.384 &  ...  & 4.644 &  ...  & 13.94 & ...   & 27.81 &  ...  & 15.51 &  ...  \\ 
\hline \hline
             &L345 n1& error &L345 n2& error & L346  & error & L351  & error & L390  & error & L394  & error &DEM L283b&error \\
\hline
\oii\ 3726  & 0.592 & 0.046 & 0.592 & 0.172 & 1.543 & 0.134 & 1.251 & 0.179 & 1.520 & 0.066 & 1.667 & 0.287 & 1.570 & 0.174\\
\oii\ 3726  & 1.487 &  ...  & 1.144 &  ...  & 3.779 & 0.201 & 3.271 & 0.546 & 3.919 & 0.452 & 4.282 & 0.958 & 3.859 & 0.325\\
\oii\ 3729  & 0.895 & 0.062 & 0.552 & 0.168 & 2.133 & 0.174 & 2.020 & 0.253 & 2.399 & 0.095 & 2.616 & 0.375 & 2.289 & 0.275\\
\neiii\ 3869&  ...  &  ...  &  ...  &  ...  & 0.033 & 0.006 & 0.072 & 0.035 &  ...  & ...   & 0.158 & 0.060 & 0.039 & 0.008\\ 
\oiii\ 4363 &  ...  &  ...  &  ...  &  ...  &  ...  &  ...  & 0.005 & 0.008 &  ...  & ...   &  ...  &  ...  &$<0.002$& 0.005\\
\hei\ 4471  &  ...  &  ...  &$<0.041$& 0.015& 0.031 & 0.004 & 0.036 & 0.008 & 0.032 & 0.003 &  ...  &  ...  & 0.031 & 0.006\\
\oiii\ 5007 &$<0.027$&  ... &$<0.012$& ...  & 0.905 & 0.043 & 1.573 & 0.078 & 0.601 & 0.009 & 1.452 & 0.090 & 1.260 & 0.064\\
\hei\ 5876  & 0.004 & 0.002 &  ...  &  ...  & 0.079 & 0.005 & 0.090 & 0.015 & 0.086 & 0.003 & 0.088 & 0.011 & 0.079 & 0.007\\
\oi\ 6300   &  ...  &  ...  & 0.009 & 0.005 &  ...  &  ...  &  ...  &  ...  & 0.010 & 0.001 & 0.013 & 0.007 &  ...  & ... \\
\siii\ 6312 &  ...  &  ...  &  ...  &  ...  &  ...  &  ...  &  ...  &  ...  & 0.012 & 0.001 &  ...  &  ...  &  ...  & ... \\
\nii\ 6584  &  ...  &  ...  & 0.295 & 0.028 &  ...  &  ...  &  ...  &  ...  & 0.265 & 0.008 & 0.242 & 0.034 &  ...  & ... \\
\sii\ 6716  &  ...  &  ...  & 0.491 & 0.043 &  ...  &  ...  &  ...  &  ...  & 0.220 & 0.007 & 0.343 & 0.041 &  ...  & ... \\
\sii\ 6731  &  ...  &  ...  & 0.349 & 0.031 &  ...  &  ...  &  ...  &  ...  & 0.157 & 0.005 & 0.231 & 0.032 &  ...  & ... \\
c(\Hbeta)   & 0.430 & 0.048 & 0.389 & 0.099 & 0.034 & 0.295 & 0.464 & 0.099 & 0.302 & 0.037 & 0.148 & 0.063 & 0.749 & 0.102 \\       
log(O/H)    &  ...  &  ...  &  ...  &  ...  &  ...  &  ...  &  -3.4 &  0.5  &  ...  &  ...  &  ...  &  ...  &  ...  & ... \\ 
log(N/O)    &  ...  &  ...  &  ...  &  ...  &  ...  &  ...  &  ...  &  ...  & -1.42 &  ...  & -1.5  &  ...  &  ...  & ... \\ 
log(He/H)   &  ...  &  ...  &  ...  &  ...  & -1.21 &  0.01 & -1.17 & 0.02  & -1.19 & 0.02  & -1.20 & 0.02  & -1.12 & 0.02\\
\Hbeta\ flux$^d$& 7.552 &  ...  & 6.041 &  ...  & 9.125 &  ...  & 27.24 &  ...  & 19.00 &  ...  & 2.035 &  ...  & 28.10 &  ...  \\ 
\enddata
\tablenotetext{a}{Flux measurements are relative to \Hbeta. }
\tablenotetext{b}{Objects with data taken on different nights are labeled such that n1 refers to 29 Jan 2008 and n2 refers to either 30 or 31 Jan 2008.} 
\tablenotetext{c}{There are two slit positions for MCELS~L~32 from each of the two nights we observed it.}
\tablenotetext{d}{The \Hbeta\ fluxes listed here are the fluxes within the 0.7" slit in units of $10^{-14}\rm\ erg\ s^{-1}\ cm^{-2}$. They are not representative of the \Hbeta\ flux from the entire nebula.} 
\end{deluxetable*}

We assign spectral types (SpT) to the stars from the rectified stellar spectra (Figure \ref{stellarspec}) following the criteria of \citet{b:Walborn_pasp90}.  The spectral type assignments are based on the independent spectral typing by four individuals and are accurate to half a spectral type.  The earliest SpT is O5.5, and the latest is B0.5. Most of the stars are luminosity class V, but our sample also contains two B supergiants (Table \ref{t:obsprops}). 

In addition, we collect single exposures of 20s and 10s in filters \emph{B} and \emph{V}, respectively.  We obtain photometry from these images using the \iraf\ task {\tt apphot}.  Ten of our targets are in common with stars in the OGLE-III survey \citep{b:Udalski_actaa08}.  We compare our measured \emph{V} magnitudes to the OGLE-III magnitudes and find a mean difference of -0.07 mag with a standard deviation of 0.15 mag.  We also have two nebulae, MCELS~L~43 and L~52, whose stars are in common with the \citet{b:Massey_apjs02} \emph{UBVR} survey of the Magellanic Clouds, and we find agreement in \emph{V} to 0.01 and 0.09 mag respectively.  As a further check, we compare the \citet{b:Massey_apjs02} and OGLE III \emph{V} magnitudes for $\sim50$ stars with \emph{V} between 13 and 14 mag.  The mean difference ($V_{\rm Massey} - V_{\rm OGLEIII}$) and standard deviation are -0.05 and 0.11 mag, respectively.  From this we see that our values are bracketed by the published literature values.  We obtain absolute \emph{V} magnitudes, $M_V$, after correcting our observed \emph{V} for the measured extinction and assuming an LMC distance modulus of 18.48 \citep{b:Westerlund_97}.

{
\begin{figure*}[ht]
\centering
\includegraphics[width=8.5in,angle=270]{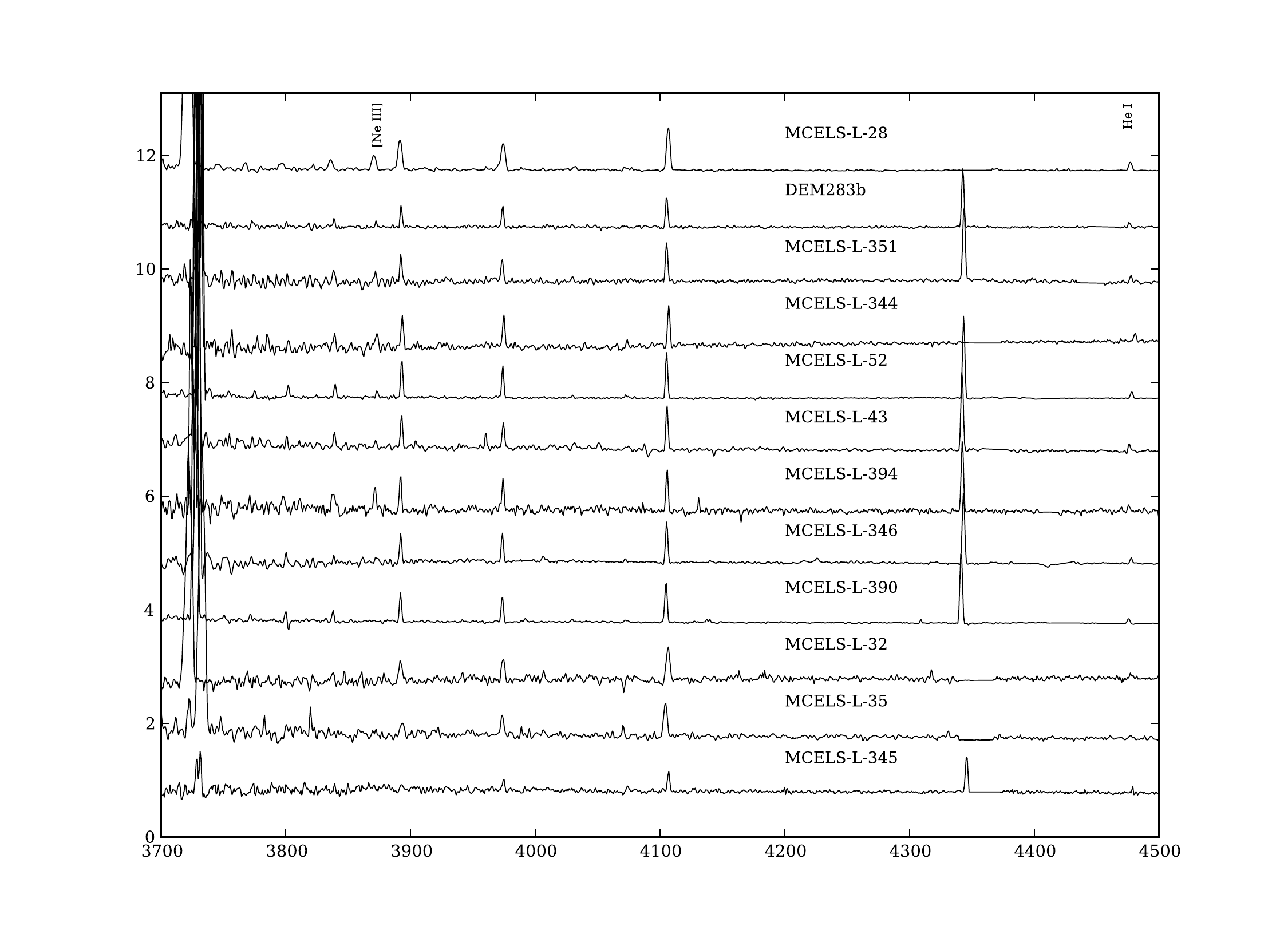}
\caption{\footnotesize Nebular spectra of the \hii\ regions in our sample for 3600--4500 \AA.  The spectra are scaled to highlight the weaker emission lines. In most of the spectra, the chip gap falls on or slightly red-ward of H$\gamma$. We assign this region with a value equal to the mean continuum level. \label{HIIspec2}} 
\end{figure*}}

\subsection{OB companions}\label{s:compan}
One important consideration is the possible contamination from OB companions in our sample.  The binary fraction for massive stars is thought to be 40-70\% \citep[e.g.,][]{b:Sana_mnras09,b:Sana_mnras11} and possibly as high as 90\% \citep{b:Kiminki_apj12}.  Any additional OB stars in the nebula will both contribute to the ionizing photon budget and affect the shape of the ionizing SED.  For OB stars, $Q_0$ changes by 0.1 to 0.6 dex for each full step in SpT \citep{b:Smith_mnras02}. Thus, the impact of a companion on $Q_0$ is maximized for equal mass stars and drops rapidly with later SpT. 

The available data put some constraints on possible companion stars.  For the objects in our sample, the nearest resolved stars have magnitudes that are $> 2$ mag fainter than our target stars, thus ruling out contributions to $Q_0$\ from companions at large distances from the central star.  However, we can still have companions on scales smaller than our spatial resolution, 1.6\arcsec\ which corresponds to $\sim 0.5$\ pc.  In fact, two of our objects, MCELS~L~346 and MCELS~L~390, are confirmed eclipsing binaries, and we discuss them in detail in \S \ref{s:indob}.  

For the rest of the sample, the observed magnitudes will include contributions from both stars, if a companion is present.  Thus, the luminosities derived from our observed magnitudes should be the sum of the luminosities from all components and will be representative of the system.  Furthermore, for a binary or companion, one would expect to have spectral type diagnostics representing a mixture of the two spectral types, except in the case of an equal mass binary.  We note that none of our stellar spectra show evidence for composite SpT.  If any of our objects are equal mass binaries, $Q_0$ from the atmosphere models would be around half that of $Q_0$ from \emph{L}(\Halpha).  We discuss $Q_0$ in more detail in \S \ref{s:q0}, but we note here that we do not find support for equal mass binaries from the $Q_0$ comparison. Thus, while we cannot conclusively rule out binaries from the rest of our sample, the data available suggest that any binaries present will not significantly affect our results.

\subsection{Method}\label{s:metdes2}

To evaluate the stellar atmosphere models, we compare the \hii\ regions in our sample to the predictions of photoionization simulations.  As discussed in the Introduction, the line emission from \hii\ regions depends primarily on the metallicity (\emph{Z}), the ionization parameter (\emph{U}), and the SED of the ionizing star.  When \emph{Z} and \emph{U} are constrained, differences between the predictions from photoionization simulations and the observed \hii\ region spectrum can be directly linked to differences in the shape of the model SED and that of the actual ionizing star.  

Photoionization calculations are performed with version 08.00 of CLOUDY \citep{b:Ferland_pasp98}.  To set up the CLOUDY simulations, we need to match the nebular abundance and ionization parameter to the observed nebulae. In MCELS~L~28 and L~351, we detect the auroral \oiii\ $\lambda$4363 line.  We use the ratio \oiii\ $\lambda4959,5007$/\oiii\ $\lambda4363$ as input for the \iraf\ task {\tt temden} to obtain the electron temperature, \Te.  With \Te\ in hand, we derive the oxygen abundance (Table \ref{t:flux}) using the \iraf\ task {\tt ionic}.  For this calculation we assume \Te(\oii) = \Te(\oiii). To set the abundance of S, C, Ne, Ar, Si and Fe, we use the relations from \citet{b:Mcgaugh_apj91}, which relate the elemental abundances to that of oxygen.  For the rest of our objects, \oiii\ $\lambda4363$ falls in the gap between CCD chips.  In these cases, we adopt the mean LMC abundances measured by \citet{b:Garnett_99}.  For all our objects, except for MCELS~L~346, L~351, and DEM~L~283b, we calculate the nitrogen abundance from the ratio \nii/\oii, as described in \citet{b:Perez-Montero_mnras05}.  Those values are shown in Table \ref{t:flux}. We note that except for MCELS~L~345, the mean difference between the calculated log(N/O) and the \citet{b:Garnett_99} values is 0.17 dex with a standard deviation of 0.11 dex.  This is comparable to the uncertainty in both the relation \citep{b:Perez-Montero_mnras05} and in the measured abundances above.
 
The ionization parameter is defined by 
\begin{equation}
U \equiv \frac{Q_0}{4\pi cn_eR^2_S } \sim (Q_0 n\epsilon^2)^{\frac{1}{3}} 
\label{eq:ion},\end{equation} where $Q_0$ is the rate of H-ionizing photons, $c$ is the speed of light, $n_e$ is the electron density, $R_S$ is the radius of the Str\"{o}mgren sphere and $\epsilon$\ is the filling factor of the gas.  Essentially, \emph{U} describes the ionizing photon density relative to the gas density.  The ionization parameter will depend on both the type of star producing the ionizing radiation and the gas distribution around the star.  As can be seen in Figure \ref{mcelsim}, the nebulae are Str\"{o}mgren spheres, which means we can assume simple spherical geometry for the simulated nebulae.  

The uncertain nebular parameters that control \emph{U} are those that describe the radial density profile of the nebula: the initial (\Rin) and outer (\Rout) radii of the cloud, hydrogen density ($n_H$), and $\epsilon$.  We use a combination of emission-line diagnostics and the MCELS \Halpha\ images to set these values.  To constrain the density distribution, we examine the MCELS images and the \Halpha\ and \oiii\ $\lambda5007$ line profiles along the slit.  The \oiii\ $\lambda$5007 and \Halpha\ profiles indicate that \Rin\ is between 0.1$\times$\Rout\ and 0.5$\times$\Rout.  We measure \Rout\ from the \Halpha\ MCELS images of the nebulae and use those values to set the inner and outer radii in the CLOUDY simulations.  To obtain $n_H$, we assume $n_H = n_e$, since hydrogen is fully ionized in a typical \hii\ region. The \oii\ $\lambda\lambda3726,3729$ and \sii\ $\lambda\lambda6716,6731$ density-sensitive doublets both indicate that nearly all our objects are in the low-density regime ($n_e < 100\ \cmcub$) below which these diagnostics are no longer sensitive to the density.  We also estimate the density using the \Halpha\ emission measure based on the \Halpha\ photometry, and find $n_e \lesssim\ 30\ \cmcub$, which is consistent with our limit from the \sii\ and \oii\ diagnostics.  Based on these constraints, we use $n_H \le \rm 100\ \cmcub$ as an upper limit for the densities in all our simulations. Additionally, the \hii\ regions in our sample are optically thick \citep{b:Pellegrini_apj12}.  Therefore, we set $n_H\rm\ and\ \epsilon$\ such that the ionizing photons are absorbed within the observed nebular size (Table \ref{t:obsprops}).  Finally all our simulated nebulae include 5 km s$^{-1}$ turbulence.

In addition to the properties discussed above, dust will affect the nebula. In particular, photoelectric heating from dust can contribute as much as 30\% of the total heating \citep{b:van-Hoof_04}.  Therefore, we include both graphite and silicate dust in our simulations. We adopt LMC gas distributions and a dust-go-gas ratio, $A_V/N(\rm \HI) = 1.2\times10^{-22}\ \rm mag\ cm^{2}$, consistent with \citet{b:Weingartner_apj01}, which were introduced into CLOUDY in \citet{b:Pellegrini_apj11}.  Finally, since our observations are based on long slit spectroscopy, we implement the CLOUDY command {\tt slit}, which predicts the line ratios that would be observed through a slit across the simulated nebula.

We run CLOUDY simulations with the same set of nebular parameters for each stellar atmosphere model.  In addition to specifying which atmosphere grid to use for the source, we must also assign the luminosity and \Teff\ of the star.  We calculate stellar luminosity from $M_V$ using 
\begin{equation} 
\log{L/\Lsun} = 0.4\times(M_{\rm bol,\odot} - (M_V + \rm BC)), 
\end{equation} 
where $M_{\rm bol, \odot}$ is the solar bolometric magnitude and BC is the the bolometric correction from \citet{b:Martins_aap05}, except for the case of MCELS~L~345, the B0.5 Iab star, where we use the bolometric correction from \citet{b:Crowther_aap06}.  The SpT-\Teff\ conversion is model-dependent and we leave the \Teff\ as a free parameter in our simulations.  We note that since the BC is temperature dependent, the input stellar luminosity changes with \Teff\ to maintain consistency.  For all our dwarf stars, we assume log($g$) = 4.0.

{
\begin{figure}[h]
\centering
\includegraphics[width=3.75in]{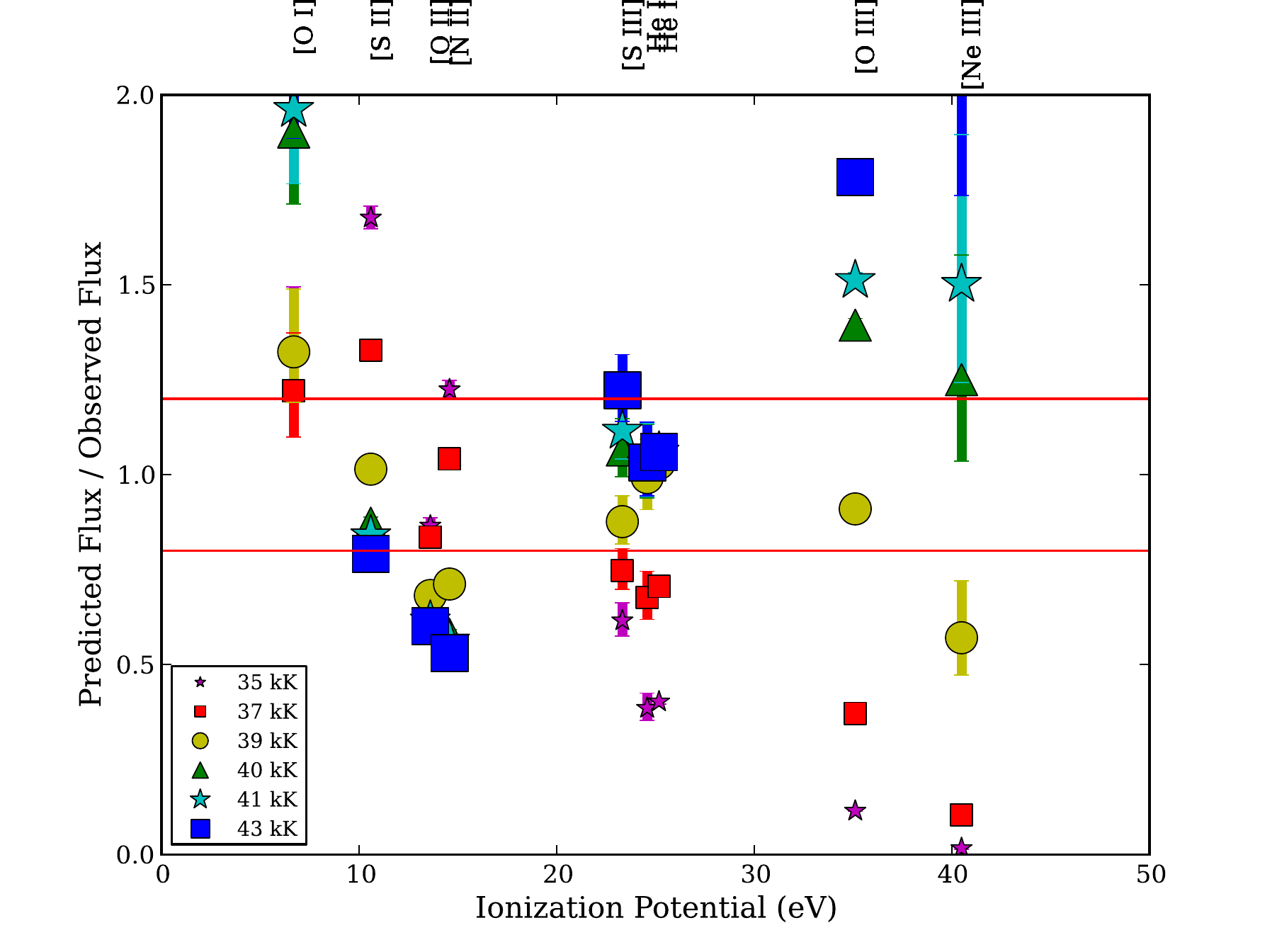}
\caption{\footnotesize Ratio of the predicted emission-line flux and the observed emission-line flux for MCELS~L~28.  The simulations shown above are ionized by WM-basic atmosphere models ranging from $\Teff = 35,000\ \rm to\ 43,000\ K$. The red horizontal lines are at $\pm 20\%$ and are representative of the observational variance.  The best-fitting model is defined as the one with the flattest overall slope, with most points lying between the red lines.  In this case, \Teff=39,000 K (yellow circles) has the best fit. \label{extemps}}
\end{figure}}

We determine the best \Teff\ for each star to be the one that produces the correct balance of ionizing flux at both high and low energies. We accomplish this task by plotting the predicted emission-line flux as a function of ionization potential.  An example is shown in Figure \ref{extemps}, in which we plot the results of six simulations that differ in \Teff.  As expected, as \Teff\ decreases from 43,000 K to 35,000 K, the predictions for \oii\ and \nii\ increase, while the predictions for \oiii\ and \neiii\ decrease. Thus, the overall slope of the points decreases with decreasing \Teff.  We define the optimal model at the turnover \Teff, where the slope is flat.  For the models shown in Figure \ref{extemps}, this would be around \Teff=39,000 K (yellow circles).  We also prioritize models that match the observed \hei\ lines in our spectrum.  The ratio of the \hei\ to \hi\ recombination lines reflects the level of He ionization in the nebula. As long as He remains partially ionized, this ratio will be primarily dependent on the SED of the ionizing source \citep{b:Kennicutt_apj00}.  We note that we do not detect \heii\ $\lambda4686$\ in any of our spectra. Therefore, we can set an upper limit on \Teff\ by requiring a non-detection of that line in the simulations as well.


\section{Photoionization Models} 
\label{s:models}

\subsection{Uniform Density Models} \label{s:singlen}

Figure \ref{mcelsim} shows that our HII regions, to first approximation, are simple Str\"{o}mgren spheres. We use the method described in \S \ref{s:metdes2} and the models described in \S \ref{s:moddesc} to first generate a grid of model \hii\ regions with uniform densities. Figures \ref{el_08c} and \ref{el_020} show the results using this prescription for MCELS~L~28 and L~43, which are ionized by an O5.5 V and an O8 V star, respectively.  These figures are representative of the results from the rest of the sample.  The left panels of the figures show  \oiii\ $\lambda5007$\ vs. \oii\ $\lambda3727$, which reflects variations of \emph{U}.  In nebulae with high \emph{U} there will be more \oiii\ relative to \oii, and the points will occupy the top left of the plot.  As \emph{U} decreases, \oii\ becomes stronger relative to \oiii, and the points will move toward the lower right.  The right panels show \oii\ $\lambda3727$ vs. \hei\ $\lambda 5876$ .  This panel illustrates how well \oii\ is predicted at the appropriate \Teff, which is traced by \hei\ \citep{b:Kennicutt_apj00}.  The solid and dotted lines show how the line ratios change when the \Rin\ and \Teff\ are changed, respectively.

{ 
\begin{figure}[h]
\centering
\includegraphics[width=3.2in]{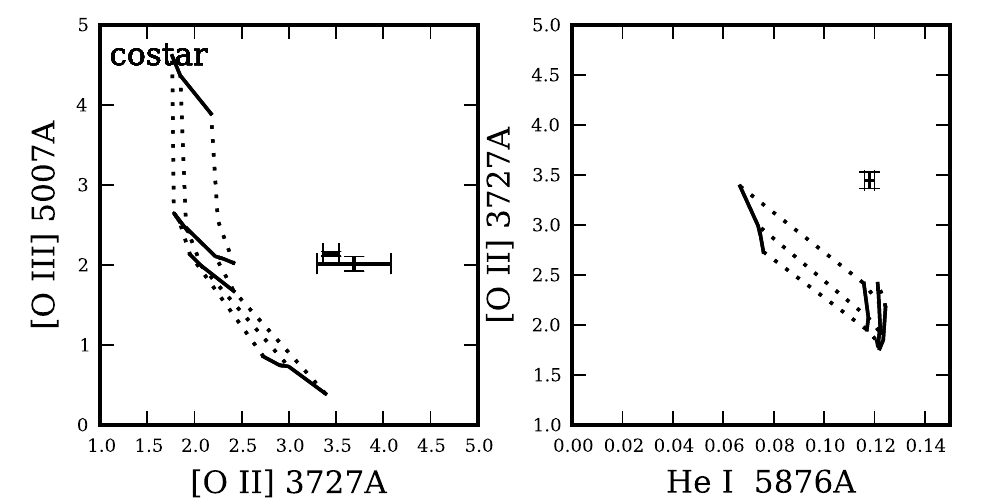} \\
\includegraphics[width=3.2in]{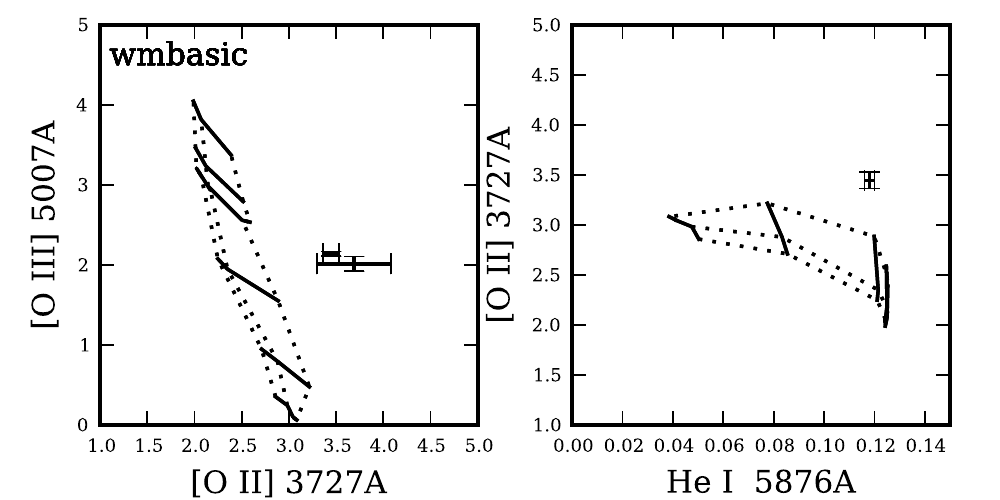} \\
\includegraphics[width=3.2in]{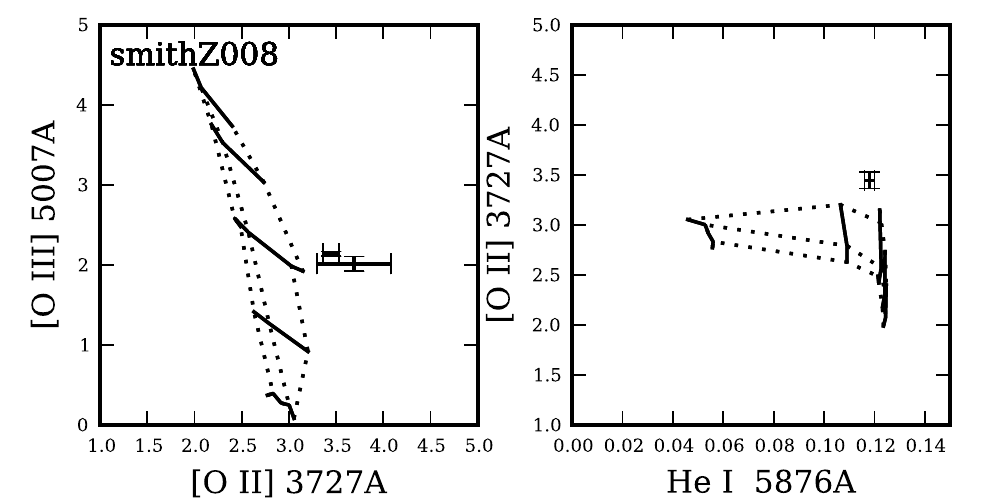} \\
\includegraphics[width=3.2in]{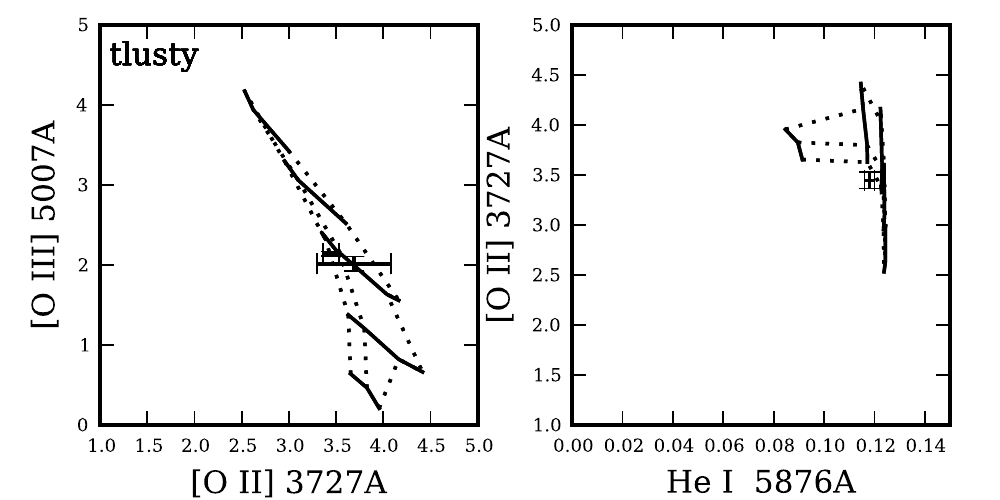}
\caption{\footnotesize Predicted line strengths of \oiii\ vs. \oii\ and \oii\ vs. \hei\ for MCELS~L~28 (O5.5 V).  Line strengths are plotted relative to \Hbeta.  The solid lines represent models with the same \Teff, but changing \Rin. The effective temperatures shown are \Teff = 35, 37, 39, 41, and 43 kK and increase from the bottom towards the top and from left to right, for \oiii\ vs. \oii\ and \oii\ vs. \hei, respectively. Note, the CoStar plot only goes up to \Teff=41 kK. Dotted lines denote models with the same \Rin\ but changing \Teff.  These lines correspond to 0.10, 0.25, and 0.50 \Rout, with the radius increasing towards the right in \oiii\ vs. \oii\ and towards the top in \hei\ v.s. \oii.  For comparison, the observed value is shown by the black point. \label{el_08c}}
\end{figure}}

{
\begin{figure}[h]
\centering
\includegraphics[width=3.2in]{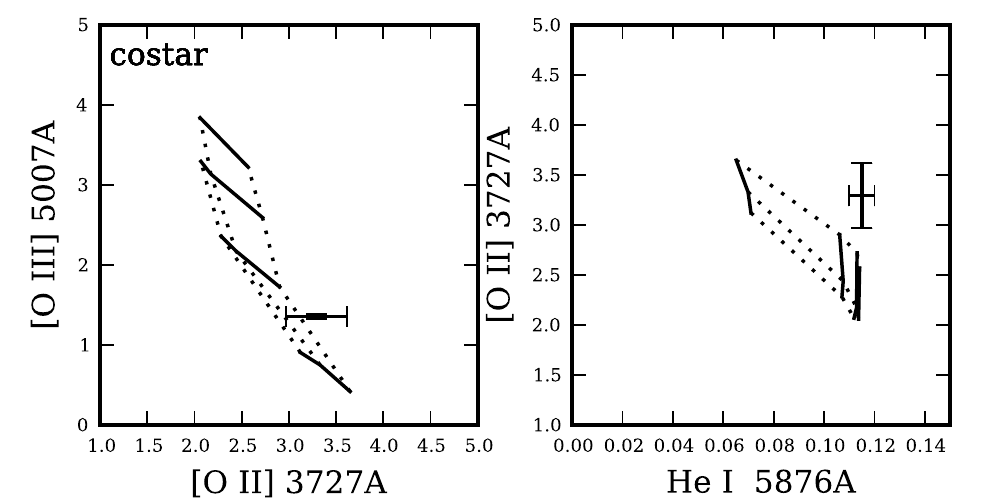} \\
\includegraphics[width=3.2in]{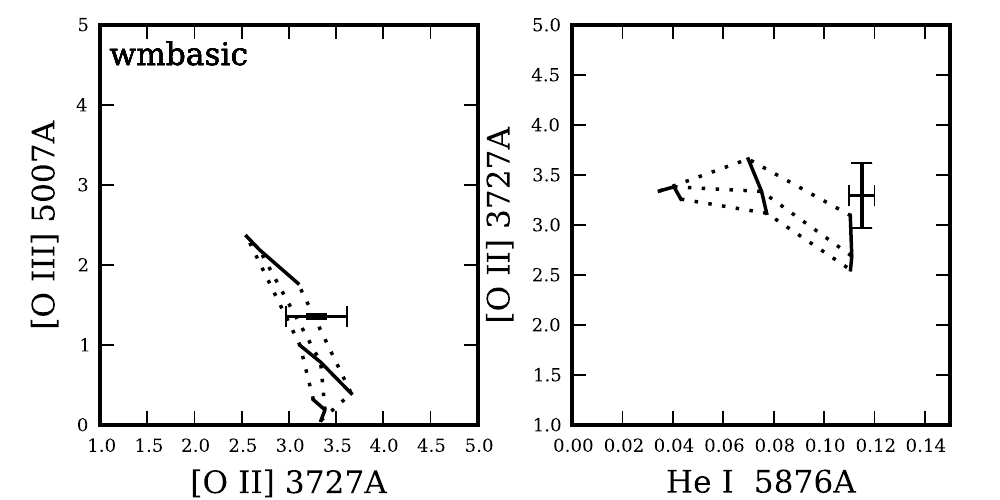} \\
\includegraphics[width=3.2in]{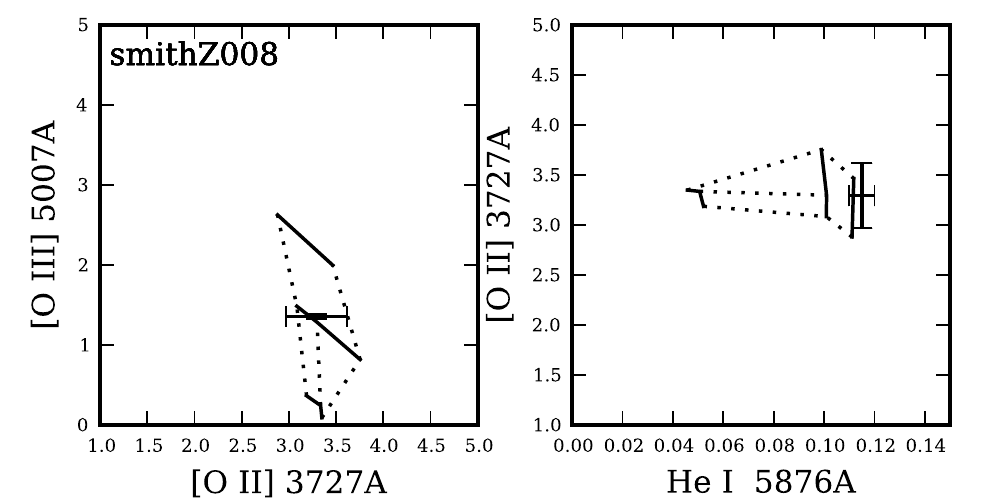} \\
\includegraphics[width=3.2in]{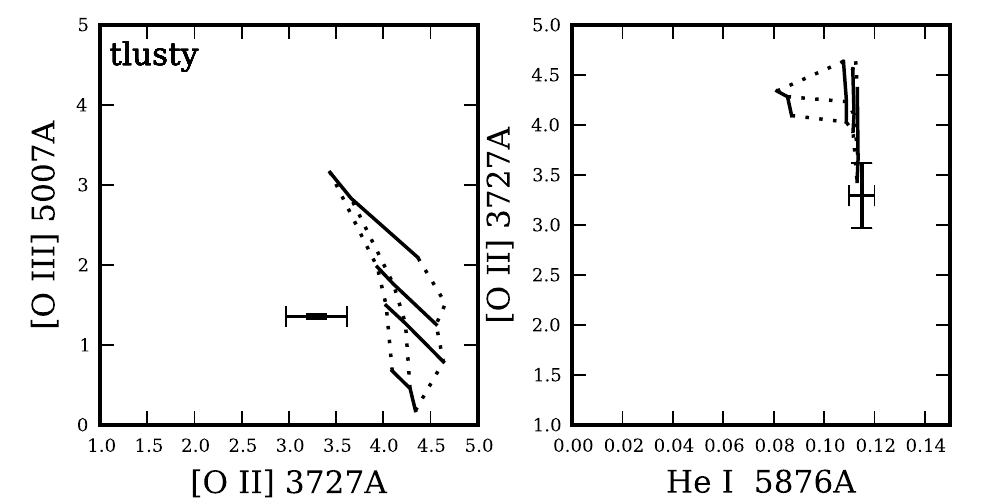}
\caption{\footnotesize Same as Figure \ref{el_08c} but for MCELS~L~43 (O8 V). The inner radii shown are 0.10, 0.25, and 0.50 \Rout.  The effective temperatures are 35, 37, 39, 40 kK. \label{el_020}}
\end{figure}}

It is immediately apparent that these models do not reproduce the observed emission-line ratios.  In the simulations using CoStar and WM-basic atmosphere models, the simulation tracks generally run below and to the left of the observed values (Figure \ref{el_08c}; left panel). This indicates that not enough \oii\ is produced when \oiii\ is well predicted and suggests that \emph{U} is too high.  Furthermore, from the righthand plots, we see that simply changing the \Teff\ is not a viable solution; the predicted \oii\ flux does not match the observations for the entire range in \Teff.  This is the case for all our objects except for MCELS~L~43 (Figure \ref{el_020}) and MCELS~L~344 (not shown).  In contrast, the simulated nebulae ionized by TLUSTY atmospheres have lower ionization parameters. The softer TLUSTY SED is able to reproduce the observed \oiii\ and \oii\ lines in  MCELS~L~28 (Figure \ref{el_08c}), MCELS~L~351, and MCELS~L~394.  For the rest of the objects, the TLUSTY SED is too soft, and the simulations have an ionization parameter that is too low (Figure \ref{el_020}).  

We rule out the adopted metallicity as the source of the discrepancy between the simulations and observations, as follows.  The metallicity of the gas strongly affects the observed line ratios.  As the metallicity decreases, there are fewer metals to provide cooling, and the nebula is hotter.  Thus, the under-prediction of the low ions seen above could indicate that the adopted metallicities are too low.  To explore how this might affect our results, we generate a CLOUDY grid for MCELS~L~28 in which we change the metallicity from log(O/H)= -3.4 \rm\ to log(O/H) = -3.9.  The results from this grid for the \oii\ and \oiii\ line strengths are shown in Figure \ref{metals}.  The measured log(O/H) = -3.51 for MCELS~L~28. 
{
\begin{figure}[h]
\centering
\subfigure[{\oiii/\oii}]{
\includegraphics[width=6.5cm]{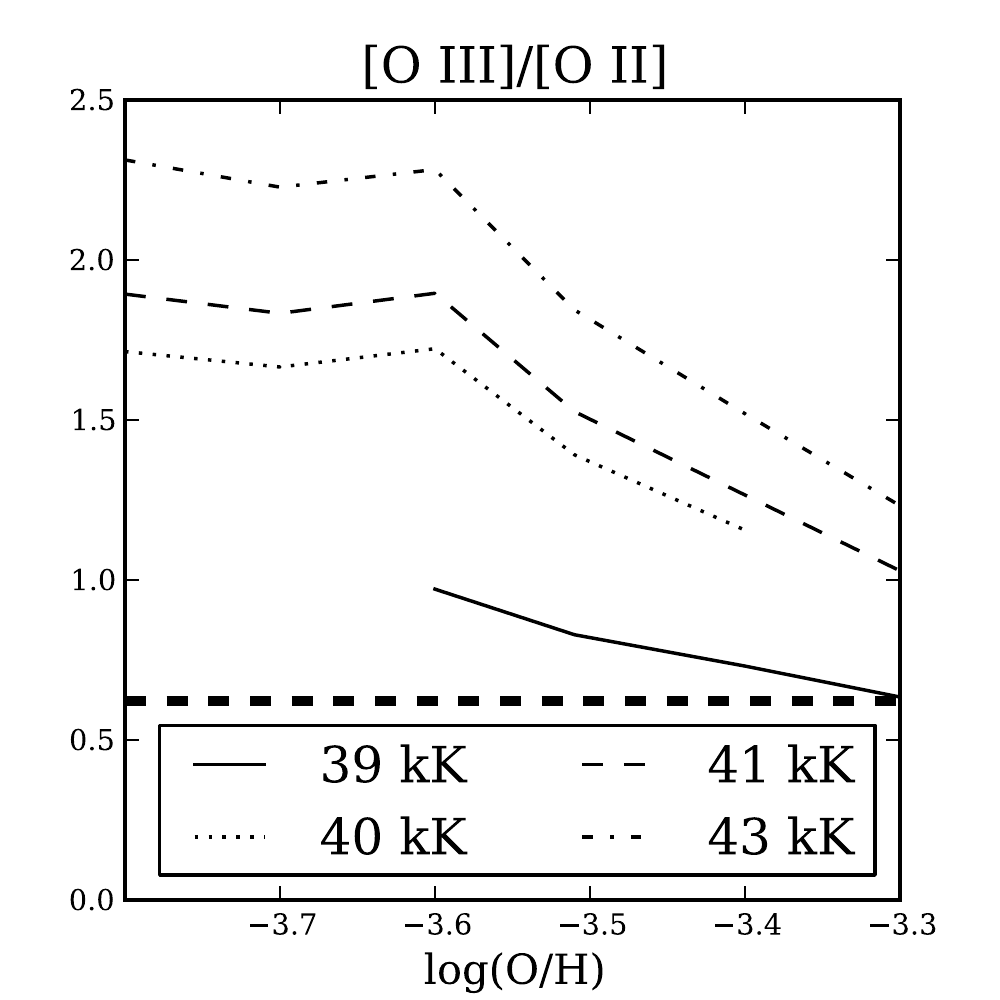} \label{metalo3o2}}
\subfigure[{\oiii\ $\lambda5007$/\Hbeta}]{
\includegraphics[width=6.5cm]{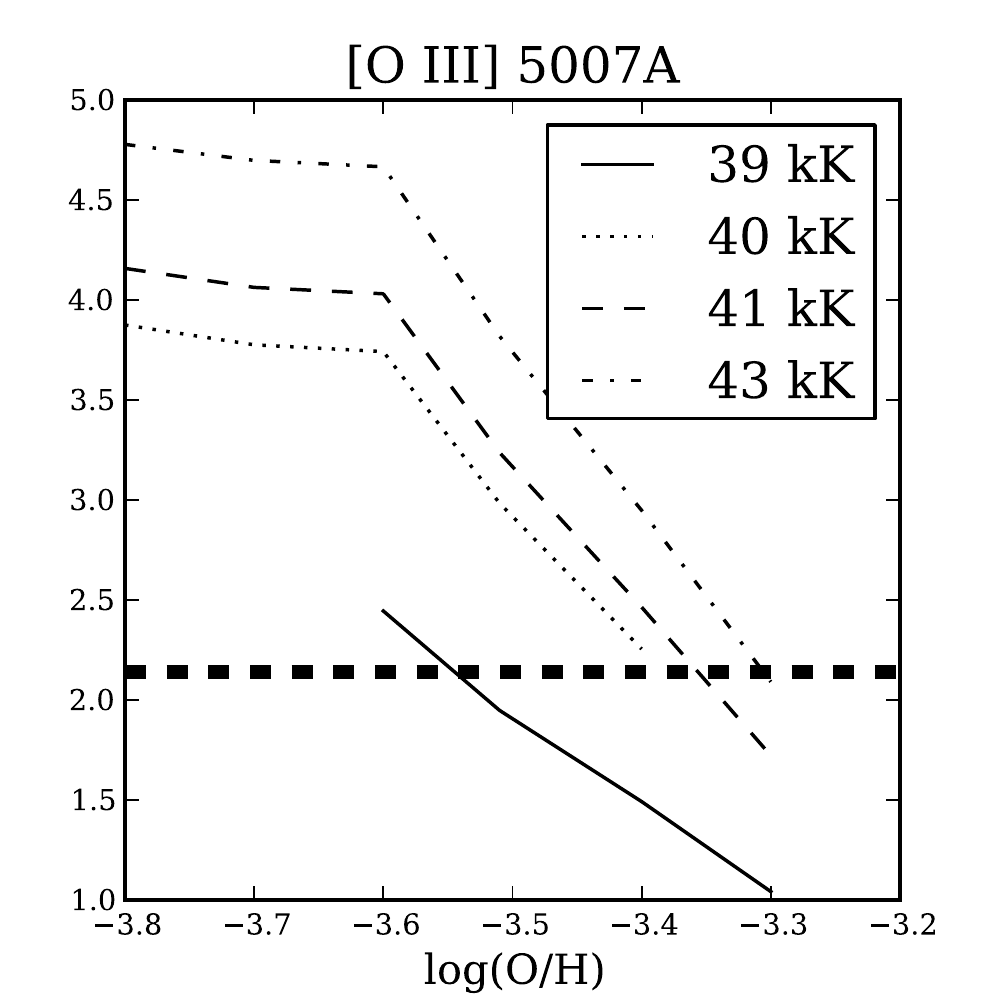} \label{metalo3}}
\subfigure[{\oii\ $\lambda3727$/\Hbeta}]{
\includegraphics[width=6.5cm]{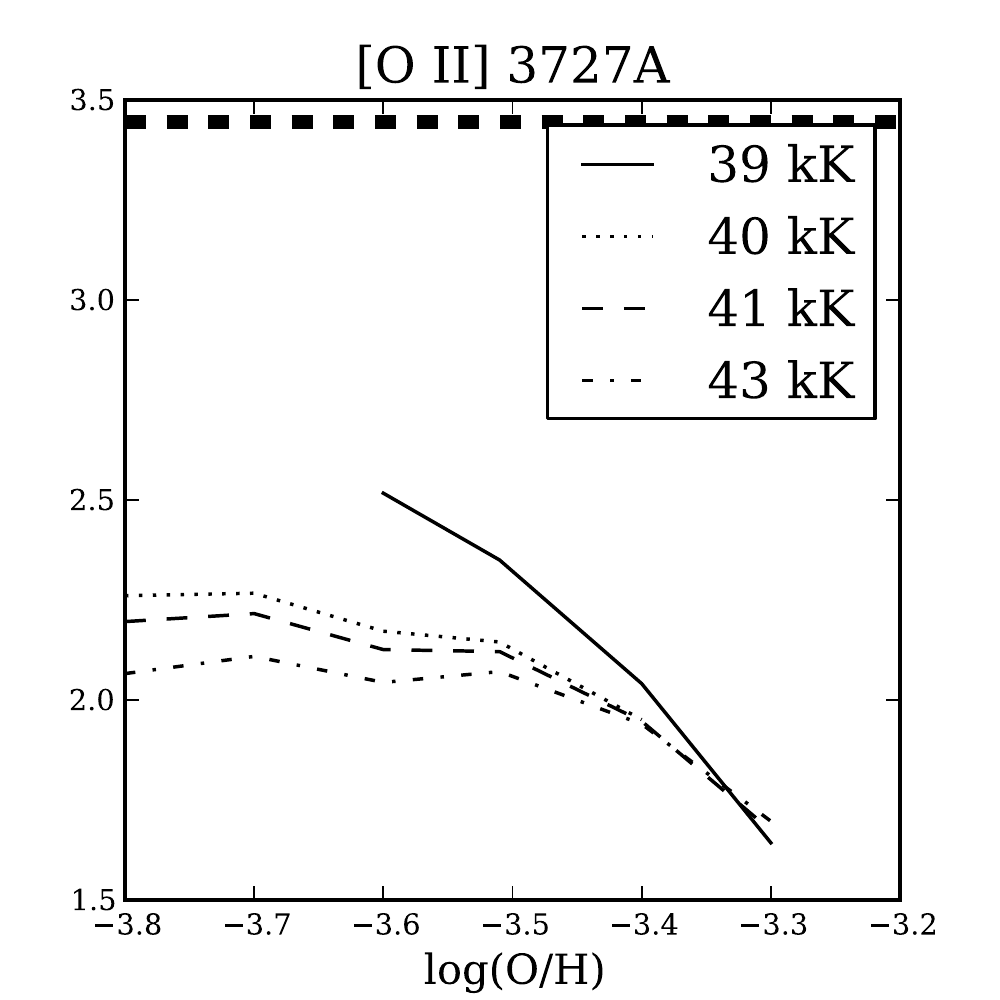} \label{metalo2}}
\caption{\footnotesize Predicted lines strengths of \oiii\ $\lambda5007$ and \oii\ $\lambda3727$ as a function of metallicity are shown for simulations of MCELS~L~28 ionized by WM-basic atmospheres with \Teff= 39--43 kK (thin lines).  These are compared to the observed values for MCELS~L~28 (thick dashed line) for which the observed metallicity is log(O/H)=-3.51 \label{metals}}
\end{figure}}

We find that changing the metallicity will not reconcile the discrepancy in the low ions.  Although increasing the metallicity does decrease the predicted \oiii/\oii\ ratio for a given \Teff\ (Figure \ref{metalo3o2}), it does so by decreasing \oiii\ (Figure \ref{metalo3}).  Meanwhile, \oii\ remains under-predicted for the whole range of metallicities (Figure \ref{metalo2}). Therefore, even though changing the metallicity might bring the overall \oiii/\oii\ closer to the observed, it cannot account for the discrepancies between the simulations and the observations for the individual lines. We note that decreasing log(O/H) by 0.1 dex, from the observed to the mean LMC value, will change the \Teff\ required to reproduce the observed emission-line fluxes by 500 K.  This is comparable to the 250--500 K uncertainty in determining \Teff\ using the emission-line ratios.  Since we do not have measured metallicities for most of the objects in our sample, this 500 K is included in the error on the best fitting \Teff, discussed below.

Another possible cause of the general under-prediction of the low ions is the nebular structure. In this case, there is not enough dense gas in our simulations at large radii to receive a diluted radiation field, which is needed to reproduce the emission from lines with low ionization potential. This could be because \Rin\ is set too close to the star, or because the nebula has a clumpy distribution of denser gas.  An increase of \Rin\ increases the \oii\ and \nii\ flux, while it simultaneously suppresses the \oiii\ and \neiii\ flux.  At face value, this would move the predictions closer to the observations. However, even if we vary the inner radius from 0.1 to 0.5 \Rout, we still are not able to simultaneously reproduce the two sets of lines (Figure \ref{el_08c}).  Furthermore, the inner radius is constrained to be $\lesssim 0.25\ \Rout\ $ in all our objects, except DEM~L~283b, based on the spatially resolved \oiii\ $\lambda5007$ line profiles and the MCELS images.  Thus, to within the $\sim$0.5 pc spatial resolution of our spectral data, these \hii\ regions do not appear to be hollow shells in projection.  Therefore, neither the metallicity nor \Rin\ can explain the discrepancy in these uniform density models.  In the next section, we explore the effects of a clumpy gas distribution.

\subsection{Density Fluctuations} \label{s:nfluc}

\begin{deluxetable*}{lcccccccccccc} 
  \tablewidth{0pt}
  \tabletypesize{\footnotesize}
  \tablecaption{Parameters of the best fit simulations. \label{t:modparams}}
  \tablehead{
     \colhead{MCELS} &
     \colhead{L 28} &
     \colhead{L 32} &
     \colhead{L 35} &
     \colhead{L 43} &
     \colhead{L 52} &
     \colhead{L 344} &
     \colhead{L 345} &
     \colhead{L 346} &
     \colhead{L 351} &
     \colhead{L 283b$^a$} &
     \colhead{L 390} &
     \colhead{L 394}} 
\startdata   
SpT          &O5.5 V& B0 V & B0 Ib           & O8 V &O6.5 V&O6.5-7 V((f))&B0.5 Iab&O9 V$^d$&O6.5 V &O6.5 V((f))&O9 V& O9 V\\ 
\Rin         & 0.25 & 0.25 &0.15$^b$,0.06$^c$& 0.10 & 0.10 & 0.10 & 0.10    &   0.25     & 0.10 & 0.25 & 0.10 & 0.10  \\
log($g$)     & 4.0  & 4.0  & 2.8$^b$,3.50$^c$& 4.0  &  4.0 & 4.0  & 2.6-2.9 &    4.0     &  4.0 &  4.0 & 4.0 & 4.0   \\
\hline
&&&&&&\textbf{CoStar}&&&& \\  
\Teff\ (kK)  & 39.5 & 34.0 &       31.0      & 37.0 & 36.0 & 37.0 &   ...   &     36.0   & 36.5 & 36.5 & 35.0  & 37.0 \\
$\epsilon$   & 0.10 & 0.10 &       0.25      & 0.10 & 0.25 & 0.10 &   ...   &    0.10    & 0.10 & 0.10 & 0.10  & 0.10 \\
\emph{n$_e$} &  75  &  75  &        25       &  50  &  25  &  60  &   ...   &    100     &  50  &  75  &  75   & 100   \\
\hline
&&&&&&\textbf{TLUSTY}&&&& \\
\Teff        & 42.5 & 33.0 &       27.5      & 39.0 & 38.5 & 39.5 &   22-24 &     37.5   & 39.5 & 39.0 & 35.0  & 39.0 \\
$\epsilon$   & 0.10 & 0.10 &       0.10      & 0.15 & 0.10 & 0.10 &   0.15  &    0.10    & 0.10 & 0.10 & 0.10  & 0.10 \\
\emph{n$_e$} &  75  &  75  &       75        &  50  &  60  &  75  &    30   &     100    &  50  & 100  &  75   & 100  \\
\hline 
&&&&&&\textbf{WM-basic}&&&& \\
\Teff (kK)   & 41.5 & 34.0 &       31.0      & 39.0 & 38.75& 39.0 &   ...   &    38.0    & 39.0 & 39.0 & 37.0 & 39.0  \\
$\epsilon$   & 0.10 & 0.10 &       0.15      & 0.10 & 0.10 & 0.10 &   ...   &    0.10    & 0.10 & 0.10 & 0.10  & 0.10 \\
\emph{n$_e$} &  75  &  75  &        25       &  50  &  60  &  60  &   ...   &    100     &  50  &  75  &  75   &  100 \\
\hline 
&&&&&&\textbf{SNC02}&&&& \\
\Teff  (kK)  & 41.5 & 34.0 &      31.0       & 39.0 & 38.5 & 39.0 &   ...   &    37.0    & 39.0 & 39.0 & 37.0  & 39.0 \\
$\epsilon$   & 0.10 & 0.10 &      0.15       & 0.10 & 0.10 & 0.10 &   ...   &    0.10    & 0.10 & 0.10 & 0.10  & 0.10 \\
\emph{n$_e$} &  75  &  75  &       25        &  50  &  60  &  60  &   ...   &    100     &  50  &  75  &  75   & 100 
\enddata
\tablenotetext{a}{L 283b refers to the DEM ID}
\tablenotetext{b}{TLUSTY simulation}
\tablenotetext{c}{CoStar, WM-basic, and SNC02}
\tablenotetext{d}{SpT is inferred and not observed, see \S\ref{s:indob}.}
\end{deluxetable*}

In \hii\ regions, non-uniform gas and temperature distributions are often invoked to explain the observed properties.  In fact, direct imaging of \hii\ regions reveals complex morphology, which includes density gradients and/or clumps and filaments with high density contrast.  Non-uniform gas density has long been invoked to explain observed parameters \citep[e.g.,][]{b:Osterbrock_apj59}.  One example is the long-standing problem reconciling the discrepant values for \Te\ that result from different diagnostics.  This has been of particular concern because the uncertainty in \Te\ translates into significant uncertainty in the derived abundances.  One suggestion is that this problem arises from temperature fluctuations in the ionized gas, which are described by the $t^2$ parameter \citep{b:Peimbert_apj67}.  These fluctuations could arise from small clumps of dense gas within the nebulae \citep[e.g.,][]{b:Viegas_mnras94,b:Williams_apj92,b:Liu_mnras93} or from fluctuations in chemical abundance rather than density \citep{b:Rubin_apjs89,b:Kingdon_apj95,b:Giammanco_aap04,b:Tsamis_mnras05}.  

We therefore explore how changing the density structure affects line emission by considering a clumpy medium.  For our analysis here, we assume that dense clumps of gas are uniformly spread throughout the nebula, with a vacuum between them \citep{b:Osterbrock_apj59}.  In reality, the clumps will be interspersed with lower density, diffuse gas.  However, the emission measure is proportional to $n_e^2$.  Therefore, most of the observed flux is coming from the dense clumps, and we can treat the inter-clump space as though it were a vacuum.  In the simulations, these clumps are treated in a statistical sense; both the volume emissivity of the gas and the optical depth along the line of sight are decreased by $\epsilon$ in each zone calculated by the code \citep{b:Ferland_pasp98}.

In our final simulation grid, \Teff\ and $\epsilon$\ are varied, with the density chosen as discussed above.  A decrease in the filling factor of the gas necessitates an increase in the density of that gas to maintain the nebular radius (Equation \ref{eq:ion}).  In this scenario, dense gas is present at small radii to experience a strong radiation field, as required for the lines with high ionization potential.  At the same time, more dense gas is present at larger radii from the star that receives a diluted radiation field to reproduce ions with low ionization potential.  Our constraints on the sizes of the simulated nebulae, combined with our upper limit on the density, effectively provide us with lower limits for $n_H$ and $\epsilon$.  We cannot decrease the density past the value that leads to correctly sized nebulae when $\epsilon$ = 1.0, and we cannot decrease $\epsilon$\ below the value that corresponds to $n_H$ = 100 \cmcub.  We note that the filling factor set by the density limit, $\epsilon\sim 0.10$ in most of our objects, is consistent with the $10^{-1} - 10^{-3}$ commonly used in the literature \citep{b:Kennicutt_apj84, b:Giammanco_aap04, b:Hunt_aap09}.  Furthermore, the nebulae in our sample exhibit substructure (Figure \ref{mcelsim}) that is consistent with this level of density fluctuations.

Table \ref{t:modparams} shows the parameters of the best fitting model for each nebula.  These parameters include \Rin\ and log($g$) for each nebula and star, as well as the \Teff, $\epsilon$, and $n_H$ that correspond to the best simulation for each atmosphere model.  Figure \ref{IPeV_set1} shows the results of these simulations.  In this Figure, we plot the ratio of the predicted emission-line flux to the observed, $F_{\rm pre}/F_{\rm obs}$, as a function of the ionization potential of the emission line.  Each model atmosphere corresponds to a different color and symbol. The error bars represent the observational error, which includes the line measurement, reddening correction and flux calibration errors. As discussed in \S\ref{s:metdes2}, these simulations are selected by finding the combination of free parameters ($\epsilon$, \Teff) that reproduces the observed \hei\ emission-lines, and yields the flattest slope with most points lying within $\pm 20\%$ of unity, denoted by the red horizontal lines.  The 20\% takes into account the observational variance, which is estimated from MCELS~L~28, L~32, L~35 and L~344, for which we have multiple nights of data and/or multiple slit positions (Figures \ref{IPeV8c}, \ref{IPeV004}, \ref{IPeVm32}, and \ref{IPeV276}).  The additive error on our selected \Teff\ and $\epsilon$\ for a given \Rin\ is $\sim 500-1000\ \rm K$ and 0.1, respectively.  The error for \Teff\ includes the contribution based on the uncertainty in our metallicity assumptions, as discussed in \S\ref{s:singlen}.

{
\begin{figure*}[h]
\centering
\subfigure[MCELS~L~28]{
\includegraphics[width=4.5cm]{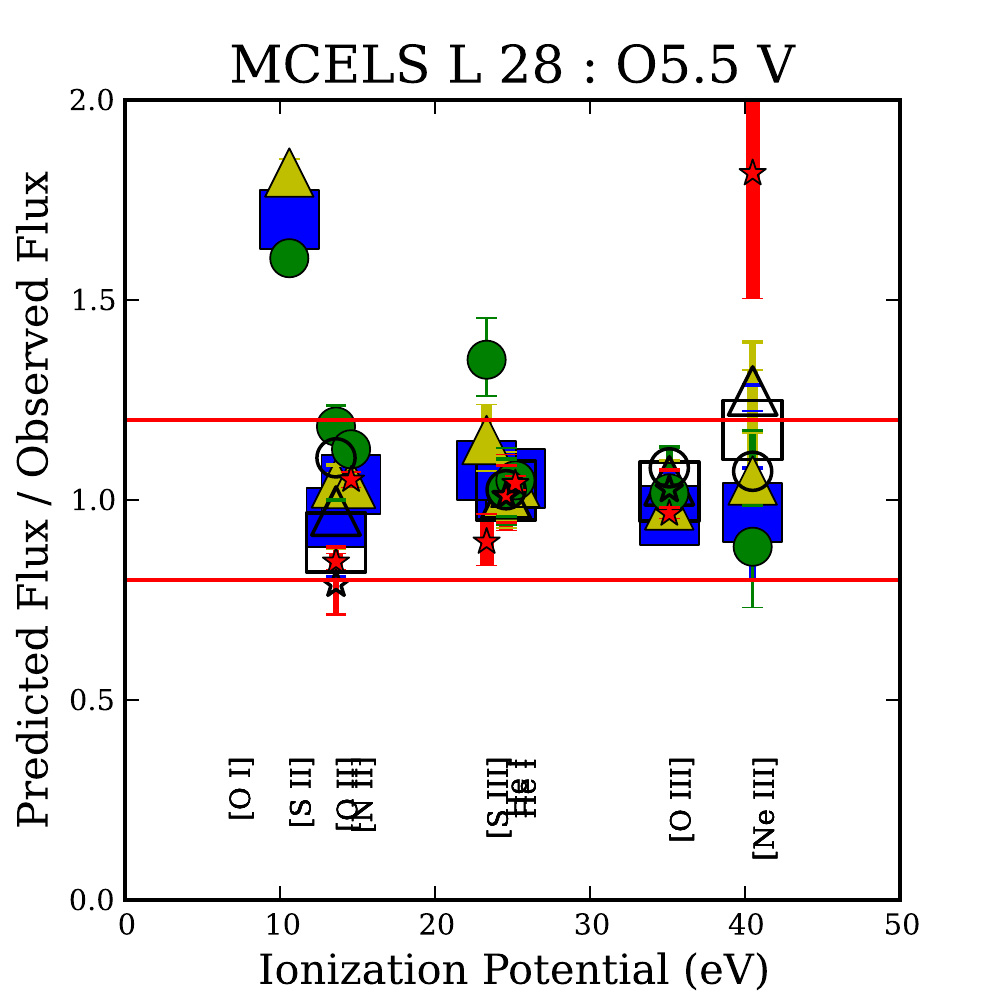} 
\label{IPeV8c}}
\subfigure[MCELS~L~52]{
\includegraphics[width=4.5cm]{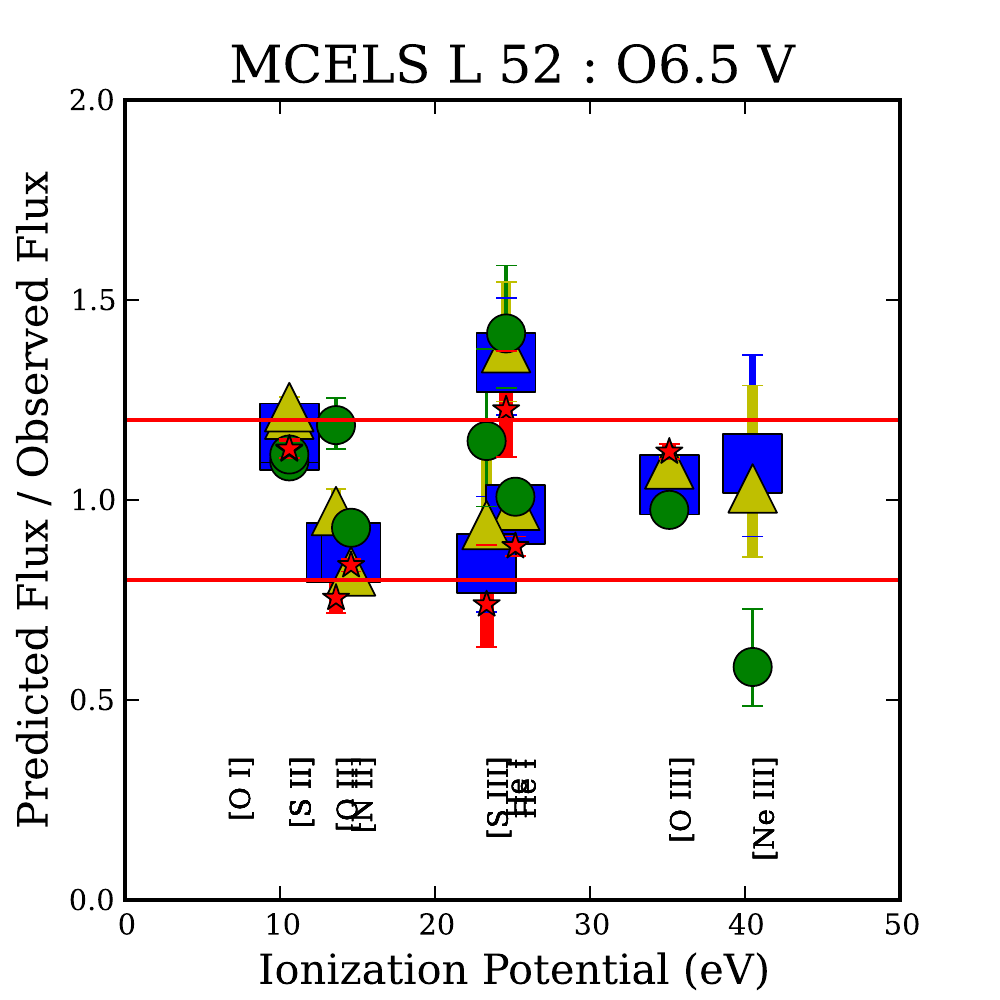}
\label{IPeV26}}
\subfigure[MCELS~L~351]{
\includegraphics[width=4.5cm]{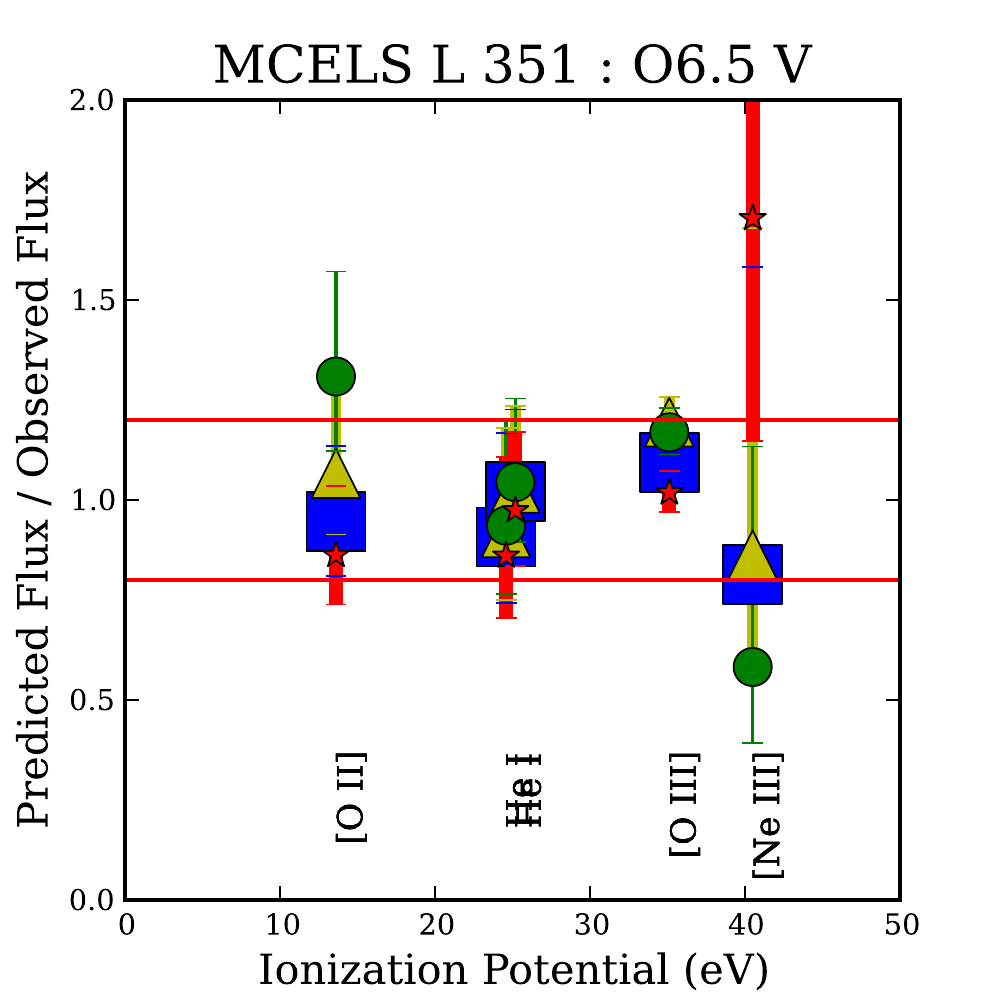}
\label{IPeV281}}
\subfigure[DEM~L~283b]{
\includegraphics[width=4.5cm]{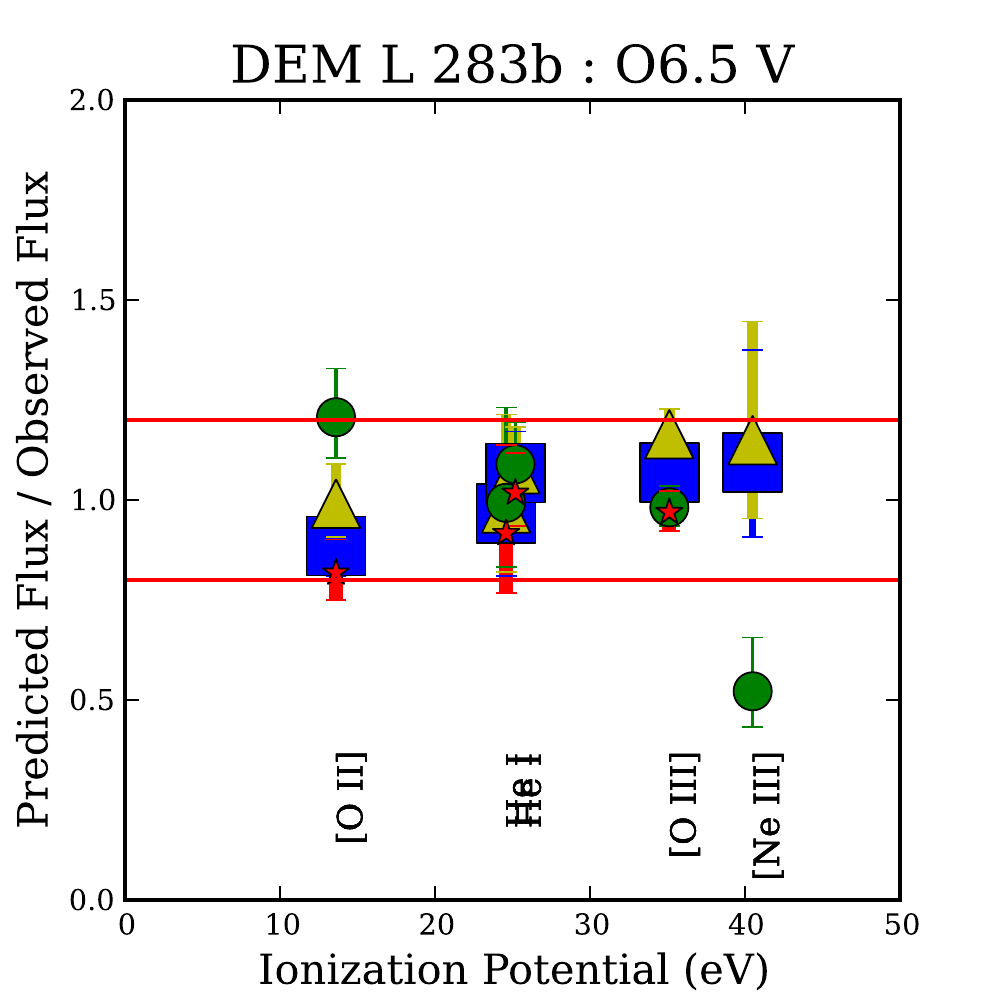}
\label{IPeV283b}}
\subfigure[MCELS~L~344]{
\includegraphics[width=4.5cm]{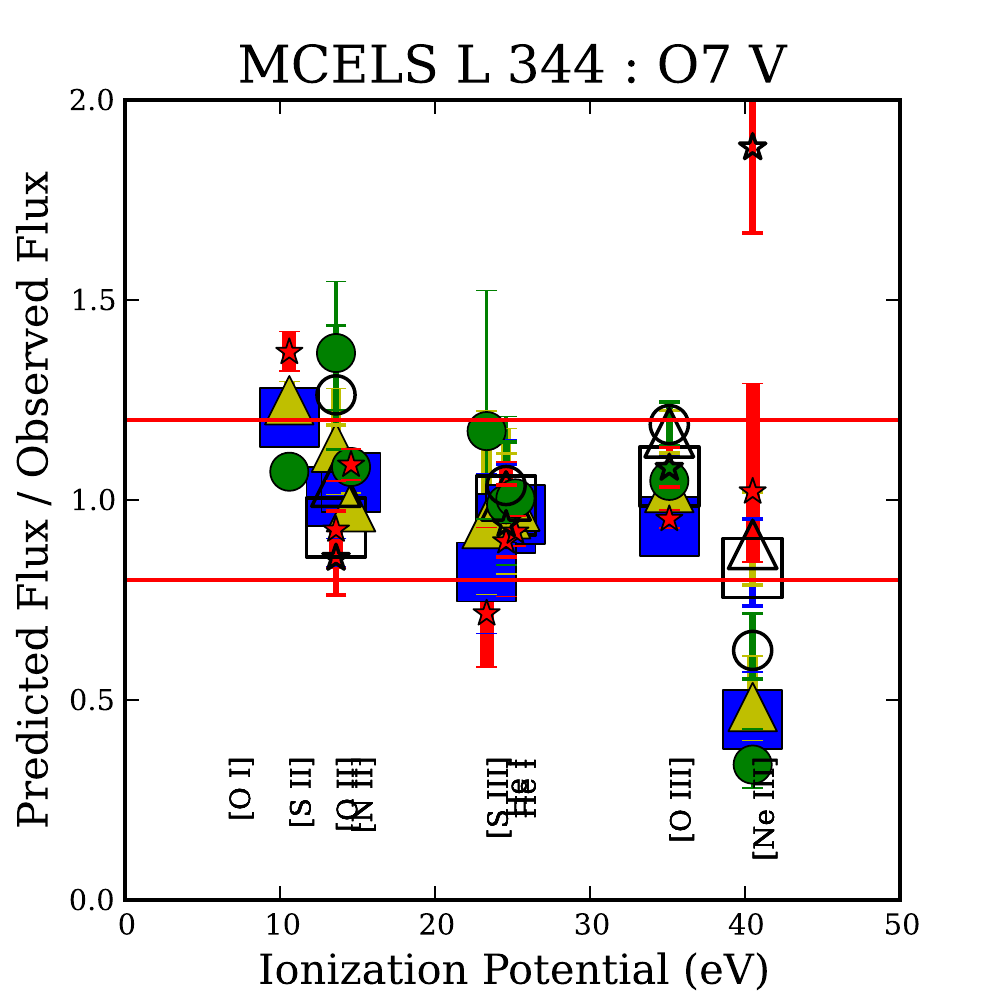}
\label{IPeV276}}
\subfigure[MCELS~L~43]{
\includegraphics[width=4.5cm]{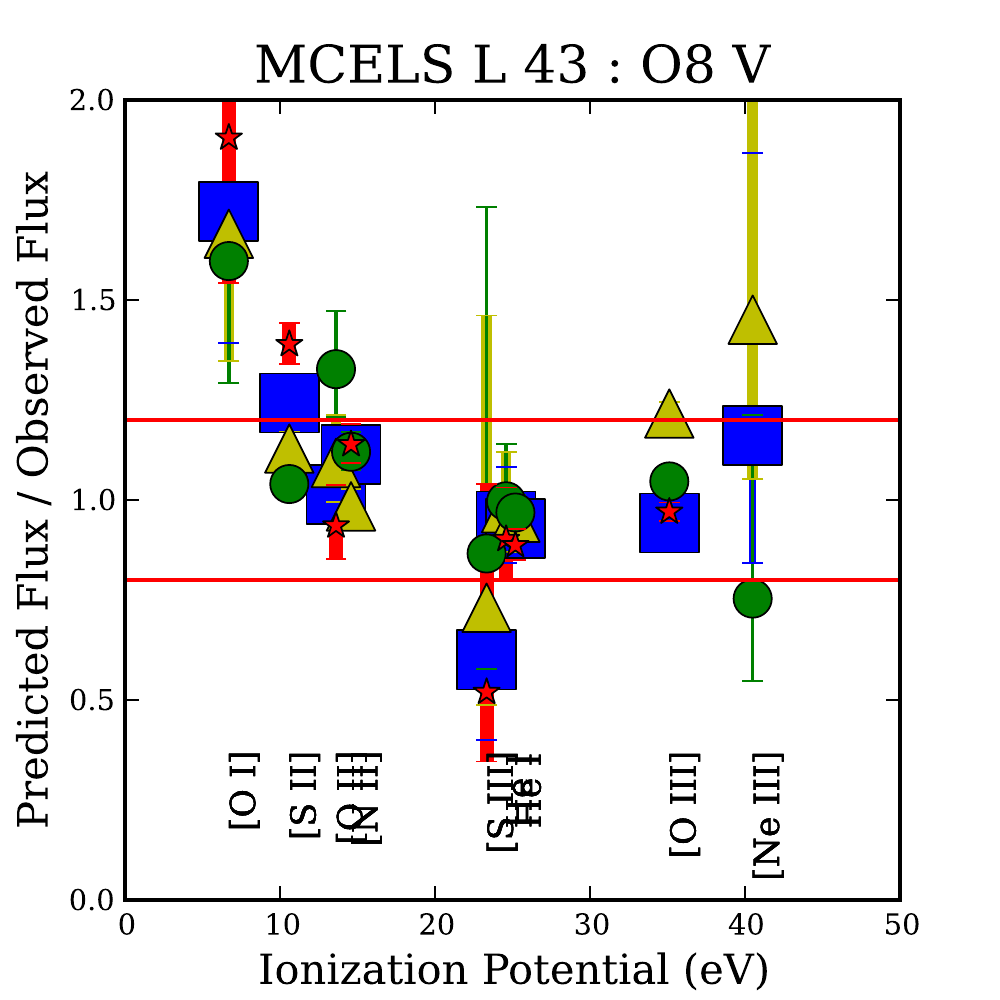}
\label{IPeV20}}
\subfigure[MCELS~L~390]{
\includegraphics[width=4.5cm]{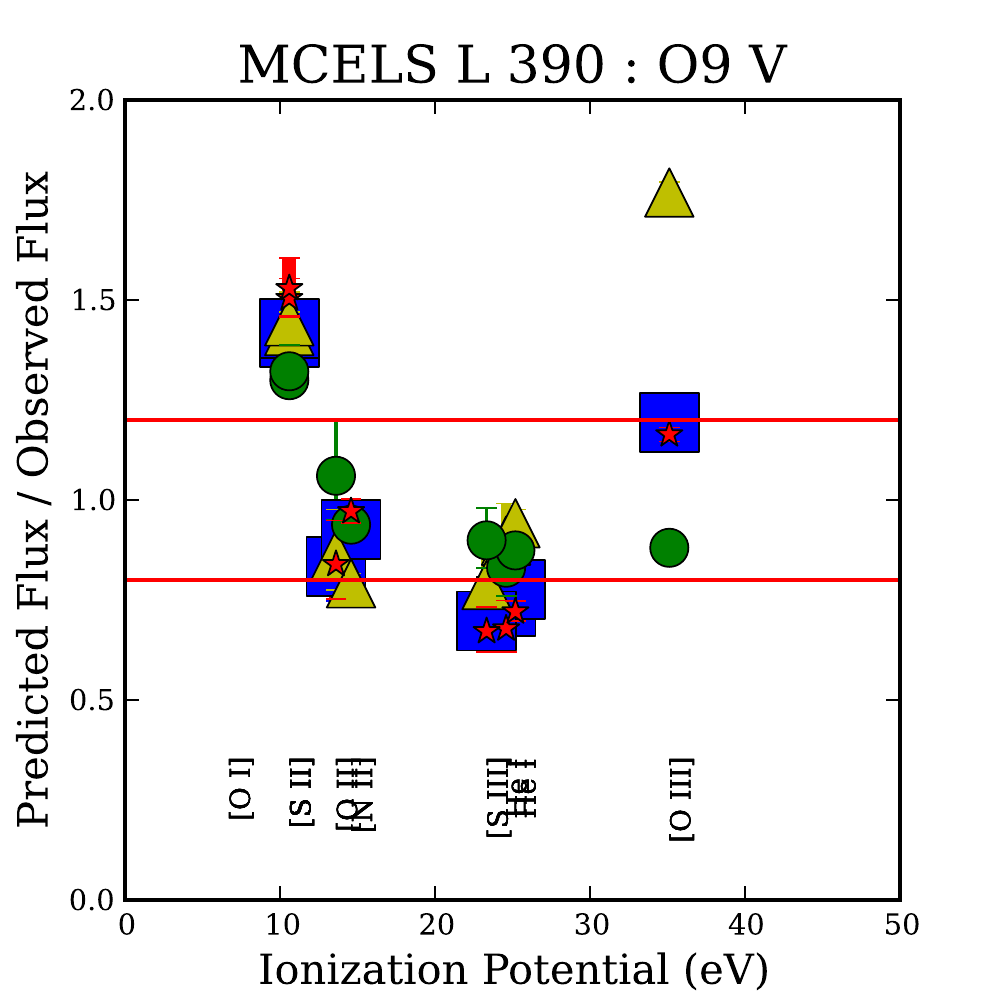}
\label{IPeV320}}
\subfigure[MCELS~L~394]{
\includegraphics[width=4.5cm]{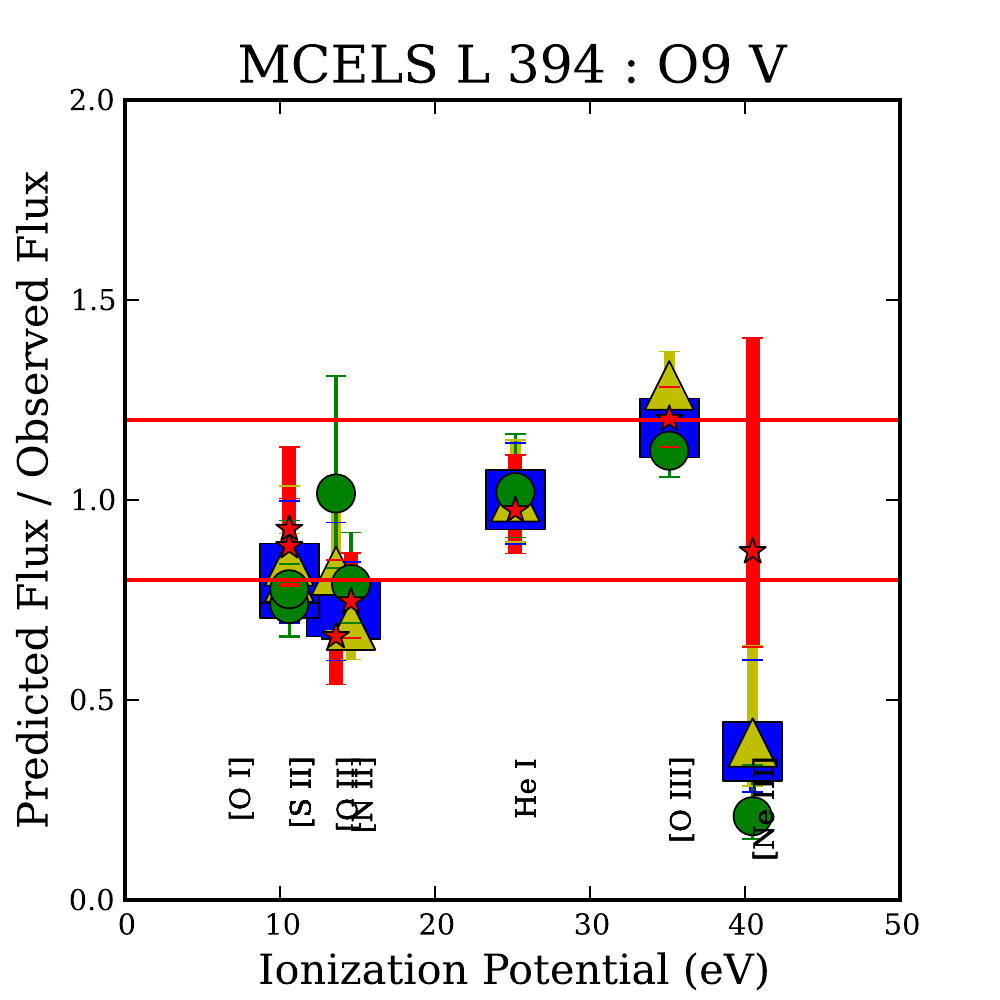}
\label{IPeV324}} 
\subfigure[MCELS~L~35 ]{
\includegraphics[width=4.5cm]{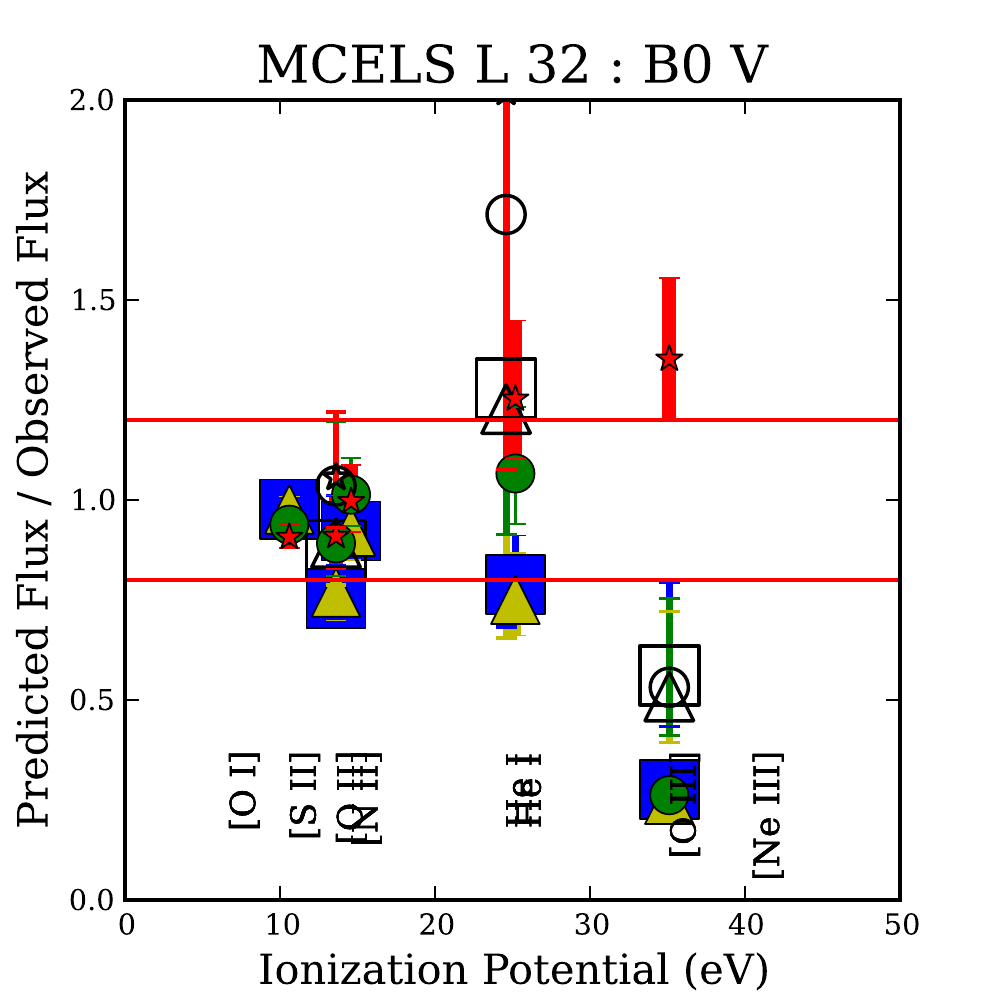}
\label{IPeV004}}
\subfigure[MCELS~L~32]{
\includegraphics[width=4.5cm]{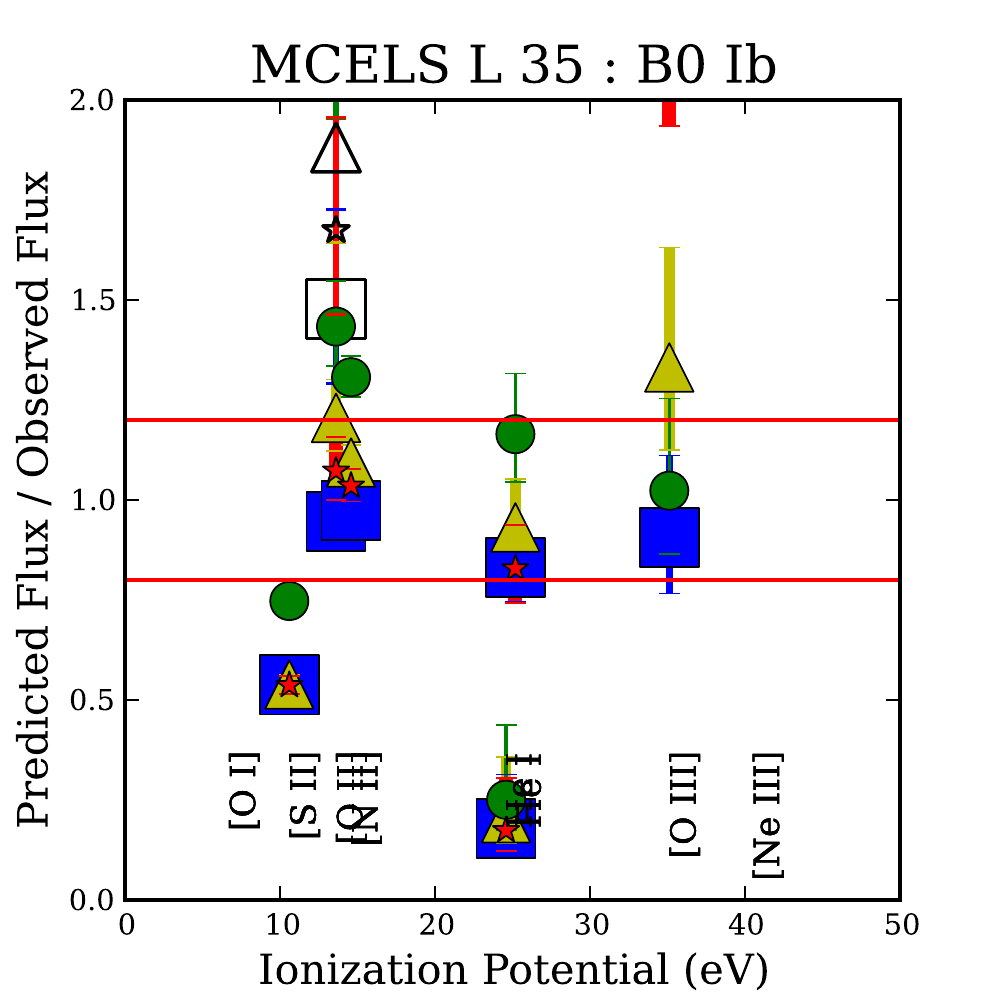}
\label{IPeVm32}}
\subfigure[MCELS~L~345]{
\includegraphics[width=4.5cm]{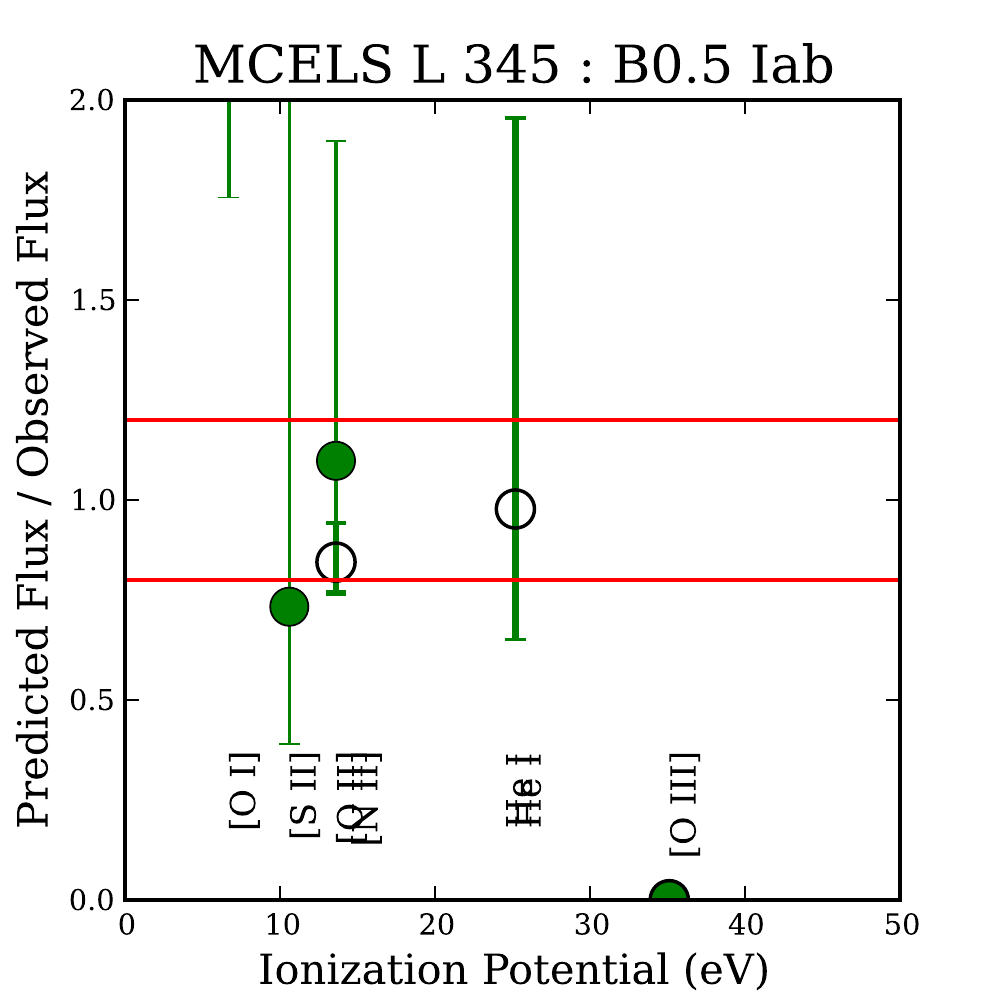}
\label{IPeV278}}
\subfigure[MCELS~L~346]{
\includegraphics[width=4.5cm]{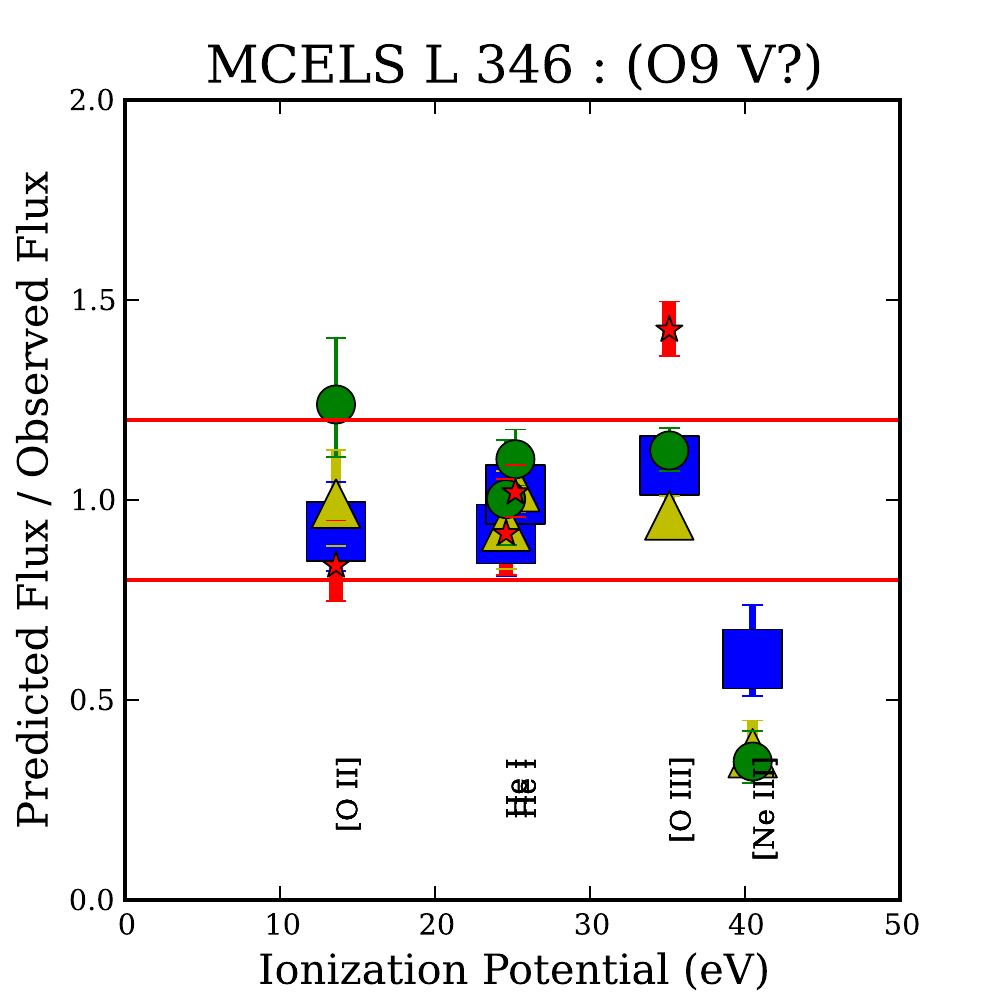}
\label{IPeV275}}
\caption{\footnotesize The ratio of the predicted flux to the observed flux is plotted as a function of the ionization potential of the emission line for the simulation that has the closest match to our observations.  The different symbols correspond to CLOUDY simulations that are ionized by different atmosphere models: CoStar, TLUSTY, WM-basic, and SNC02 are denoted by the red stars, green circles, blue squares, and yellow triangles, respectively. The filled data points use the observed flux from the night of 2008 Jan 30 or 31, while the hollow data points are from 2008 Jan 29.\label{IPeV_set1}}
\end{figure*}}

{
\begin{figure}[hb]
\includegraphics[width=8.5cm]{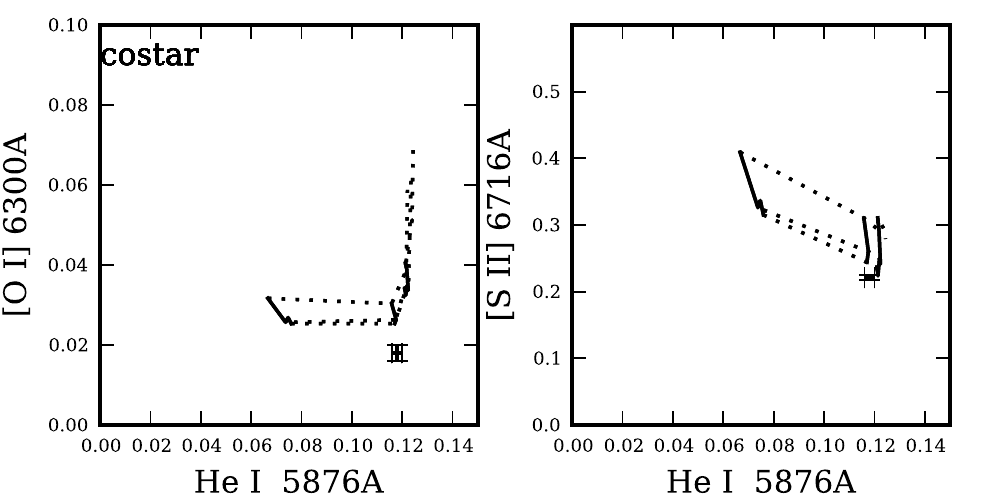}
\includegraphics[width=8.5cm]{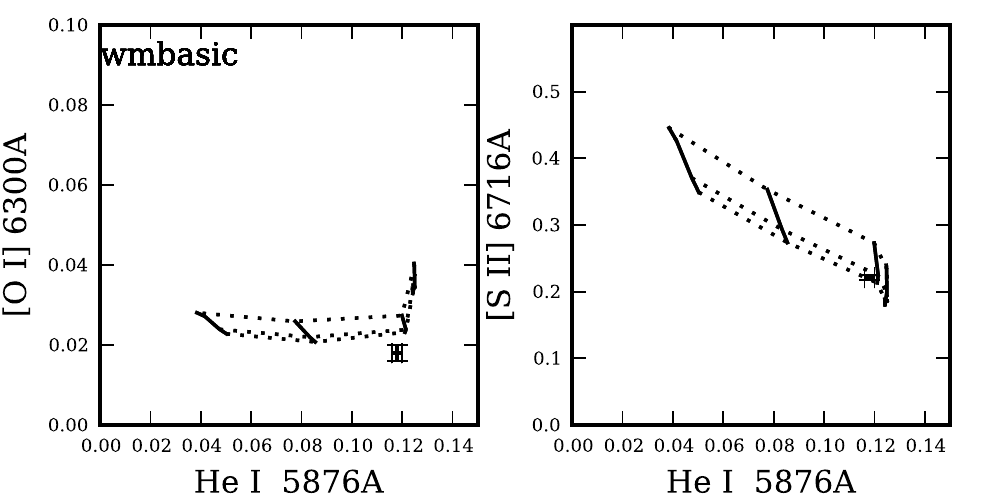}
\includegraphics[width=8.5cm]{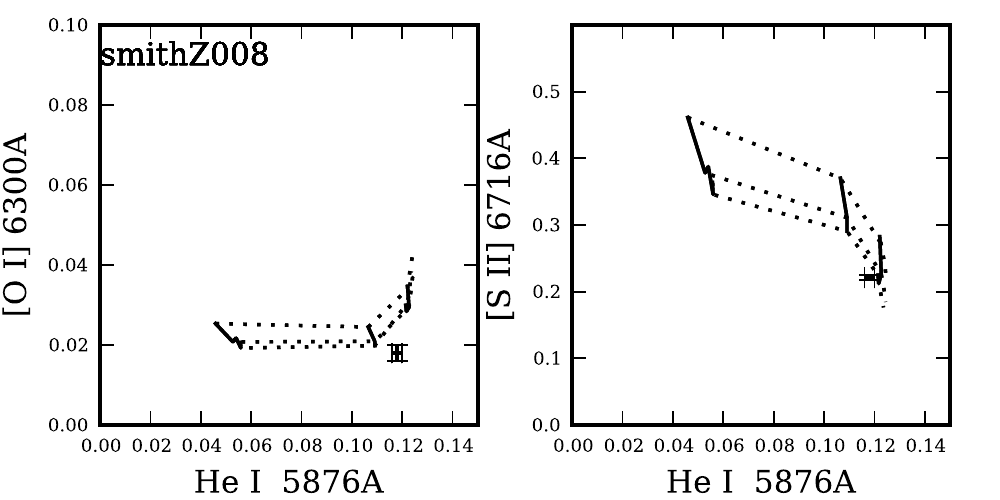}
\includegraphics[width=8.5cm]{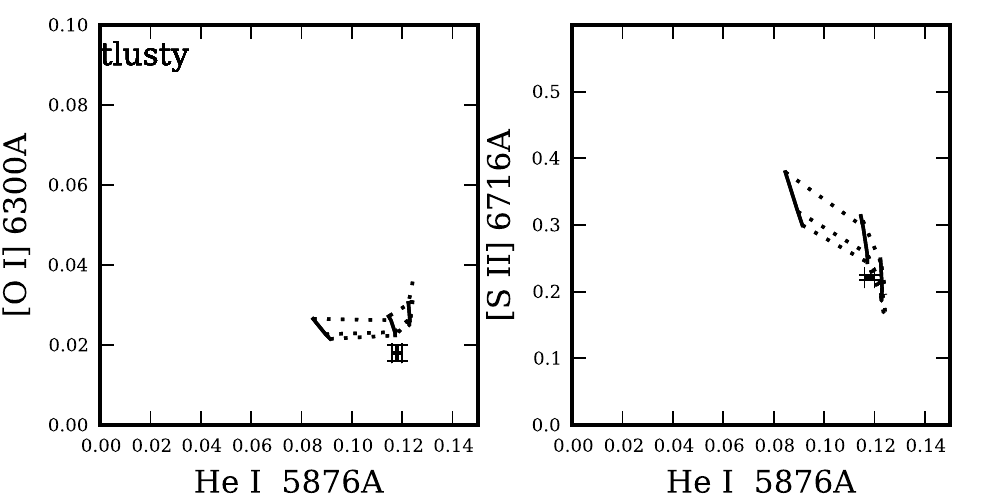}
\caption{\footnotesize \oi\ $\lambda6300$ and \sii\ $\lambda6716$ vs \hei\ $\lambda6876$ for MCELS~L~28.  Dotted lines are models with constant \Rin=0.10, 0.25, and 0.50 \Rout\ and solid lines have constant \Teff = 35, 37, 39 kK, except WM-basic, which also includes \Teff = 40 kK.  The values of \Rin\ increase going upward and \Teff\ increases from left to right. \label{O1S2}}
\end{figure}}

While they are included in Figure \ref{IPeV_set1} for completeness, we find that the \oi\ $\lambda6300$ and \sii\ $\lambda6716,6731$ emission lines are not good diagnostics for distinguishing between the SEDs.  These lines are primarily dependent on nebular conditions, and less so on the particular atmosphere chosen.  This can be seen in Figure \ref{O1S2}, where we have plotted the predictions for \hei\ $\lambda5876$ against those of \oi\ and \sii\ from a simulation grid that varies $\epsilon$\ and \Teff\ for MCELS~L~28.  Given the same nebular conditions, \oi\ and \sii\ are insensitive to different ionizing SEDs.  This is particularly true for \sii, which changes by up to 25\% for a large range in \Teff.  A similar behavior is observed for \oi, although we do see significant changes in \oi\ at high \Teff.  This might be explained by their relevant excitation mechanisms. In the case of \sii, the ionization potential of this line is 10.60 eV, which is below that of H. In this regime, the differences between atmosphere models are small, and so the line will be more sensitive to changes in the gas density and morphology. \oi\ $\lambda6300$, on the other hand, is produced via charge-exchange and, in the conditions of a typical nebula, depends critically on the density of H$^{+}$ \citep{b:Osterbrock_06}.

\subsection{Rate of Ionizing Photons \label{s:q0}}

One important comparison is between the rate of ionizing photons produced by the observed nebulae, \Qha, and that of the best fitting atmosphere model, \Qsed.  To calculate \Qha, we obtain \Halpha\ photometry from the MCELS, \Halpha\ emission-line image (Table \ref{t:obsprops}).  The rate of ionizing photons calculated from this photometry is listed in column 2 of Table \ref{t:ionphot}.  We note that the MCELS narrowband images are not continuum subtracted.  As a result, the error on \Qha\ $\sim 20\%$\ \citep{b:Pellegrini_apj12}.  Columns 3-6 show the ratio of \Qsed\ and the measured \Qha. 

It is immediately apparent that the \Qsed\ for dwarf stars is offset from \Qha\ by +5, -12, -12, and -37 \% on average for TLUSTY, WM-basic, SNC02, and CoStar, respectively. This excludes MCELS L394, for which \Qsed\ is 3--5 times higher than \Qha\ (see \S \ref{s:indob}), and the two supergiants.  From this diagnostic, TLUSTY, best represents the ionizing source.  However, both WM-basic and SNC02 also do well, particularly within the 20\% error on the \Halpha\ photometry.  We note that the offset does depend on atmosphere model, but we find no trend with spectral type or $A_V$.  

One important factor that affects this comparison is the assumed gravity. In our simulations, we use log($g$)= 4.0 for all luminosity class V stars. However, these stars may have log($g$) in the range from 3.8 to 4.1 \citep[e.g.,][]{b:Herrero_aap99,b:Massey_apj05,b:Martins_aap05}.  Decreasing log($g$) by just 0.1 dex can increase \Qsed\ by as much as a factor of 2.  This change more than compensates for any discrepancy shown in Table \ref{t:ionphot}, and supports that \Qsed\ is consistent with \Qha, within the uncertainty.

We note that \Teff\ will affect \Qsed. Both the bolometric correction and the shape of the SED play a role in determining $Q_0$, and both properties depend on \Teff.  As noted in \S \ref{s:nfluc}, the error on \Teff\ is between 500 and 1000 K. If \Teff\ is too low by 1000 K, \Qsed\ would increase by 10--15\%.  However, a change in \Teff\ of 1000 K significantly affects the ionization structure of the nebula, which worsens many of the fits shown in Figure \ref{IPeV_set1}. Thus, uncertainty in \Teff\ does not significantly contribute to any offset in $\Qsed/\Qha$.  

In addition to the general offsets in \Qsed/\Qha, we also find differences between model grids.  In general, TLUSTY has the highest \Qsed/\Qha, with most values near unity, and CoStar has the lowest, with many values near 0.6. WM-basic and SNC02 fall between them.  The trend between atmosphere models reflects the important role that the relative hardness of the SED plays in the temperature selection process, which in turn affects $Q_0$.  For a given effective temperature, different atmosphere models will produce different $Q_0$ \citep[e.g.,][]{b:Voges_aj08,b:Simon-Diaz_mnras08}. On top of that, we are sensitive to the shape of the ionizing SED because our atmosphere-model selection criteria depend on matching the emission-line ratios.  This means that for a softer atmosphere model (e.g., TLUSTY), we select simulations with higher effective temperatures than we would for simulations using harder atmosphere models (e.g., CoStar).  This effect leads to the trend seen above: the best-fitting models for TLUSTY have higher $Q_0$ than those of CoStar.  Therefore, while the systematic offset in $Q_0$ is most likely caused by uncertainty in log($g$), it may also suggest that the WM-basic, SNC02, and CoStar  atmosphere models are generally harder than the observed stars.  However, we note that this is opposite from what we find based on the \neiii\ line, as we discuss in \S\ref{s:res_mod}.  

{
\begin{deluxetable*}{ccl|cccc}
  \tablewidth{0pt}
  \tabletypesize{\small}
  \tablecaption{$Q_0$\ Comparison \label{t:ionphot}}
  \tablehead{\multicolumn{3}{c}{} & 
	  \multicolumn{4}{c}{\Qsed/\Qha} \\
	  \colhead{MCELS} &
	  \colhead{SpT} &
	  \colhead{\Qha} &
	  \colhead{CoStar} &
	  \colhead{TLUSTY} &
	  \colhead{SNC02} &
	  \colhead{WM-basic}}
\startdata
L 28&O5.5 V&48.94&0.51&0.95&0.67&0.71\\ 
L 32&B0 V&47.91&1.01&1.27&1.08&0.84\\
L 35&B0 Ib&48.20&0.95&2.40&1.62&0.95\\
L 43&O8 V&48.47&0.53&0.97&0.79&0.74\\
L 52&O6.5 V&48.90&0.39&0.80&0.64&0.62\\
L 344&O7-6.5 V((f))&49.02&0.54&0.99&0.80&0.75\\
L 345&B0.5 Iab&48.18&\nodata&0.16-0.47&\nodata&\nodata\\
L 346&O9 V$^b$&48.48&1.06&1.84&1.27&1.36\\
L 351&O6.5 V&49.37&0.27&0.55&0.42&0.39\\
L 390&O9 V&48.65&0.66&1.01&1.16&1.60\\
L 394&O9 V&47.99&2.94&5.36&4.37&4.08\\
L 283b$^a$&O6.5 V((f))&48.51&0.70&1.44&1.10&1.02
\enddata
\tablenotetext{a}{Refers to the DEM catalog number}
\tablenotetext{b}{Eclipsing binary for which the SpT of the ionizing source inferred and not observed, see \S\ref{s:indob}}
\end{deluxetable*}}

\subsection{Individual Objects}\label{s:indob}

In this section we briefly comment on the individual objects in our sample.  Unless otherwise noted, we assume the mean LMC abundances from \citet{b:Garnett_99}.  Table \ref{t:flux} lists the Helium abundances, which we calculate from the \hei\ lines using the relations from \citet{b:Benjamin_apj02}.  As a reference, Table \ref{t:modparams} shows the parameters of our best fitting models.

\textbf{MCELS L 28} is ionized by OGLE-III 8225 \citep{b:Udalski_actaa08}, which is the earliest spectral type in our sample, an O5.5 V star (Figures \ref{im456} and \ref{IPeV8c}).  In this nebula, we detect the auroral \oiii\ $\lambda4363$ emission line, from which we derive $\log(\rm O/H) = -3.51$.  The simulations with the best fit have $\epsilon$=0.10 and $n_H$ = 75 \cmcub. As noted in \S\ref{s:singlen}, based only on the oxygen lines, the TLUSTY simulation shows a good fit using $\epsilon$=1.0 and $n_H = 15$ \cmcub. However, for that simulation, both \neiii\ and \nii\ are under-predicted.  Figure \ref{IPeV8c} clearly shows this is not the case in the $\epsilon=0.10$ models.  If we had selected the $\epsilon=1$ model as our final best fit, the corresponding \Teff\ would be 39,000 K.  This is 3,500 K cooler than the adopted \Teff=42,500 K selected with the lower filling factor.  This \Teff\ difference highlights the importance of using as many emission-line diagnostics as available to determine the fits.  The best fitting \Teff\ are 39500, 42500, 41500, and 41500 K for CoStar, TLUSTY, WM-basic, and SNC02, respectively.

\textbf{MCELS L 52} is ionized by M2002 19696 \citep{b:Massey_apjs02}, an O6.5 V star (Figures \ref{im9} and \ref{IPeV26}).  We find excellent agreement for lines with ionization potential (IP) below 40 eV, and all the atmosphere models produce consistent results. For \neiii, whose IP = 40.96, this agreement does not hold. The simulation using CoStar, the hardest atmosphere, over-predicts \neiii, and TLUSTY under-predicts it.  Interestingly, we find somewhat different nebular parameters produce the best fitting simulations between CoStar and the other atmosphere models.  For CoStar, a slightly higher filling factor and lower density, 0.25 and 25 \cmcub, respectively, work best.  Whereas, $\epsilon=0.10$ and $n_H=60\ \cmcub$ produce better fits with the other atmosphere models. We note that this difference is close to our error in determining $\epsilon$ of 0.10, and, thus, might not be significant.   The best fitting \Teff\ are 36000, 38500, 38750, and 38500 K for CoStar, TLUSTY, WM-basic, and SNC02, respectively.

\textbf{MCELS L 351} is ionized by OGLE-III 43846, which is also an O6.5 V star and, with a radius of 11 pc, it is the largest nebula in our sample (Figures \ref{im25} and \ref{IPeV281}).  We detect \oiii\ $\lambda4363$\ in this object, from which we calculate \Te\ and log(O/H) (Table \ref{t:flux}). However, our \oiii\ $\lambda4363$\ detection is at a very poor signal-to-noise, and our result is consistent with the mean LMC metallicity. Therefore, we use the mean LMC metallicity in the CLOUDY simulations.  As for MCELS~L~52, we find excellent agreement between simulations and observations for emission lines with IP $<40$\ eV. The best fitting \Teff\ are 36500, 39500, 39000, and 39000 K for CoStar, TLUSTY, WM-basic, and SNC02, respectively.

\textbf{DEM L 283b} is ionized by OGLE-III 44989, an O6.5 V((f)) star (Figures \ref{im34} and \ref{IPeV283b}).  Since we do not have IMACS photometry for this object, we use the OGLE-III \emph{V}-band magnitude to calculate the stellar luminosity.  From the \oiii\ $\lambda5007$ line profile, we assign \Rin\ = 0.25 \Rout. Overall, we find good fits for all atmosphere models for lines with IP $< 40$ eV and $\epsilon = 0.10$. The best fitting \Teff\ are 36500, 39000, 39000, and 39000 K for CoStar, TLUSTY, WM-basic, and SNC02, respectively.

\textbf{MCELS L 344} is ionized by OGLE-III 45830, an O6.5-7 V((f)) star (Figures \ref{im3133} and \ref{IPeV276}).  We note that the nebular structure diverges from a Str\"{o}mgren sphere to the east in Figure \ref{im3133}.  However, our slit position runs N-S and covers the regular portion of the nebula.  Overall we find excellent fits for models where IP $< 40$ eV.  The measured \neiii\ flux shows a 45\% difference between the spectra from 2008 Jan 29 and 31, which results in a large scatter for \neiii\ in Figure \ref{IPeV276}.  This apparent discrepancy is caused by the large observational uncertainty.  Interestingly, for Jan 31, the simulation using CoStar is the only one to match the observations, and the others under-predict \neiii\ by $> 50\%$.  The best fitting \Teff\ are 37000, 39500, 39000, and 39000 K for CoStar, TLUSTY, WM-basic, and SNC02, respectively.

\textbf{MCELS L 43} is ionized by M2002 17251, an O8 V star (Figures \ref{im7} and \ref{IPeV20}).  As noted in \S \ref{s:singlen}, MCELS L 43 is one of the few objects for which \oiii/\oii\ is well predicted by simulations using $\epsilon = 1$ with atmosphere models from WMbasic, SNC02 and CoStar. However, the right panel of Figure \ref{el_020} shows that these models cannot simultaneously match the \hei\ lines, which are a more reliable \Teff\ indicator.  At the temperatures required to match \hei, \oiii\ $\lambda5007$ is over predicted by a factor of 2--3.  The fit improves in all emission lines when we decrease the filling factor to 0.10 (Figure \ref{IPeV20}).  The best fitting \Teff\ are 37000, 39000, 39000, and 39000 K for CoStar, TLUSTY, WM-basic, and SNC02, respectively.  

\textbf{MCELS L 390} is ionized by OGLE-III 28307, an O9 V star (Figures \ref{im2122} and \ref{IPeV320}). For this object, the simulation using CoStar reproduces the relative amounts of \oiii\ and \neiii, while the others do not. This is in contrast to the rest of our sample, for which the CoStar models are too hard.  This star is an eclipsing binary, according to the OGLE-III catalog of eclipsing binaries \citep{b:Graczyk_actaa11}. However, the second star only contributes 10\% of the \emph{I} band flux.  Futhermore, Table \ref{t:ionphot} shows agreement between \Qsed\ and \Qha.  Therefore, the second star is not significant.  The best fitting \Teff\ are 35000, 35000, 37000, and 37000 K for CoStar, TLUSTY, WM-basic, and SNC02, respectively.

\textbf{MCELS L 394} is ionized by OGLE-III 28239, a star whose spectral type is between an O9 and a B1.5 V, based on the noisy stellar spectrum (Figure \ref{im2122} and \ref{IPeV324}).  However, we assign this star the earlier SpT in this range because the observed \neiii\ rules out stars with SpT of B0 V or later.  Even with $\epsilon$ = 0.10, it is still challenging to reproduce the \sii, \oii\ and \nii\ lines.  As seen in Figure \ref{IPeV324} the predictions lie at the edge of our $20\%$\ tolerance limit. Additionally, all the atmosphere models, except for CoStar, do not produce enough flux at the energies needed to reproduce the observed \neiii\ emission-line (Figure \ref{IPeV324}). The best fitting \Teff\ are 37000, 39000, 39000, and 39000 K for CoStar, TLUSTY, WM-basic, and SNC02, respectively.

Unlike the other objects, \Qha\ in MCELS~L~394 is a factor of a few less than \Qsed\ (Table \ref{t:ionphot}).  This could indicate that the nebula is optically thin.  If the nebula were optically thin, we expect to observe less \oii\ and \sii\ than if the nebula were optically thick.  Based on our modeling, however, we see the opposite. We observe too much \oii\ and \sii\ relative to the optically thick model.  Therefore, while  the \Halpha\ photometry may suggest an optically thin nebula, it is clear that this does not adequately explain all the observations. 

\textbf{MCELS L 32} is ionized by OGLE-III 8229, a B0 V star (Figure \ref{im456} and \ref{IPeV004}).  \neiii\ is not detected in this nebula, and thus provides a firm upper limit for the \Teff\ of the ionizing star.  Both of the grids generated with the WM-basic code include atmosphere models down to \Teff = 30,000 K, which is adequate for this B0 star. However, the CoStar grid in CLOUDY does not have dwarf B star models. The coolest CoStar dwarf, \Teff=34,000 K, is still a bit too hard to match the observed emission lines.   We optimized the \Teff\ to reproduce the \hei\ lines, with the result that \oiii\ is under-predicted.  The best fitting \Teff\ are 34000, 33000, 34000, and 34000 K for CoStar, TLUSTY, WM-basic, and SNC02, respectively.
 
\textbf{MCELS L 35} is ionized by OGLE-III 8203, one of our two supergiants, a B0 Ib star (Figure \ref{im456} and \ref{IPeVm32}).  In addition to varying the \Teff\ and $\epsilon$, we vary the surface gravity between log($g) = 2.5 - 3.5$, as appropriate for supergiants \citep{b:Crowther_aap06,b:Trundle_aap04}.  TLUSTY is the only atmosphere code which includes this whole range of gravities and temperatures. The lowest gravity in the available grids for WM-basic and SNC02 are 3.0 and 2.95, respectively.  In the case of CoStar, the lowest available gravity for \Teff$>$26,000 K is log($g$) = 2.86.  For TLUSTY, we find that simulations using log($g) = 2.6-3.0$ all yield good fits, although at different temperatures. For example, for log($g) = 2.6$, the best fitting \Teff = 27,250 K, whereas for log($g) = 3.0$, \Teff = 28,500 K.  For WM-basic, SNC02 and CoStar we use log($g) = 3.5$ and find \Teff = 31,000 K for all three atmospheres. 

\textbf{MCELS L 345} is ionized by OGLE-III 50093, the other supergiant in our sample, a B0.5 Iab star (Figure \ref{im3133} and \ref{IPeV278}). Based on the \nii/\oii\ ratio, this object has a much higher nitrogen abundance than the rest of our sample, log($\rm N/O) = -0.77$ compared to the mean LMC value of log($\rm N/O) = -1.5$. This higher abundance is confirmed by the simulations because those that use the mean value are unable to reproduce the \nii/\oii\ line ratio. \oiii\ $\lambda5007$ is not detected in this object, which places an upper limit of \Teff=26,000 K.  As discussed for MCELS L 32, we are only able to test the TLUSTY atmosphere models for this object.  Note, the points along the bottom of Figure \ref{IPeV278} are lines for which we only have upper limits on the line strength, and the simulations are consistent with the data.  We find the best fitting \Teff\ falls in the range of 22,000 - 24,000 K.

\textbf{MCELS L 346} is ionized by an eclipsing binary, MACHO 81.9725.16 \citep{b:Alcock_aj97}.  The spectral type determined from our observed data is B1-1.5 V (Figure \ref{stellarspec}).  However, we do detect \neiii\ in this object, despite the late SpT.  The strengths of the \neiii\ and \oiii\ lines suggest that this nebula is ionized by a star of much earlier spectral type.  Based on the \oiii/\Hbeta\ and \neiii/\Hbeta\ line ratios compared to the rest of our sample, the ionizing star in this nebula likely has a SpT between O8 V and O9.5 V.  This analysis is further supported by the measured log(\Qha)=48.48, which falls into the [47.99-48.65] range spanned by the O8 -- 9 V stars in our sample.  Considering the short period of this binary, 1.2 days, it is possible that we observed the nebula when the earlier type star was eclipsed by a later type companion.  Since the early type star will dominate the ionizing SED, we model this object with a single atmosphere model. 

We note that one side of this nebula has an irregular morphology, while the other is more Str\"{o}mgren-like (Figure \ref{im24}).  Therefore, we only use the emission-line ratios from the portion of the slit that lies along the regular half of the nebula.  The best fitting models are shown in Figure \ref{IPeV275}.  These models have \Rin = 0.25 \Rout, $\epsilon$ = 0.10, and $n_H = 100$\ \cmcub.  We find excellent fits for WM-basic, TLUSTY, and SNC02 except at \neiii\ $\lambda3869$, while CoStar over-predicts both \oiii\ and \neiii.  The best fitting \Teff\ are 36,000, 37,500, 38,000, and 37,000 K for CoStar, TLUSTY, WM-basic, and SNC02, respectively.

\section{Discussion} 
\label{s:results}

\subsection{Comparison of Different Atmosphere Models} 
\label{s:res_mod}

In the previous section we discussed the simulations for each object individually.  However, the question at hand is: how well do the atmosphere models reproduce the ionizing population, in general?   In this section, we discuss the trends we find across our sample. To look at the results collectively, we plot $F_{\rm pre}/F_{\rm obs}$ for individual emission lines as a function of the spectral type of the ionizing star. These plots are shown in Figure \ref{ions}. 

With the exception of \neiii\ $\lambda3869$ (Figure \ref{NeIIIfit}), \hii\ region simulations using CoStar are consistent with the other atmosphere models and the observations (Figure \ref{ions}).  For high ionization potential lines, such as \neiii\ shown in Figure \ref{NeIIIfit}, simulations with the CoStar atmospheres over-produce the line emission.  Other studies have similarly found that the CoStar SED is too hard at high energies  \citep[e.g.,][]{b:Oey_apjs00,b:Morisset_aap04,b:Mokiem_aap04}.  Additionally, we find that a small range of \Teff\ reproduces the observed nebular emission for most of our range in SpT (Figure \ref{sptteff}).  This implies that the SEDs generated by CoStar are less sensitive to changes in the atmospheres between stars of different SpT.  

{
\begin{figure*}[h]
\centering
\subfigure[\oii\ 3727\ang]{
\includegraphics[width=7.5cm]{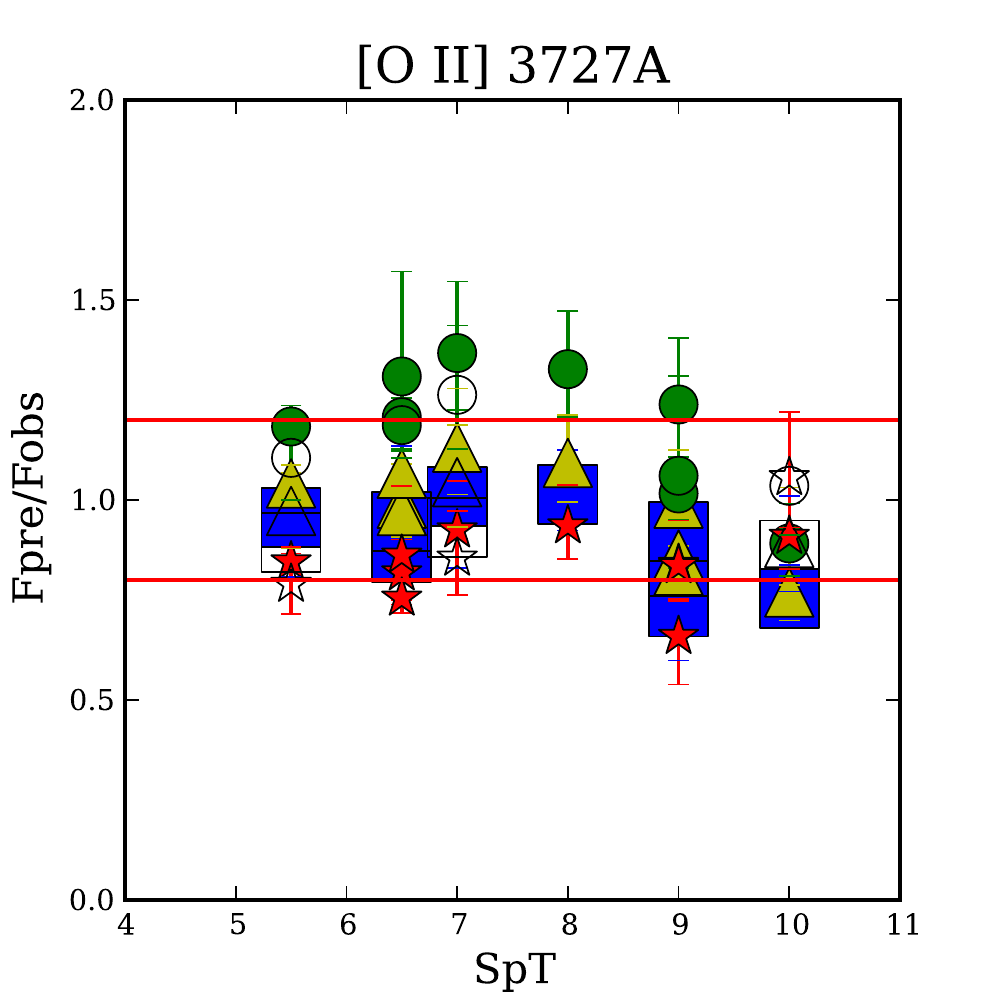}
\label{OIIfit}}   
\subfigure[\nii\ 6584\ang]{
\includegraphics[width=7.5cm]{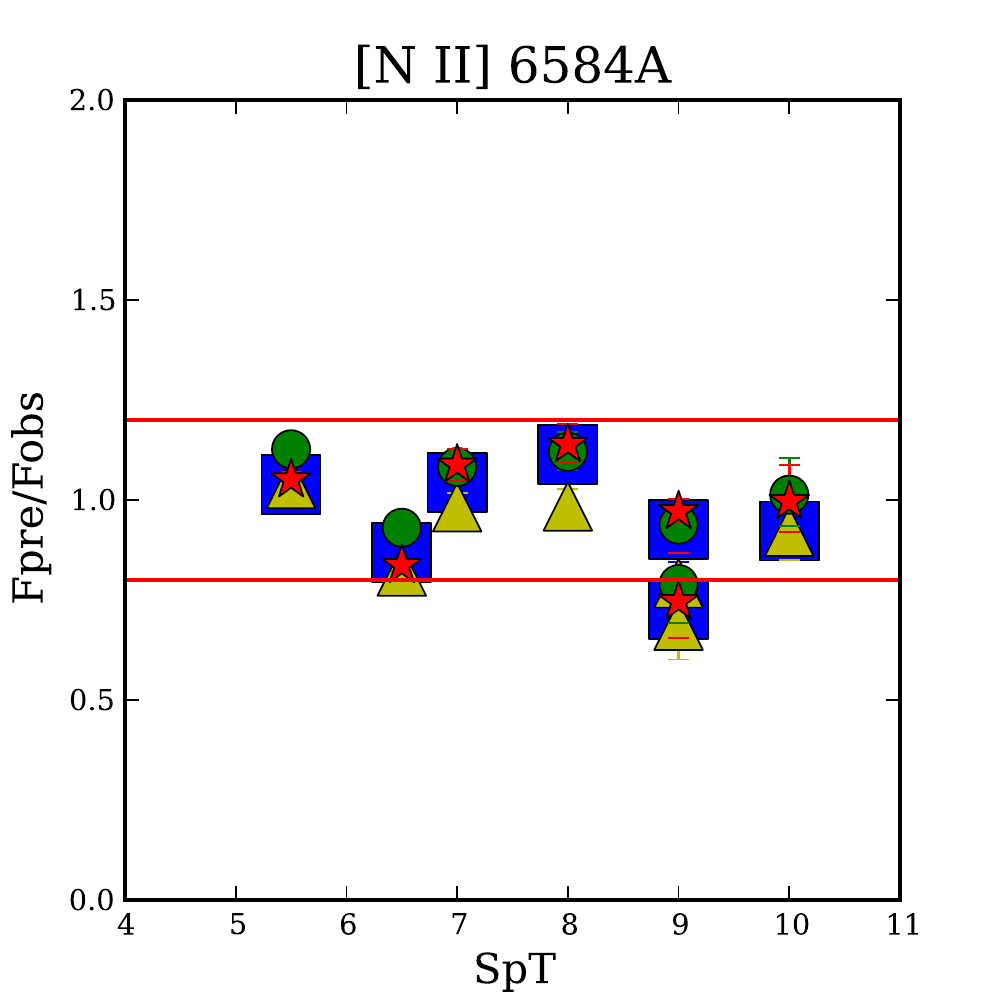}
\label{NIIfit}}
\\
\subfigure[\oiii\ 5007\ang]{
\includegraphics[width=7.5cm]{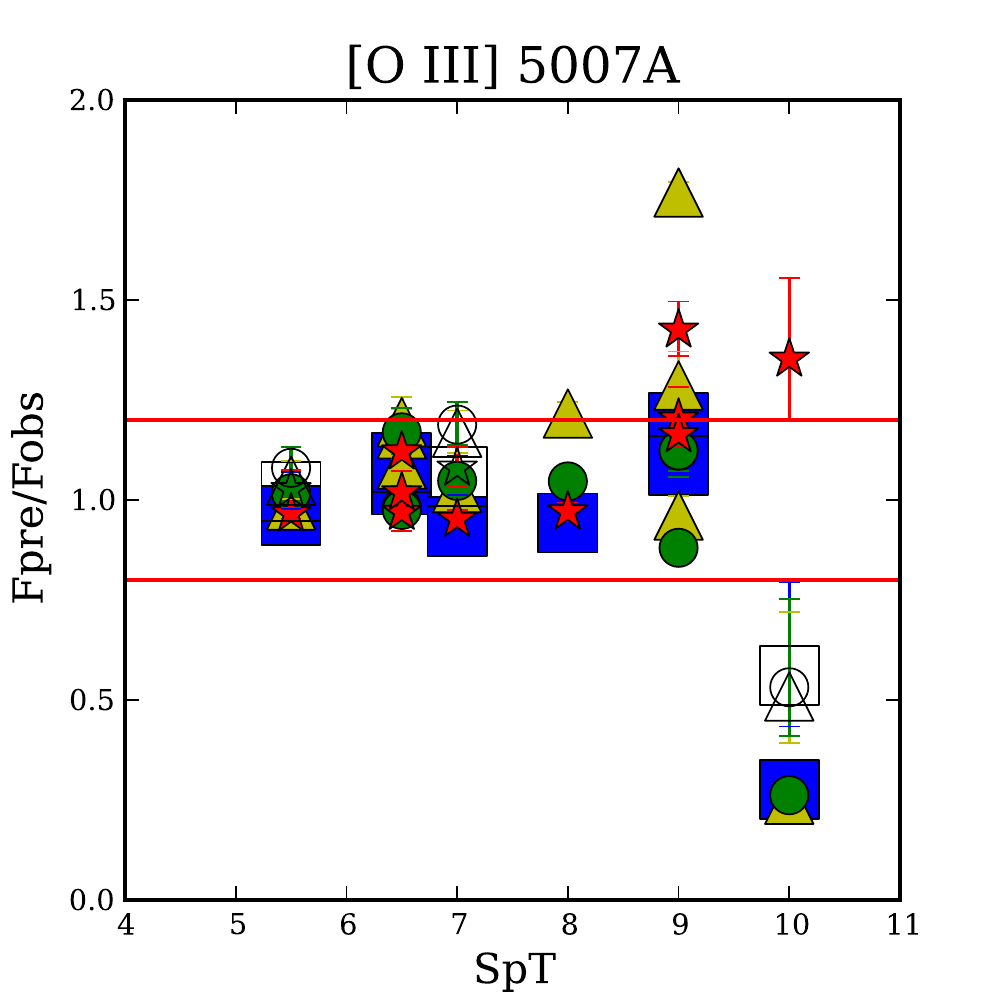}
\label{OIIIfit}}   
\subfigure[\neiii\ 3869\ang]{
\includegraphics[width=7.5cm]{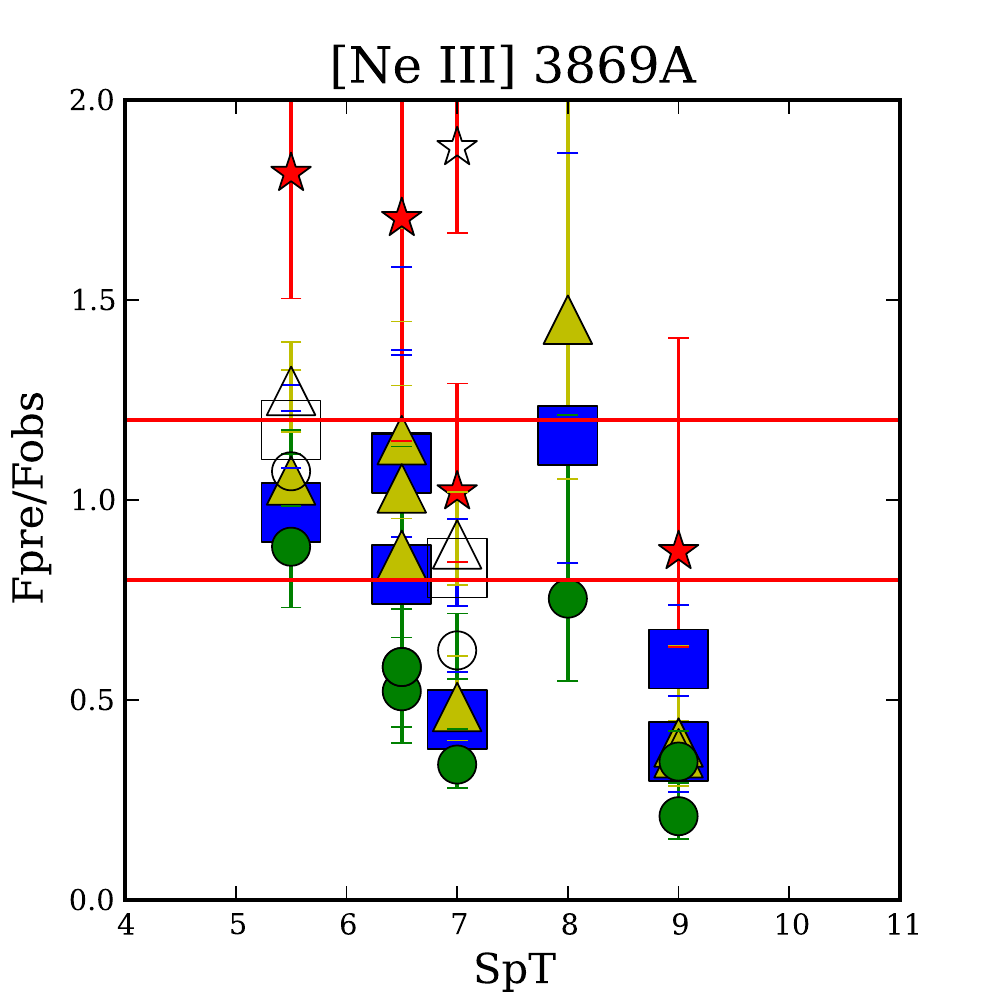}
\label{NeIIIfit}}
\caption{\footnotesize $F_{\rm pre}$ / $F_{\rm obs}$ as a function of spectral type for \oii\ $\lambda3727$, \nii\ $\lambda6584$, \oiii\ $\lambda5007$ and \neiii\ $\lambda3869$ in panels \emph{a,b,c} and \emph{d}, respectively. Colors and symbols are as for Figure \ref{IPeV_set1}. \label{ions}}   
\end{figure*}}

{
\begin{figure*}[h]
\centering
\subfigure[\Teff=41,000 K]{
\includegraphics[width=8.5cm]{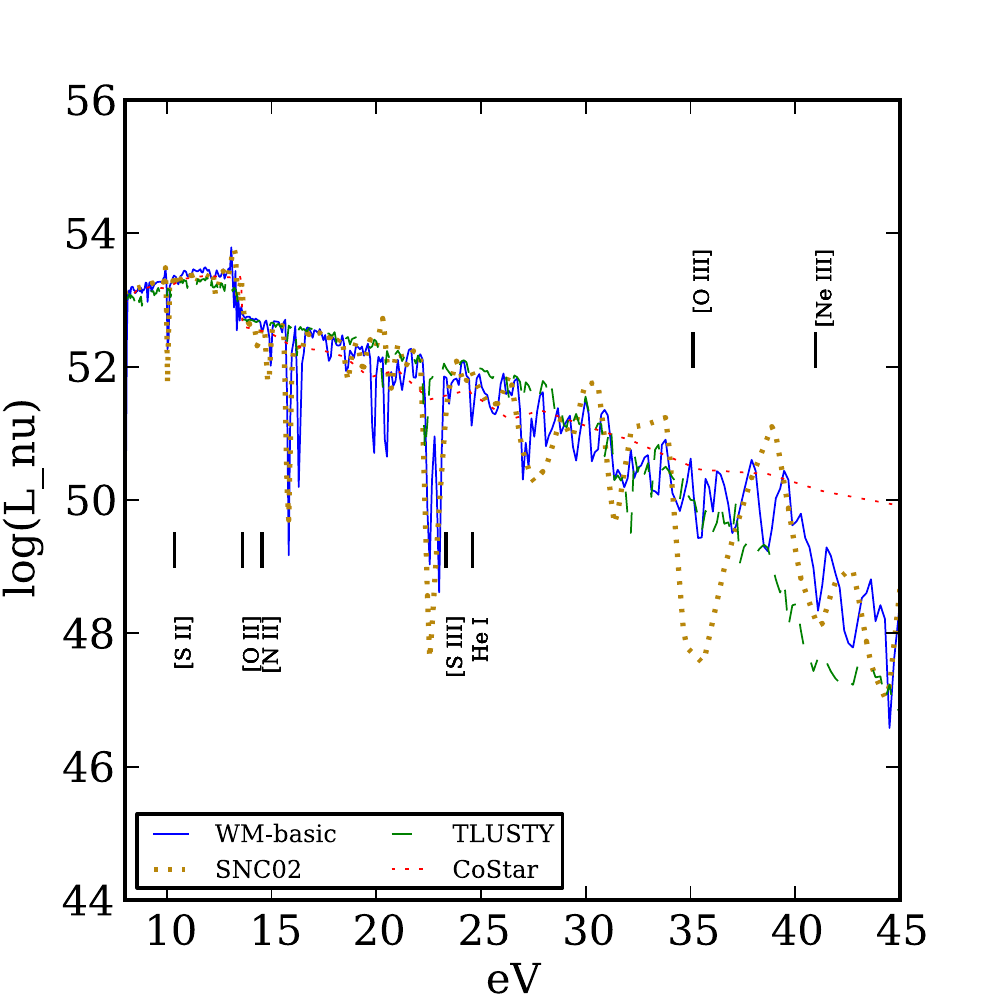}}
\subfigure[\Teff=35,000 K]{
\includegraphics[width=8.5cm]{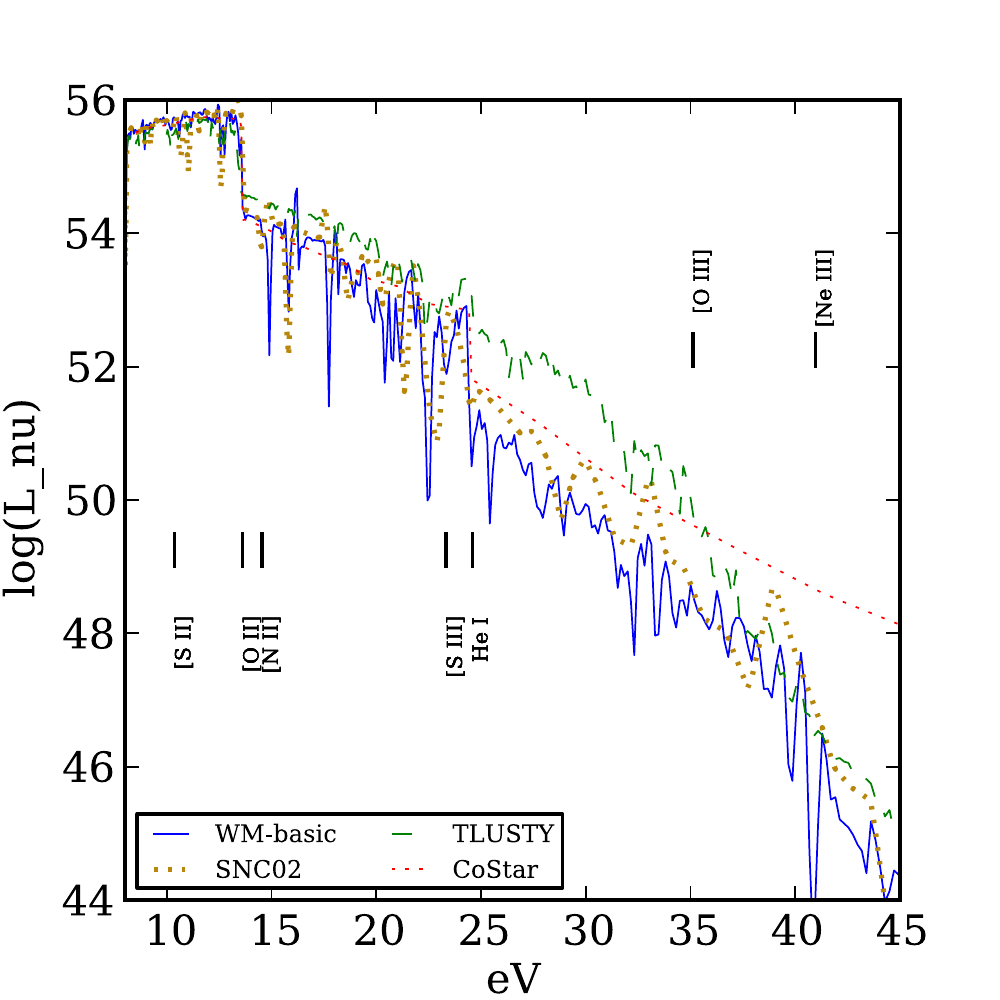}}
\caption{\footnotesize SEDs for stars with \Teff = 41,000 K and log($L_{\rm bol})=37.09$ (top) and \Teff = 35,000 K and log($L_{\rm bol})=37.78$ (bottom) for each stellar atmosphere model. \label{seds}}
\end{figure*}
}

As with CoStar, the TLUSTY models reproduce most of the emission lines up to the ionization potential of Ne$^{+2}$.  At the IP of Ne$^{+2}$, 41 eV, the modeled \hii\ region spectrum under-predicts the line emission, and it does so to a greater degree than the other atmosphere models.  This suggests that the TLUSTY atmosphere is not producing enough flux at high energies and reflects differences in the incident SEDs, as shown in Figure \ref{seds}.  Figure \ref{seds} compares the SEDs produced by different stellar atmosphere codes for two temperatures, \Teff=35,000 K and \Teff=41,000 K.  At both \Teff, the high energy slope of the TLUSTY SED is steeper than the other atmosphere models.  Similarly, \citet{b:Simon-Diaz_mnras08} show that for ionizing SEDs with $30,000\leq \Teff \leq 40,000$\ K, TLUSTY produces less flux at high energies than model atmospheres with expanding winds.  Using IR emission lines, \citet{b:Morisset_aap04} also find that plane-parallel model atmospheres, TLUSTY included, are too soft between 27 eV and 41 eV to reproduce the observed nebular emission.  This confirms that the treatment of the expanding wind plays an important role in reproducing the flux of ionizing photons with higher energies.  \citet{b:Gabler_aap89} showed that self-consistent treatment of the wind was particularly important for the continuum near the \heii\ ionizing edge at 54.4 eV.  It was also shown by \citet{b:Sellmaier_aap96} that the \neiii\ problem, in which photoionization models consistently under-predicted \neiii\ $\lambda3869$ \citep[e.g.,][]{b:Simpson_apj95,b:Simpson_apj86,b:Mathis_apj85,b:Rubin_apj91}, is not an issue in simulations using non-LTE atmospheres that have a wind extension.  We note that while there is a general trend for \neiii\ to be under predicted, the scatter is large and the degree to which \neiii\ is under predicted in our TLUSTY simulations is much less than that seen for earlier generations of stellar atmosphere models.

At the opposite energy regime, simulations using TLUSTY yield more \oii\ emission than the other atmosphere models, and typically over-predict \oii\ 3727\ang\ by 15-30\% (Figure \ref{OIIfit}).  While TLUSTY is softer \citep[][]{b:Simon-Diaz_mnras08}, it is surprising that this trend is not seen in the other lines with similar ionization potential, such as \nii\ $\lambda6584$ (Figure \ref{NIIfit}).  One explanation for this could be the relative shapes of the SEDs above the ionization potentials of N$^{+2}$ (29.6 eV) and O$^{+2}$ (35.1 eV), since the flux in that energy range controls the ionization from O$^+$ to O$^{+2}$ and N$^+$ to N$^{+2}$.  For energies greater than 35 eV, the luminosity of the TLUSTY SED drops relative to other models, at least for $\Teff < 40,000$\ K \citep{b:Simon-Diaz_mnras08}.  Thus, the slight excess of \oii\ may result from the softer SED.

Simulations using atmospheres generated with the WM-basic code are the closest to the observed values for most emission lines (Figure \ref{ions}\emph{a-c}).  However, Figure \ref{NeIIIfit} shows that the predictions for \neiii\ $\lambda3869$ have more scatter than what is seen for the emission lines of lower IP.  The ratio of predicted and observed flux for \neiii\ ranges from 0.7 to 1.7, with most points under-predicted by 20-30\%.  As expected, the atmosphere models presented in the SNC02 grid are consistent with the WM-basic grid implemented in CLOUDY,  although we find that they occasionally require a slightly different \Teff.  This is likely due to the different stellar properties (stellar radius, mass loss rate and wind terminal velocity) used in generating the atmosphere model grid, as discussed in \S \ref{s:moddesc}.  We note that the temperature differences, when they occur, are in the range of 250--750 K.  Although the predictions from the two grids are close, the WM-basic grid does slightly better than the SNC02 grid in reproducing \oiii\ for late SpT stars. 

\subsection{Comparison between SpT and \Teff}\label{s:SpTEf}

The calibration between spectral type and effective temperature is an important consequence of the atmosphere modeling.  Ordinarily, the calibration is based on careful fitting of photospheric absorption lines in model SEDs to reproduce high-resolution stellar spectra.  Therefore, the SpT-\Teff\ calibration depends on the stellar atmosphere models used in the fitting.  This bias is especially apparent when one compares the calibrations based on atmospheres codes without line blanketing \citep[e.g.,][]{b:Vacca_apj96} to calibrations based on atmospheres with line blanketing \citep[e.g.,][]{b:Martins_aap05,b:Massey_apj05}.  The latter calibrations assign effective temperatures that are several thousand degrees cooler than the previous calibrations.

{
\begin{figure*}[h]
\centering
\includegraphics[width=15cm]{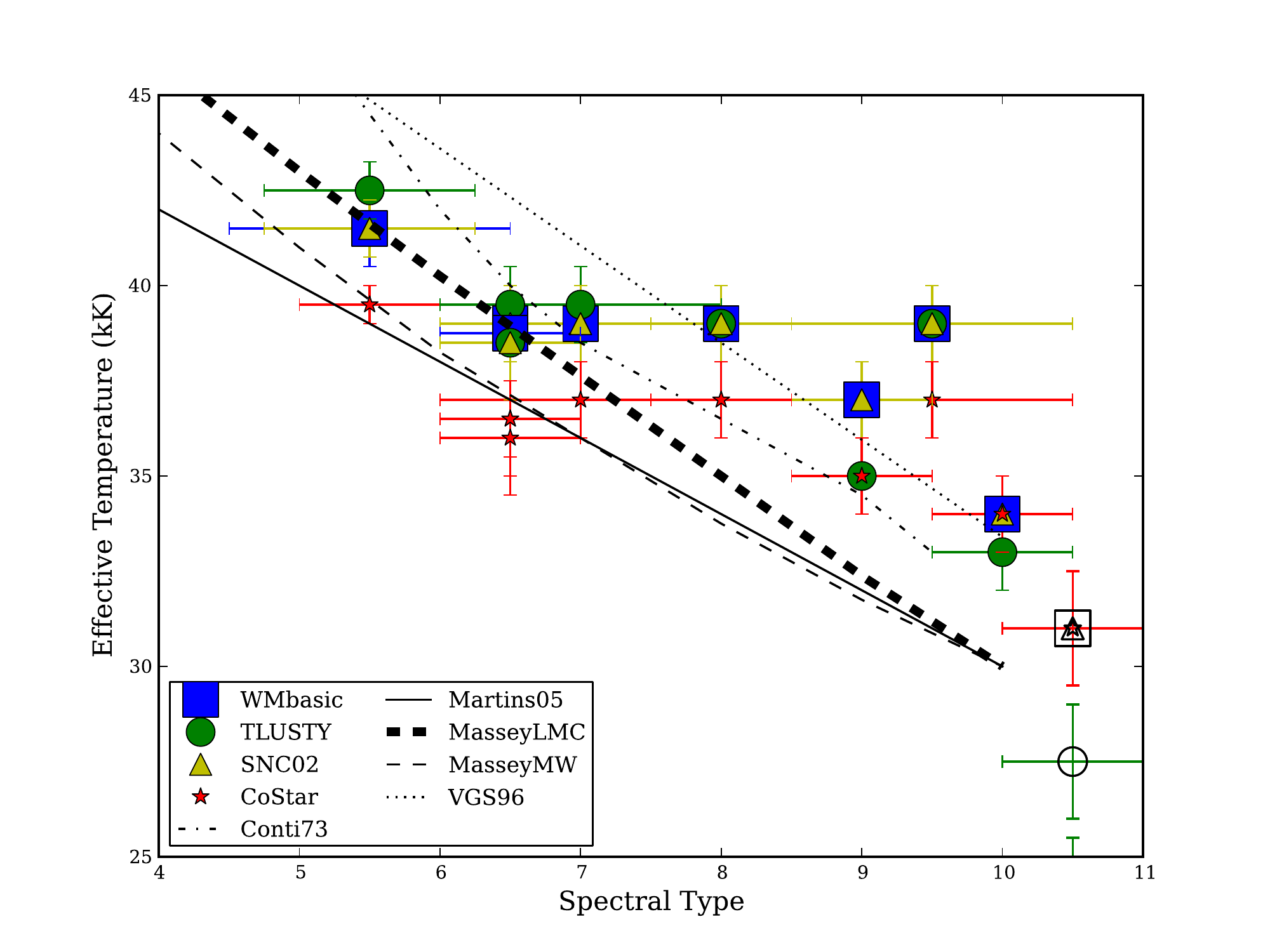}
\caption{\footnotesize A comparison of the \Teff\ from this paper to different SpT-\Teff\ calibrations.  Symbols represent temperatures derived from this work, while the lines are SpT-\Teff\ calibrations for dwarf stars in the literature.  Dash-dotted, solid, thin- and thick-dashed, and dotted correspond to calibrations from \citet{b:Conti_apj73}, \citet{b:Martins_aap05}, \citet{b:Massey_apj05}, and  \citet{b:Vacca_apj96} respectively.  We note that the thick dashed line is a calibration for the LMC while the rest are derived from Galactic stars. The hollow symbols correspond to the luminosity class I stars.  \label{sptteff}}   
\end{figure*}
}

In contrast to this method, the effective temperatures from Table \ref{t:modparams} depend on reproducing the ionization balance in the nebula that the star ionizes.  The points in Figure \ref{sptteff} show the effective temperatures that we generate as a function of the SpTs from Table \ref{t:obsprops}.  For comparison, the lines show calibrations from the literature for dwarf stars.  The typical error on the literature calibrations is $\sim 1000$\ K \citep{b:Martins_aap05}.  Figure \ref{sptteff} does not show MCELS~L~346 since the SpT of the ionizing star is uncertain.  MCELS~L~394, for which the stellar spectrum is noisy, is placed at SpT = O9.5 with a large error bar.  Based on the nebular spectrum, however, this object is probably at the upper end of the shown error bar.  

We find that, in general, WM-basic, SNC02, and TLUSTY predict similar \Teff, while CoStar predicts cooler \Teff.  This trend is consistent with our findings in \S \ref{s:q0}; the CoStar SEDs are harder than the other atmosphere models, and therefore require a lower \Teff\ to reproduce the observed emission-line spectra.  For the earliest stars, the best fitting \Teff\ selected for TLUSTY is slightly hotter than that of WM-basic and SNC02.  At late SpT we see a trend toward the opposite.  Since the treatment of the winds is the biggest difference between the TLUSTY and WM-basic atmospheres, this may reflect changes in the wind properties toward late SpT stars.

We note that the two supergiants in our sample have \Teff\ between 24 kK and 31 kK. This large range spans, and is consistent with, the expected \Teff\ for Magellanic class I stars 25-32 kK \citep[e.g.,][]{b:Mokiem_aap07,b:Trundle_aap04}.

The effective temperatures presented in Figure \ref{sptteff} are consistent with the relations in the literature, within the errors. However, we note that the points appear to follow a shallower slope.  This is primarily caused by the stars of later spectral type. The O5.5 - O7 V stars follow the LMC calibration of \citet{b:Massey_apj05} (thick-dashed line), but the later spectral types prefer the hotter temperatures of the \citet{b:Vacca_apj96} calibration (dotted line).  

One contributing factor to the \Teff\ trend is the assumed log($g$).  We calculate an expected variation in log($g$) for dwarf stars across our spectral range of 0.05 dex, based on the modeled stellar parameters from \citet{b:Schaerer_aap93}. If there is an error in our spectroscopically determined luminosity class, it will affect the best fitting \Teff\ selected.  Specifically, if any of the stars classified as dwarfs are actually giants or supergiants, the \Teff\ will be 1000-3000 K cooler than the \Teff\ chosen for a dwarf star.  The difference between the log($g$) appropriate for dwarfs and supergiants increases for later SpT.  Therefore, changes to the SED caused by uncertainty in log($g$) will be strongest for stars of late spectral type \citep[e.g.,][]{b:Martins_aap05}.  We find 0-1000 K decrease in \Teff\ for MCELS~L~28, an O5.5 V, going from log($g$)=4.0 to log($g$)=3.0, but a 2000-3000 K decrease in \Teff\ for MCELS~L~20, an O8 V. 

Another consideration is the metallicity for which these SpT-\Teff\ calibrations are valid, since lower metallicity stars will have higher \Teff\ \citep{b:Massey_apj05,b:Mokiem_aap04}.  All of the literature calibrations, except for the thick-dashed line, are calibrations that use solar metallicity stars.  Thus, we expect most of these calibrations to lie at cooler \Teff\ than the points in Figure \ref{sptteff}.  While this is the case for the calibrations based on line blanketed models, which we see by comparing the thick-dashed line to the thin-dashed and solid lines, we note the scatter between the points is larger than the difference in \Teff\ between the LMC and the MW calibrations. Furthermore, the scatter is comparable to the difference between calibrations based on atmospheres with and without line blanketing.  

The conventional approach to obtaining the SpT-\Teff\ calibration is based on photospheric lines and is sensitive to the SED at the photosphere.  Our method, however, depends on the ionization balance of the nebula and is also sensitive to changes in the SED due to layers outside the photosphere.  Thus, SpT-\Teff\ relations derived from nebular data have the potential to probe factors in the upper layers of the atmosphere models that do not match the conditions in the ionizing stars.  Our sample is small, and as seen earlier, there are a variety of uncertainties, limiting its usefulness in this regard, but with larger samples and better data, this approach promises an alternative and important way to refine SpT-\Teff\ calibrations.

\section{Conclusions} \label{s:conc}

Diagnostics from \hii\ region emission-line spectra are used extensively to determine the physical conditions of the ISM of galaxies, as well as understand their stellar populations and their chemical evolution.  The shape of the ionizing SED is one of the primary properties that determines the \hii\ region emission-line spectrum.  However, the intrinsic shapes of massive star SEDs are uncertain, and we depend on atmosphere models to describe them. It is therefore crucial to understand and quantify how well massive star atmosphere models represent the ionizing sources. 

In this study, we compare a sample of single-star \hii\ regions from the nearby LMC to photoionization simulations ionized by atmosphere models from widely-available CoStar, TLUSTY, and WM-basic atmosphere grids.  We select our sample of small, Str\"{o}mgren sphere-like \hii\ regions from narrowband MCELS images.  We obtain spectra and photometry of this sample using IMACS on the Magellan Baade Telescope at Las Campanas Observatory. We assign spectral types to the ionizing stars in our sample and find a range from O5.5 V to B0 V with 10 of our 12 \hii\ regions ionized by dwarf stars.  While we choose \hii\ regions that are likely to be ionized by a single star, contamination by OB companions cannot be entirely ruled out.  Two of the objects in our sample, MCELS~L~346 and MCELS~L~390, are confirmed eclipsing binaries.  MCELS~L~346 contains a nearly equal mass binary, and is modeled with two ionizing stars.  The rest of our sample show no obvious signs of OB companions, such as composite spectral types or large discrepancies between \Qha\ and \Qsed, and as discussed in \S \ref{s:compan} the binary status will not affect our results.

We evaluate the stellar atmosphere models by matching the emission-line spectra predicted by CLOUDY photoionization simulations with that observed in the nebulae.  Our nebular simulations show that uniform Str\"{o}mgren spheres cannot reproduce the observations. In these cases, \sii, \oii\ and \nii\ are under-predicted, which suggests that the ionization parameter is too high in these simulations. We explore the effects of changing \Rin\ and \emph{Z}, but find that neither parameter can resolve this issue.  Additionally, we find that \sii\ and \oi\ depend strongly on the nebular parameters, but only weakly on the particular SED used in the simulation.   In contrast to the uniform Str\"{o}mgren spheres, simulations that assume a clumpy medium reproduce the observed data.  The clumpy medium is described by specifying a filling factor of dense gas and assuming that the remaining volume is a vacuum. In general, we find that a filling factor of $\sim 0.10$\ produces simulations that agree with observations.  

We compare the rate of ionizing photons predicted by the best fitting atmosphere, assuming a constant log($g$), with the rate calculated from \emph{L}(\Halpha).  Within the errors due to uncertainty in log($g$), which can be as high as a factor of two, we find that the two values are consistent.   However, we do find a systematic offset between the rate of ionizing photons predicted by different atmosphere models.  TLUSTY produces the highest $Q_0$, while WM-basic and SNC02 have slightly lower rates.  In contrast, the $Q_0$ from CoStar is a factor of two lower than the other atmosphere models.  The hardness of the CoStar SED can explain this offset.  We select the best-fitting atmosphere models by matching the ionization balance in the nebula. Since the CoStar models have harder SEDs, the best-fitting CoStar atmospheres are a few thousand degrees cooler and produce fewer ionizing photons as a result.

We evaluate the stellar atmosphere models by comparing the predicted to observed emission line ratios.  For emission lines with IP $\leq 36$\ eV, we find that simulations ionized by WM-basic and SNC02 atmospheres, reproduce the observations.  Simulations using TLUSTY are also consistent, with the exception of \oii\ $\lambda3727$.   For this line, TLUSTY over-predicts the amount of \oii\ $\lambda3727$ in the nebula. Surprisingly, despite having the most approximate treatment of line blanketing, simulations using CoStar reproduce the emission lines up to [O III], at 35.1 eV, but they do so at lower \Teff\ than the other model atmospheres. 

At energies $> 36$\ eV, the predictions from different atmosphere models diverge from each other and the observations.  We find significant scatter between the predictions and observations of \neiii\ $\lambda3869$ from all atmosphere models.  CoStar, the hardest atmosphere over-predicts the \neiii\ line by more than a factor of two in most of our objects. In contrast, the other atmosphere models range from over- to under-predicting \neiii\ by 70\%.  Simulations ionized by TLUSTY, with its softer SED and plane-parallel geometry, systematically produce less \neiii\ than simulations using WM-basic and SNC02. 

Finally, we compare the SpT-\Teff\ calibrations in the literature with the best fitting \Teff\ from the simulations in this study. Our results for stars with SpT earlier than O8 V fall along the LMC calibration of \citet{b:Massey_apj05}, while stars later than O8 V seem to be hotter than the calibrations predict.   The SpT-\Teff\ calibrations from the literature are based on photospheric lines, while the effective temperatures obtained from this work depend the ionization balance of the nebulae.  The comparison of these two methods has the potential to reflect differences between the outer atmospheres of the models and the actual ionizing stars. However, there are a variety of uncertainties and our sample is small, which limits its usefulness in this regard.  Further study with a larger sample and detailed stellar modeling is needed to determine if the deviations at late SpT are significant.

This work uses single-star HII regions to test the predictions of stellar atmosphere models.  By using single-star Str\"{o}mgren spheres, we dramatically reduce the free parameters involved in evaluating atmosphere models with observed nebulae.  Future directions that could improve on the groundwork presented here include leveraging a larger range in \Teff\ by including stars earlier than O5.5 V, using a detailed fit to high resolution stellar spectra to generate the input model SEDs, and evaluating model atmospheres at different metallicities.

\acknowledgments

This work was funded by AST-0806476. We thank Andrew Graus and Joel Lamb for assistance with spectral typing and Zuzana Srostlik for help with the observations.  We are grateful to the U-M FANG research group and Lee Hartmann, Sylvain Veilleux, Timothy McKay, Lisa, Kewley, Gary Ferland, and Joachim Puls for helpful discussions while preparing this manuscript.  We thank the anonymous referee for useful comments that have improved this paper.


\clearpage


\begin{thebibliography}{87}
\expandafter\ifx\csname natexlab\endcsname\relax\def\natexlab#1{#1}\fi

\bibitem[{{Alcock} {et~al.}(1997){Alcock}, {Allsman}, {Alves}, {Axelrod},
  {Becker}, {Bennett}, {Cook}, {Freeman}, {Griest}, {Lacy}, {Lehner},
  {Marshall}, {Minniti}, {Peterson}, {Pratt}, {Quinn}, {Rodgers}, {Stubbs},
  {Sutherland}, \& {Welch}}]{b:Alcock_aj97}
{Alcock}, C., {Allsman}, R.~A., {Alves}, D., {et~al.} 1997, \aj, 114, 326

\bibitem[{{Baldwin} {et~al.}(1981){Baldwin}, {Phillips}, \&
  {Terlevich}}]{b:Baldwin_pasp81}
{Baldwin}, J.~A., {Phillips}, M.~M., \& {Terlevich}, R. 1981, \pasp, 93, 5

\bibitem[{{Benjamin} {et~al.}(2002){Benjamin}, {Skillman}, \&
  {Smits}}]{b:Benjamin_apj02}
{Benjamin}, R.~A., {Skillman}, E.~D., \& {Smits}, D.~P. 2002, \apj, 569, 288

\bibitem[{{Bresolin} {et~al.}(1999){Bresolin}, {Kennicutt}, \&
  {Garnett}}]{b:Bresolin_apj99}
{Bresolin}, F., {Kennicutt}, Jr., R.~C., \& {Garnett}, D.~R. 1999, \apj, 510,
  104

\bibitem[{{Cardelli} {et~al.}(1989){Cardelli}, {Clayton}, \&
  {Mathis}}]{b:Cardelli_apj89}
{Cardelli}, J.~A., {Clayton}, G.~C., \& {Mathis}, J.~S. 1989, \apj, 345, 245

\bibitem[{{Conti}(1973)}]{b:Conti_apj73}
{Conti}, P.~S. 1973, \apj, 179, 181

\bibitem[{{Copetti} {et~al.}(1986){Copetti}, {Pastoriza}, \&
  {Dottori}}]{b:Copetti_aap86}
{Copetti}, M.~V.~F., {Pastoriza}, M.~G., \& {Dottori}, H.~A. 1986, \aap, 156,
  111

\bibitem[{{Crowther} {et~al.}(2006){Crowther}, {Lennon}, \&
  {Walborn}}]{b:Crowther_aap06}
{Crowther}, P.~A., {Lennon}, D.~J., \& {Walborn}, N.~R. 2006, \aap, 446, 279

\bibitem[{{Crowther} {et~al.}(1999){Crowther}, {Pasquali}, {De Marco},
  {Schmutz}, {Hillier}, \& {de Koter}}]{b:Crowther_aap99}
{Crowther}, P.~A., {Pasquali}, A., {De Marco}, O., {et~al.} 1999, \aap, 350,
  1007

\bibitem[{{Davies} {et~al.}(1976){Davies}, {Elliott}, \&
  {Meaburn}}]{b:Davies_memras76}
{Davies}, R.~D., {Elliott}, K.~H., \& {Meaburn}, J. 1976, \memras, 81, 89

\bibitem[{{Dufour}(1975)}]{b:Dufour_apj75}
{Dufour}, R.~J. 1975, \apj, 195, 315

\bibitem[{{Edmunds} \& {Pagel}(1984)}]{b:Edmunds_mnras84}
{Edmunds}, M.~G., \& {Pagel}, B.~E.~J. 1984, \mnras, 211, 507

\bibitem[{{Ercolano} {et~al.}(2007){Ercolano}, {Bastian}, \&
  {Stasi{\'n}ska}}]{b:Ercolano_mnras07}
{Ercolano}, B., {Bastian}, N., \& {Stasi{\'n}ska}, G. 2007, \mnras, 379, 945

\bibitem[{{Esteban} {et~al.}(1993){Esteban}, {Smith}, {V'{i}lchez}, \&
  {Clegg}}]{b:Esteban_aap93}
{Esteban}, C., {Smith}, L.~J., {V'{i}lchez}, J.~M., \& {Clegg}, R.~E.~S. 1993,
  \aap, 272, 299

\bibitem[{{Ferland} {et~al.}(1998){Ferland}, {Korista}, {Verner}, {Ferguson},
  {Kingdon}, \& {Verner}}]{b:Ferland_pasp98}
{Ferland}, G.~J., {Korista}, K.~T., {Verner}, D.~A., {et~al.} 1998, \pasp, 110,
  761

\bibitem[{{Gabler} {et~al.}(1989){Gabler}, {Gabler}, {Kudritzki}, {Puls}, \&
  {Pauldrach}}]{b:Gabler_aap89}
{Gabler}, R., {Gabler}, A., {Kudritzki}, R.~P., {Puls}, J., \& {Pauldrach}, A.
  1989, \aap, 226, 162

\bibitem[{{Garnett}(1999)}]{b:Garnett_99}
{Garnett}, D.~R. 1999, in IAU Symposium, Vol. 190, New Views of the Magellanic
  Clouds, ed. {Y.-H.~Chu, N.~Suntzeff, J.~Hesser, \& D.~Bohlender}, 266

\bibitem[{{Giammanco} {et~al.}(2004){Giammanco}, {Beckman}, {Zurita}, \&
  {Rela{\~n}o}}]{b:Giammanco_aap04}
{Giammanco}, C., {Beckman}, J.~E., {Zurita}, A., \& {Rela{\~n}o}, M. 2004,
  \aap, 424, 877

\bibitem[{{Giveon} {et~al.}(2002){Giveon}, {Sternberg}, {Lutz}, {Feuchtgruber},
  \& {Pauldrach}}]{b:Giveon_apj02}
{Giveon}, U., {Sternberg}, A., {Lutz}, D., {Feuchtgruber}, H., \& {Pauldrach},
  A.~W.~A. 2002, \apj, 566, 880

\bibitem[{{Gonzalez-Delgado} {et~al.}(1994){Gonzalez-Delgado}, {Perez},
  {Tenorio-Tagle}, {Vilchez}, {Terlevich}, {Terlevich}, {Telles},
  {Rodriguez-Espinosa}, {Mas-Hesse}, {Garcia-Vargas}, {Diaz}, {Cepa}, \&
  {Castaneda}}]{b:Gonzalez-Delgado_apj94}
{Gonzalez-Delgado}, R.~M., {Perez}, E., {Tenorio-Tagle}, G., {et~al.} 1994,
  \apj, 437, 239

\bibitem[{{Gordon} {et~al.}(2003){Gordon}, {Clayton}, {Misselt}, {Landolt}, \&
  {Wolff}}]{b:Gordon_apj03}
{Gordon}, K.~D., {Clayton}, G.~C., {Misselt}, K.~A., {Landolt}, A.~U., \&
  {Wolff}, M.~J. 2003, \apj, 594, 279

\bibitem[{{Graczyk} {et~al.}(2011){Graczyk}, {Soszy{\'n}ski}, {Poleski},
  {Pietrzy{\'n}ski}, {Udalski}, {Szyma{\'n}ski}, {Kubiak}, {Wyrzykowski}, \&
  {Ulaczyk}}]{b:Graczyk_actaa11}
{Graczyk}, D., {Soszy{\'n}ski}, I., {Poleski}, R., {et~al.} 2011, AcA, 61,
  103

\bibitem[{{Hamuy} {et~al.}(1994){Hamuy}, {Suntzeff}, {Heathcote}, {Walker},
  {Gigoux}, \& {Phillips}}]{b:hamuy_pasp94}
{Hamuy}, M., {Suntzeff}, N.~B., {Heathcote}, S.~R., {et~al.} 1994, \pasp, 106,
  566

\bibitem[{{Henize}(1956)}]{b:Henize_apjs56}
{Henize}, K.~G. 1956, \apjs, 2, 315

\bibitem[{{Herrero} {et~al.}(1999){Herrero}, {Corral}, {Villamariz}, \&
  {Mart{\'{\i}}n}}]{b:Herrero_aap99}
{Herrero}, A., {Corral}, L.~J., {Villamariz}, M.~R., \& {Mart{\'{\i}}n}, E.~L.
  1999, \aap, 348, 542

\bibitem[{{Hubeny} \& {Lanz}(1995)}]{b:Hubeny_apj95}
{Hubeny}, I., \& {Lanz}, T. 1995, \apj, 439, 875

\bibitem[{{Hunt} \& {Hirashita}(2009)}]{b:Hunt_aap09}
{Hunt}, L.~K., \& {Hirashita}, H. 2009, \aap, 507, 1327

\bibitem[{{Hunter} \& {Massey}(1990)}]{b:Hunter_aj90}
{Hunter}, D.~A., \& {Massey}, P. 1990, \aj, 99, 846

\bibitem[{{Kaler}(1978)}]{b:Kaler_apj78}
{Kaler}, J.~B. 1978, \apj, 220, 887

\bibitem[{{Kennicutt}(1983)}]{b:Kennicutt_apj83}
{Kennicutt}, Jr., R.~C. 1983, \apj, 272, 54

\bibitem[{{Kennicutt}(1984)}]{b:Kennicutt_apj84}
---. 1984, \apj, 287, 116

\bibitem[{{Kennicutt} {et~al.}(2000){Kennicutt}, {Bresolin}, {French}, \&
  {Martin}}]{b:Kennicutt_apj00}
{Kennicutt}, Jr., R.~C., {Bresolin}, F., {French}, H., \& {Martin}, P. 2000,
  \apj, 537, 589

\bibitem[Kewley 
\& Dopita(2002)]{b:Kewley_ApJS02} Kewley, L.~J., \& Dopita, M.~A.\ 2002, \apjs, 142, 35 


\bibitem[{{Kiminki} \& {Kobulnicky}(2012)}]{b:Kiminki_apj12}
{Kiminki}, D.~C., \& {Kobulnicky}, H.~A. 2012, \apj, 751, 4

\bibitem[{{Kingdon} \& {Ferland}(1995)}]{b:Kingdon_apj95}
{Kingdon}, J.~B., \& {Ferland}, G.~J. 1995, \apj, 450, 691

\bibitem[{{Kudritzki} \& {Hummer}(1990)}]{b:Kudritzki_araa90}
{Kudritzki}, R.~P., \& {Hummer}, D.~G. 1990, \araa, 28, 303

\bibitem[{{Kudritzki} \& {Puls}(2000)}]{b:Kudritzki_araa00}
{Kudritzki}, R.-P., \& {Puls}, J. 2000, \araa, 38, 613

\bibitem[{{Lamers} {et~al.}(1995){Lamers}, {Snow}, \&
  {Lindholm}}]{b:Lamers_apj95}
{Lamers}, H.~J.~G.~L.~M., {Snow}, T.~P., \& {Lindholm}, D.~M. 1995, \apj, 455,
  269

\bibitem[{{Lanz} \& {Hubeny}(2003)}]{b:Lanz_apjs03}
{Lanz}, T., \& {Hubeny}, I. 2003, \apjs, 146, 417

\bibitem[{{Leitherer} {et~al.}(1999){Leitherer}, {Schaerer}, {Goldader},
  {Gonz{\'a}lez Delgado}, {Robert}, {Kune}, {de Mello}, {Devost}, \&
  {Heckman}}]{b:Leitherer_apjs99}
{Leitherer}, C., {Schaerer}, D., {Goldader}, J.~D., {et~al.} 1999, \apjs, 123,
  3

\bibitem[{{Liu} \& {Danziger}(1993)}]{b:Liu_mnras93}
{Liu}, X.-W., \& {Danziger}, J. 1993, \mnras, 263, 256

\bibitem[Mart{\'{\i}}n-Hern{\'a}ndez et 
al.(2002)]{b:Martin-Hernandez_aap02} Mart{\'{\i}}n-Hern{\'a}ndez, N.~L., Vermeij, R., Tielens, A.~G.~G.~M., van der Hulst, J.~M., \& Peeters, E.\ 2002, \aap, 389, 286 

\bibitem[{{Martins} {et~al.}(2005){Martins}, {Schaerer}, \&
  {Hillier}}]{b:Martins_aap05}
{Martins}, F., {Schaerer}, D., \& {Hillier}, D.~J. 2005, \aap, 436, 1049

\bibitem[{{Massey}(2002)}]{b:Massey_apjs02}
{Massey}, P. 2002, \apjs, 141, 81

\bibitem[{{Massey} {et~al.}(2005){Massey}, {Puls}, {Pauldrach}, {Bresolin},
  {Kudritzki}, \& {Simon}}]{b:Massey_apj05}
{Massey}, P., {Puls}, J., {Pauldrach}, A.~W.~A., {et~al.} 2005, \apj, 627, 477

\bibitem[{{Mathis} {et~al.}(1985){Mathis}, {Chu}, \&
  {Peterson}}]{b:Mathis_apj85}
{Mathis}, J.~S., {Chu}, Y.-H., \& {Peterson}, D.~E. 1985, \apj, 292, 155

\bibitem[{{McGaugh}(1991)}]{b:Mcgaugh_apj91}
{McGaugh}, S.~S. 1991, \apj, 380, 140

\bibitem[{{Mokiem} {et~al.}(2004){Mokiem}, {Mart{\'{\i}}n-Hern{\'a}ndez},
  {Lenorzer}, {de Koter}, \& {Tielens}}]{b:Mokiem_aap04}
{Mokiem}, M.~R., {Mart{\'{\i}}n-Hern{\'a}ndez}, N.~L., {Lenorzer}, A., {de
  Koter}, A., \& {Tielens}, A.~G.~G.~M. 2004, \aap, 419, 319

\bibitem[{{Mokiem} {et~al.}(2007){Mokiem}, {de Koter}, {Evans}, {Puls},
  {Smartt}, {Crowther}, {Herrero}, {Langer}, {Lennon}, {Najarro}, {Villamariz},
  \& {Vink}}]{b:Mokiem_aap07}
{Mokiem}, M.~R., {de Koter}, A., {Evans}, C.~J., {et~al.} 2007, \aap, 465, 1003

\bibitem[Morisset et al.(2002)]{b:Morisset_aap02} Morisset, C., Schaerer, D., Mart{\'{\i}}n-Hern{\'a}ndez, N.~L., et al.\ 2002, \aap, 386, 558

\bibitem[{{Morisset} {et~al.}(2004){Morisset}, {Schaerer}, {Bouret}, \&
  {Martins}}]{b:Morisset_aap04}
{Morisset}, C., {Schaerer}, D., {Bouret}, J.-C., \& {Martins}, F. 2004, \aap,
  415, 577

\bibitem[{{Oey} {et~al.}(2000){Oey}, {Dopita}, {Shields}, \&
  {Smith}}]{b:Oey_apjs00}
{Oey}, M.~S., {Dopita}, M.~A., {Shields}, J.~C., \& {Smith}, R.~C. 2000, \apjs,
  128, 511

\bibitem[{{Osterbrock} \& {Flather}(1959)}]{b:Osterbrock_apj59}
{Osterbrock}, D., \& {Flather}, E. 1959, \apj, 129, 26

\bibitem[{{Osterbrock} \& {Ferland}(2006)}]{b:Osterbrock_06}
{Osterbrock}, D.~E., \& {Ferland}, G.~J. 2006, {Astrophysics of gaseous nebulae
  and active galactic nuclei}

\bibitem[{{Pauldrach} {et~al.}(2001){Pauldrach}, {Hoffmann}, \&
  {Lennon}}]{b:Pauldrach_aap01}
{Pauldrach}, A.~W.~A., {Hoffmann}, T.~L., \& {Lennon}, M. 2001, \aap, 375, 161

\bibitem[{{Pauldrach} {et~al.}(1998){Pauldrach}, {Lennon}, {Hoffmann},
  {Sellmaier}, {Kudritzki}, \& {Puls}}]{b:Pauldrach_98}
{Pauldrach}, A.~W.~A., {Lennon}, M., {Hoffmann}, T.~L., {et~al.} 1998, in
  Astronomical Society of the Pacific Conference Series, Vol. 131, Properties
  of Hot Luminous Stars, ed. I.~{Howarth}, 258

\bibitem[{{Peimbert}(1967)}]{b:Peimbert_apj67}
{Peimbert}, M. 1967, \apj, 150, 825

\bibitem[{{Pellegrini} {et~al.}(2011){Pellegrini}, {Baldwin}, \&
  {Ferland}}]{b:Pellegrini_apj11}
{Pellegrini}, E.~W., {Baldwin}, J.~A., \& {Ferland}, G.~J. 2011, \apj, 738, 34

\bibitem[{{Pellegrini} {et~al.}(2012){Pellegrini}, {Oey}, {Winkler}, {Points},
  {Smith}, {Jaskot}, \& {Zastrow}}]{b:Pellegrini_apj12}
{Pellegrini}, E.~W., {Oey}, M.~S., {Winkler}, P.~F., {et~al.} 2012, \apj

\bibitem[{{P{\'e}rez-Montero} \& {D{\'{\i}}az}(2005)}]{b:Perez-Montero_mnras05}
{P{\'e}rez-Montero}, E., \& {D{\'{\i}}az}, A.~I. 2005, \mnras, 361, 1063

\bibitem[{{Prinja} {et~al.}(1990){Prinja}, {Barlow}, \&
  {Howarth}}]{b:Prinja_apj90}
{Prinja}, R.~K., {Barlow}, M.~J., \& {Howarth}, I.~D. 1990, \apj, 361, 607

\bibitem[{{Rigby} \& {Rieke}(2004)}]{b:Rigby_apj04}
{Rigby}, J.~R., \& {Rieke}, G.~H. 2004, \apj, 606, 237

\bibitem[{{Rubin}(1989)}]{b:Rubin_apjs89}
{Rubin}, R.~H. 1989, \apjs, 69, 897

\bibitem[{{Rubin} {et~al.}(1991){Rubin}, {Simpson}, {Haas}, \&
  {Erickson}}]{b:Rubin_apj91}
{Rubin}, R.~H., {Simpson}, J.~P., {Haas}, M.~R., \& {Erickson}, E.~F. 1991,
  \apj, 374, 564

\bibitem[{{Sana} {et~al.}(2009){Sana}, {Gosset}, \& {Evans}}]{b:Sana_mnras09}
{Sana}, H., {Gosset}, E., \& {Evans}, C.~J. 2009, \mnras, 400, 1479

\bibitem[{{Sana} {et~al.}(2011){Sana}, {James}, \& {Gosset}}]{b:Sana_mnras11}
{Sana}, H., {James}, G., \& {Gosset}, E. 2011, \mnras, 416, 817

\bibitem[{{Schaerer} \& {de Koter}(1997)}]{b:Schaerer_aap97}
{Schaerer}, D., \& {de Koter}, A. 1997, \aap, 322, 598

\bibitem[{{Schaerer} \& {Schmutz}(1994)}]{b:Schaerer_aap94}
{Schaerer}, D., \& {Schmutz}, W. 1994, \aap, 288, 231

\bibitem[Schaerer et al.(1993)]{b:Schaerer_aap93} Schaerer, D., Meynet, G., Maeder, A., \& Schaller, G.\ 1993, \aaps, 98, 523 

\bibitem[{{Sellmaier} {et~al.}(1996){Sellmaier}, {Yamamoto}, {Pauldrach}, \&
  {Rubin}}]{b:Sellmaier_aap96}
{Sellmaier}, F.~H., {Yamamoto}, T., {Pauldrach}, A.~W.~A., \& {Rubin}, R.~H.
  1996, \aap, 305, L37

\bibitem[{{Sim{\'o}n-D{\'{\i}}az} \&
  {Stasi{\'n}ska}(2008)}]{b:Simon-Diaz_mnras08}
{Sim{\'o}n-D{\'{\i}}az}, S., \& {Stasi{\'n}ska}, G. 2008, \mnras, 389, 1009

\bibitem[{{Simpson} {et~al.}(1995){Simpson}, {Colgan}, {Rubin}, {Erickson}, \&
  {Haas}}]{b:Simpson_apj95}
{Simpson}, J.~P., {Colgan}, S.~W.~J., {Rubin}, R.~H., {Erickson}, E.~F., \&
  {Haas}, M.~R. 1995, \apj, 444, 721

\bibitem[{{Simpson} {et~al.}(1986){Simpson}, {Rubin}, {Erickson}, \&
  {Haas}}]{b:Simpson_apj86}
{Simpson}, J.~P., {Rubin}, R.~H., {Erickson}, E.~F., \& {Haas}, M.~R. 1986,
  \apj, 311, 895

\bibitem[{{Smith} {et~al.}(2002){Smith}, {Norris}, \&
  {Crowther}}]{b:Smith_mnras02}
{Smith}, L.~J., {Norris}, R.~P.~F., \& {Crowther}, P.~A. 2002, \mnras, 337,
  1309 (SNC02)

\bibitem[{{Smith} {et~al.}(2005){Smith}, {Points}, {Chu}, {Winkler},
  {Aguilera}, {Leiton}, \& {MCELS Team}}]{b:Smith_05}
{Smith}, R.~C., {Points}, S.~D., {Chu}, Y.-H., {et~al.} 2005, in Bulletin of
  the American Astronomical Society, Vol.~37, American Astronomical Society
  Meeting Abstracts, 1200

\bibitem[{{Stasi{\'n}ska} \& {Leitherer}(1996)}]{b:Stasinska_apjs96}
{Stasi{\'n}ska}, G., \& {Leitherer}, C. 1996, \apjs, 107, 661

\bibitem[{{Stasi{\'n}ska} \& {Schaerer}(1997)}]{b:Stasinska_aap97}
{Stasi{\'n}ska}, G., \& {Schaerer}, D. 1997, \aap, 322, 615

\bibitem[{{Stoy}(1933)}]{b:Stoy_mnras33}
{Stoy}, R.~H. 1933, \mnras, 93, 588

\bibitem[{{Trundle} {et~al.}(2004){Trundle}, {Lennon}, {Puls}, \&
  {Dufton}}]{b:Trundle_aap04}
{Trundle}, C., {Lennon}, D.~J., {Puls}, J., \& {Dufton}, P.~L. 2004, \aap, 417,
  217

\bibitem[{{Tsamis} \& {P{\'e}quignot}(2005)}]{b:Tsamis_mnras05}
{Tsamis}, Y.~G., \& {P{\'e}quignot}, D. 2005, \mnras, 364, 687

\bibitem[{{Udalski} {et~al.}(2008){Udalski}, {Soszynski}, {Szymanski},
  {Kubiak}, {Pietrzynski}, {Wyrzykowski}, {Szewczyk}, {Ulaczyk}, \&
  {Poleski}}]{b:Udalski_actaa08}
{Udalski}, A., {Soszynski}, I., {Szymanski}, M.~K., {et~al.} 2008, AcA, 58,
  89

\bibitem[{{Vacca} {et~al.}(1996){Vacca}, {Garmany}, \& {Shull}}]{b:Vacca_apj96}
{Vacca}, W.~D., {Garmany}, C.~D., \& {Shull}, J.~M. 1996, \apj, 460, 914

\bibitem[{{van Hoof} {et~al.}(2004){van Hoof}, {Weingartner}, {Martin}, {Volk},
  \& {Ferland}}]{b:van-Hoof_04}
{van Hoof}, P.~A.~M., {Weingartner}, J.~C., {Martin}, P.~G., {Volk}, K., \&
  {Ferland}, G.~J. 2004, in Astronomical Society of the Pacific Conference
  Series, Vol. 313, Asymmetrical Planetary Nebulae III: Winds, Structure and
  the Thunderbird, ed. M.~{Meixner}, J.~H. {Kastner}, B.~{Balick}, \&
  N.~{Soker}, 380

\bibitem[{{Viegas} \& {Clegg}(1994)}]{b:Viegas_mnras94}
{Viegas}, S.~M., \& {Clegg}, R.~E.~S. 1994, \mnras, 271, 993

\bibitem[{{V\'{i}lchez} \& {Pagel}(1988)}]{b:Vilchez_mnras88}
{V\'{i}lchez}, J.~M., \& {Pagel}, B.~E.~J. 1988, \mnras, 231, 257

\bibitem[{{Voges} {et~al.}(2008){Voges}, {Oey}, {Walterbos}, \&
  {Wilkinson}}]{b:Voges_aj08}
{Voges}, E.~S., {Oey}, M.~S., {Walterbos}, R.~A.~M., \& {Wilkinson}, T.~M.
  2008, \aj, 135, 1291

\bibitem[{{Walborn} \& {Fitzpatrick}(1990)}]{b:Walborn_pasp90}
{Walborn}, N.~R., \& {Fitzpatrick}, E.~L. 1990, \pasp, 102, 379

\bibitem[{{Weingartner} \& {Draine}(2001)}]{b:Weingartner_apj01}
{Weingartner}, J.~C., \& {Draine}, B.~T. 2001, \apj, 548, 296

\bibitem[{{Westerlund}(1997)}]{b:Westerlund_97}
{Westerlund}, B.~E. 1997, {The Magellanic Clouds}

\bibitem[{{Williams}(1992)}]{b:Williams_apj92}
{Williams}, R.~E. 1992, \apj, 392, 99

\bibitem[{{Zanstra}(1927)}]{b:Zanstra_apj27}
{Zanstra}, H. 1927, \apj, 65, 50

\end{thebibliography}
\end{document}